\documentclass{article}
\usepackage{arxiv}

\usepackage[utf8]{inputenc} % allow utf-8 input
\usepackage[T1]{fontenc}    % use 8-bit T1 fonts
\usepackage{hyperref}       % hyperlinks
\usepackage{url}            % simple URL typesetting
\usepackage{booktabs}       % professional-quality tables
\usepackage{amsfonts}       % blackboard math symbols
\usepackage{nicefrac}       % compact symbols for 1/2, etc.
\usepackage{microtype}      % microtypography
\usepackage{lipsum}		% Can be removed after putting your text content
\usepackage{graphicx}
\usepackage[numbers]{natbib}
\usepackage{doi}
\usepackage{amsmath}

\usepackage{algorithm}
\usepackage{algcompatible}
\usepackage{algpseudocode}
\usepackage[%
    prefixes-as-symbols=false,
    ]{siunitx}
\usepackage{tikz}
\usetikzlibrary{shapes,arrows}
\usetikzlibrary{decorations.pathreplacing}
\usetikzlibrary {arrows.meta}
\usetikzlibrary{calc}
\usepackage{pgffor}
\usepackage{tikz-3dplot}
\usepackage{listings}
\usetikzlibrary{decorations.pathreplacing}
\usepackage[braket, qm]{qcircuit}
\usepackage{listings}
\usepackage{cleveref}
\usepackage{lineno}
\usepackage{xcolor}
\usepackage{xspace}
\usepackage{upgreek}
\usepackage{subcaption}
\usepackage{svg}
\usepackage{multirow}
\usepackage{ourcommands}
\usepackage{ifthen}
\usepackage[colorinlistoftodos,prependcaption,textsize=tiny]{todonotes}

\usepackage{pifont}

\title{Fully Quantum Lattice Gas Automata Building Blocks for Computational Basis State Encodings}

\author{ \href{https://orcid.org/0000-0002-8102-6389}{\includegraphics[scale=0.06]{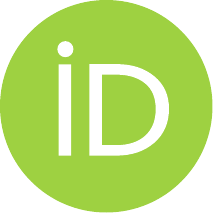}\hspace{1mm}C\u{a}lin A.~Georgescu}\\
	Delft University of Technology\\
	Mekelweg 4, 2628CD, Delft\\
	\texttt{c.a.georgescu@tudelft.nl} \\
	\And
    \href{https://orcid.org/0000-0001-7751-9060}{\includegraphics[scale=0.06]{icons/orcid.pdf}\hspace{1mm}Merel A.~Schalkers}\\
	Delft University of Technology\\
	Mekelweg 4, 2628CD, Delft\\
	\texttt{m.a.schalkers@tudelft.nl} \\
	\And
    \href{https://orcid.org/0000-0003-0802-945X}{\includegraphics[scale=0.06]{icons/orcid.pdf}\hspace{1mm}Matthias M\"{o}ller}\\
	Delft University of Technology\\
	Mekelweg 4, 2628CD, Delft\\
	\texttt{m.moller@tudelft.nl} \\
}

\renewcommand{\shorttitle}{Fully Quantum Lattice Gas Automata Building Blocks}

\hypersetup{
pdftitle={Fully Quantum Lattice Gas Automata Building Blocks},
pdfsubject={quant-ph},
pdfauthor={C\u{a}lin A.~Georgescu, Merel A.~Schalkers, Matthias M\"{o}ller},
pdfkeywords={Quantum computing, Lattice Boltzmann Method,
Quantum software},
}

\begin{document}
\maketitle
\begin{abstract}
Lattice Gas Automata (LGA) is a classical method for simulating physical phenomena,
including Computational Fluid Dynamics (CFD).
Quantum LGA (QLGA) is the family of methods that implement
LGA schemes on quantum computers.
In recent years, QLGA has garnered attention from researchers thanks to its potential
of efficiently modeling CFD processes by either reducing memory requirements
or providing simultaneous representations of exponentially many LGA states.
In this work, we introduce novel building blocks for QLGA algorithms
that rely on computational basis state encodings.
We address every step of the algorithm, from initial conditions to measurement,
and provide detailed complexity analyses that account for all discretization
choices of the system under simulation.
We introduce multiple ways of instantiating initial conditions,
efficient boundary condition implementations
for novel geometrical patterns, a novel collision operator that models
less restricted interactions than previous implementations,
and quantum circuits that extract quantities of interest out of the quantum state.
For each building block, we provide intuitive examples and open-source implementations
of the underlying quantum circuits.
\end{abstract}

\keywords{{Quantum Computing \and Lattice Gas Automata \and Computational Fluid Dynamics}}

% \linenumbers
\section{Introduction \label{sec:sp-1-1-intro}}

Computational Fluid Dynamics (CFD) has become an indispensable tool
across a broad spectrum of disciplines and industries.
From civil engineering \cite{wijesooriya2023technical}
to aerospace applications \cite{mani2023perspective},
and medicine \cite{zingaro2024electromechanics},
CFD enables the analysis of complex fluid systems
for which theory is intractable and experimentation is impractical.
With such a broad range of applications,
the capabilities that practitioners demand of CFD technologies
have increased significantly.
For decades, the capacity of CFD methods
has grown in tandem with the scale of the available hardware,
leading to a surge of practical use cases.
However, even at the scale of today's hardware capabilities,
many CFD applications that would benefit countless industries remain out of reach
due to sheer computational demands.
Motivated by these practical limitations,
and further fueled by growing concerns about the decreased rate
of computational hardware advancement \citep{theis2017end},
researchers began investigating alternatives to classical CFD methods.
One emerging and promising alternative that has received increasing
attention in recent years is quantum computing.

Quantum Computing (QC) \cite{nielsen2010quantum} relies on
the exploitation of quantum mechanical phenomena to
manipulate basic units of information called quantum bits (qubits).
The appeal of QC stems from two fundamental properties of
quantum information -- superposition and entanglement.
Unlike classical bits, a register of $n$ qubits can
carry a superposition of  $2^n$ basis states,
which can be processed simultaneously in a process known as quantum parallelism.
This allows for certain kinds of classical information to be compressed
into logarithmically many qubits,
which together with entanglement
may enable significant computational speedups compared to classical algorithms.
These advantages of QC have drawn the attention of CFD researchers,
leading to the creation of Quantum CFD (QCFD) methods.

The still nascent QCFD research field has branched out into several directions.
Variational Quantum Algorithms (VQAs) cast the equations of fluid dynamics
onto an optimization framework with the aim of classically
optimizing shallow parameterized
quantum circuits suitable for near-term noisy quantum devices.
Such approaches have proven successful in targeting
the advection-diffusion \cite{demirdjian2022variational} equation,
the quasi-1D Navier-Stokes equations \cite{kyriienko2021solving},
and the Poisson equation \cite{sato2021variational}, among others.
Recently, quantum realizations of physics-informed neural networks
\cite{raissi2019physics} such as \cite{berger2025trainable} and \cite{panichi2025quantum}
have also shown promising results for different partial differential equations.
Though less demanding on the quantum hardware, such approaches incur
the drawback of quantum-classical communication in the parameter optimization pipeline.
Further hindrances of variational methods include costly function evaluations
and the barren pleateau problem \cite{mcclean2018barren}.

A different branch of QCFD research pursues the application of Quantum Linear Solvers (QLSs).
QLS algorithms evolve quantum states that encode the solutions of
linear algebraic systems of equations, as first introduced by \citet{harrow2009quantum}
through a routine that has become known as HHL.
Extensions of the HHL algorithm have been applied to develop
quantum implementations
of the Finite Element Method \cite{montanaro2016quantum}
and the Finite Volume Method \cite{chen2022quantum},
as well as to solve the Poisson equation \cite{cao2013quantum}.
The greatest appeal of QLS-based algorithms is their exponentially
lower computational time with which the solution state can be prepared.
However, three obstacles that QLSs face include
(i) preparing the necessary starting state to encode
the appropriate system of equations (the input problem),
(ii) extracting the information out of the prepared quantum state (the output problem),
and (iii)  the high dependence of the algorithm on the
structure of the matrix encoding the system \cite{harrow2009quantum, aaronson2015read}.
These remain open challenges in the pursuit of practical quantum advantage.

In addition to the development of VQA- and QLS-based solvers,
QCFD research has branched out into a third, 
separate direction -- lattice-based discrete velocity methods.
Classically, this family of methods includes Lattice Gas Automata (LGA) \cite{wolf2004lattice}
and the Lattice Boltzmann Method (LBM) \cite{kruger2017lattice}.
The foundation of these CFD methods lies in the kinetic theory of
gases, and their implementations follow different principles than the two other branches of QCFD.
Both LGA and the LBM follow a time-marching procedure that tracks the evolution of a system
of fictitious particles through phase space.
Both LGA and LBM consist of a repeating
routine of two cornerstone steps.
First, particles travel across discrete \emph{velocity channels}
in a step called \emph{streaming}, followed by
inter-particle interactions, called \emph{collision} or \emph{scattering}.
Two features of these steps that make LGA and the LBM promising
candidates for QC implementations are the linearity of streaming and the locality of collision.
The former allows for straightforward implementations
of particle transport steps in several quantum encodings,
while the latter makes it easy to take advantage of quantum parallelism,
since collision operators can be applied simultaneously throughout the lattice.
Where the LGA and the LBM fundamentally differ is in their model of particles.
LGA particles are typically boolean and follow an exclusionary principle, that is, 
at most one particle can occupy a velocity channel at a time.
In contrast, the LBM operates at a larger, mesoscopic scale, which
does not track individual particle trajectories
but rather the behavior of \emph{populations} of particles
interpreted through a global probability distribution.
Historically, the LBM has superseded LGA as a fluid dynamics simulation tool, in large part due
to the latter's susceptibility to statistical noise.
Replacing a discrete number of boolean particles with a smooth distribution function
innately circumvents the requirement of stochastic ensemble averaging that LGA 
algorithms suffer from.
For this reason, much recent research has focused on porting LBM
primitives to the quantum computing paradigm.
In doing so, Quantum LBM (QLBM) research has itself
taken three different directions.

The first direction is based on the
Linear Combinations of Unitaries (LCU)  \cite{childs2012hamiltonian} paradigm,
in which complex operators are expressed as sums of unitary operators controlled
on the state of ancillary qubits.
These approaches include the work of \citet{budinski2021quantum, ljubomir2022quantum},
 \citet{wawrzyniak2024quantum}, and \citet{tiwari2025algorithmic}.
 Though these methods allow the lattice information to be encoded in logarithmically many 
 qubits, this compression comes at the cost of an inexact time-evolution operator.
 Due to this inexactness, each time step of the algorithm introduces
states that are orthogonal to the representation of the physical system, an, in turn,
significantly increase the number of measurements
 required to extract meaningful information out of the quantum state.
 In practice, applying these methods requires
 a procedure resembling quantum state tomography
 \cite{xu2025improved} to implement rejection sampling,
 as well as re-initialization of the quantum
 state after each time step \cite{wawrzyniak2024quantum, xu2025improved}.
 Expressing the unitary matrices that perform collision and reinitialization
 into hardware-native gates introduces further substantial classical overhead.
 
The second direction of QLBM research has focused on implementing the separate QLBM
streaming and collision steps, without the need for the inexact LCU decomposition.
This includes the works of \citet{todorova2020quantum}, \citet{budinski2023efficient},
and \citet{schalkers2024efficient}, which
detail the implementation of the streaming step,
while \citet{steijl2020quantum} and \citet{moawad2022investigating} implement the nonlinear collision
step by means of quantum arithmetic.
However, \citet{schalkers2024importance} and \citet{fonio2023quantum},
have shown that the quantum encodings previously
used to implement streaming cannot model collision,
and vice-versa.

The third direction of QLBM research has been  studying applications
where the governing Lattice Boltzmann Equation (LBE)
either does not necessitate any nonlinear terms in the computation of the
collision step, or can be practically linearized.
The former includes the works of \citet{xu2025improved} and \citet{wawrzyniak2025dynamic},
who design circuits for the application of the QLBM to the advection-diffusion equation.
The linearization approach by means of Carleman linearization was studied by Itani et al.
\citep{itani2022analysis, itani2023qalb}
and Sanavio et al. \cite{sanavio2024lattice, sanavio2025carleman},
while \citet{kumar2024decomposition}
propose a linear decomposition of the collision operator matrix.
To the best of our knowledge, however, expressing the linearized version of the 
LBE in a manner that is efficiently implementable
through the one- and two-qubit gates that are native to quantum hardware
while targeting nonlinear flow regimes
remains an open challenge.

In light of the challenges facing QLBM development, researchers have simultaneously
turned their attention to the several variations of algorithms within the LGA umbrella.
In what follows, we introduce the LGA algorithm and the features that make
it a prime candidate for quantum implementations, as well as 
the current state of the art and the challenges we address in this work.

\subsection{Classical and Quantum Lattice Gas Automata \label{subsec:sp-1-2-intro-lga}}

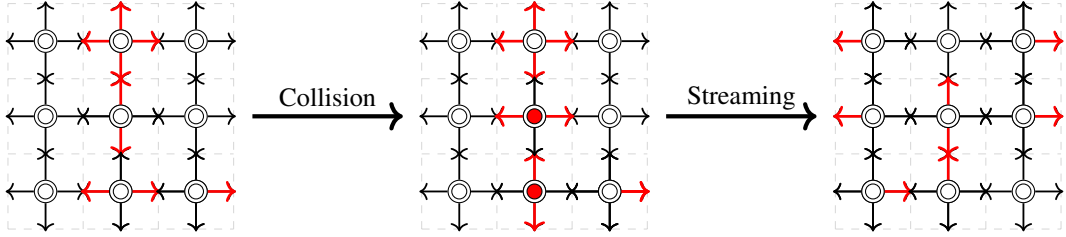
\begin{figure}
	\centering
\begin{tikzpicture}[scale=0.25,shift={(8,0)}]
	% Grid lines
	\draw[
	help lines,
	line width=0.4pt,
	color=gray!30,
	dashed
	] (-8, -8) grid[step={($(4, 4) - (0, 0)$)}] (4, 4);

	\foreach \x in {-6,-2,2} {
		\foreach \y in {-6,-2,2} {
			\dtwoqfour{(\x,\y)}{0.1cm}{2}{0}{1.5}{white}
		}
	}
	
	\dtwoqfourcolor{(-2,2)}{0.1cm}{2}{0}{1.5}{white}{red}{red}{red}{red}
	\dtwoqfourcolor{(-2,-2)}{0.1cm}{2}{0}{1.5}{white}{black}{red}{black}{red}
	\dtwoqfourcolor{(-2,-6)}{0.1cm}{2}{0}{1.5}{white}{red}{black}{red}{black}
	\dtwoqfourcolor{(2,-6)}{0.1cm}{2}{0}{1.5}{white}{red}{black}{black}{black}
	
	\draw[->, ultra thick, black] (5, -2) -- (13, -2) node at (9, -1) {Collision};
	
	% Picture 2
	\begin{scope}[shift={(22,0)}]
		
		% Grid lines
		\draw[
		help lines,
		line width=0.4pt,
		color=gray!30,
		dashed
		] (-8, -8) grid[step={($(4, 4) - (0, 0)$)}] (4, 4);

		\foreach \x in {-6,-2,2} {
			\foreach \y in {-6,-2,2} {
				\dtwoqfour{(\x,\y)}{0.1cm}{2}{0}{1.5}{white}
			}
		}
		
	\dtwoqfourcolor{(-2,2)}{0.1cm}{2}{0}{1.5}{white}{red}{red}{red}{red}
	\dtwoqfourcolor{(-2,-6)}{0.1cm}{2}{0}{1.5}{red}{black}{red}{black}{red}
	\dtwoqfourcolor{(-2,-2)}{0.1cm}{2}{0}{1.5}{red}{red}{black}{red}{black}
	\dtwoqfourcolor{(2,-6)}{0.1cm}{2}{0}{1.5}{white}{red}{black}{black}{black}
	
	\dtwoqfourcolor{(2,-6)}{0.1cm}{2}{0}{1.5}{white}{red}{black}{black}{black}
		
		\draw[->, ultra thick, black] (5, -2) -- (13, -2) node at (9, -1) {Streaming};
	\end{scope}

	% Picture 3
	\begin{scope}[shift={(44,0)}]
		
		% Grid lines
		\draw[
		help lines,
		line width=0.4pt,
		color=gray!30,
		dashed
		] (-8, -8) grid[step={($(4, 4) - (0, 0)$)}] (4, 4);

		\foreach \x in {-6,-2,2} {
			\foreach \y in {-6,-2,2} {
				\dtwoqfour{(\x,\y)}{0.1cm}{2}{0}{1.5}{white}
			}
		}
	\dtwoqfourcolor{(-6,2)}{0.1cm}{2}{0}{1.5}{white}{black}{black}{red}{black}
	\dtwoqfourcolor{(2,2)}{0.1cm}{2}{0}{1.5}{white}{red}{black}{black}{black}
	\dtwoqfourcolor{(-6,-2)}{0.1cm}{2}{0}{1.5}{white}{black}{black}{red}{black}
	\dtwoqfourcolor{(-2,-2)}{0.1cm}{2}{0}{1.5}{white}{black}{red}{black}{red}
	\dtwoqfourcolor{(2,-2)}{0.1cm}{2}{0}{1.5}{white}{red}{black}{black}{black}
	\dtwoqfourcolor{(-6,-6)}{0.1cm}{2}{0}{1.5}{white}{red}{black}{black}{black}
	\dtwoqfourcolor{(-2,-6)}{0.1cm}{2}{0}{1.5}{white}{black}{red}{black}{black}

	\end{scope}
\end{tikzpicture}
	\caption{Overview of the LGA collision and streaming steps.\label{fig:sp-1-0-lga-stream-collide}}
\end{figure}

The first theory of Cellular Automata (CA) traces back to the seminal work
of John von Neumann and Arthur Burks around the
middle of the $20^{\text{th}}$ century \cite{v1966theory}.
CA are models of physical systems in which space is discretized
into cells that have a finite number of states, and which evolve and
interact with one another according to a set of rules in discrete time.
Despite their apparent simplicity, CA are a foundational model of computation
that can accurately model numerous physical processes \cite{chopard2012cellular}.
For the purposes of this work, we use the term Lattice Gas Automata
to refer to the subset of CA that model hydrodynamical processes.
LGA models have been successfully applied to
core problems in hydrodynamics, including the Navier-Stokes equations,
as well as related applications to magneto-hydrodynamics \cite{benzi1992lattice, boghosian1999lattice}.
Two of the best-known variations of LGA have become known under the 
HPP \cite{hardy1973time, hardy1976molecular} and
FHP \cite{frisch1986lattice} acronyms, and model two-dimensional lattices.
\Cref{fig:sp-1-0-lga-stream-collide} shows the time evolution
of an LGA system that adheres to HPP model, which prescribes 
$4$ velocity channels in a two-dimensional lattice.
Velocity channels that are occupied are highlighted in red,
whereas empty channels are black.
The LGA loop consists of collision and streaming.
Collision locally redistributes the occupancy of velocity channels for each lattice site,
such that mass and momentum are conserved.
Gridpoints where this collision occurs and is non-trivial
are highlighted in red in the middle panel of \Cref{fig:sp-1-0-lga-stream-collide} .
Streaming simply propagates the occupancy state of each channel
to the neighboring gridpoints, according to the velocity discretization.

The first rough link between CA and quantum computing can be traced back to
Richard Feynman's idea of utilizing quantum computers for the purpose of simulating
physical processes in 1982 \cite{feynman1982simulating}.
It was not until the beginning of the following decade that the foundation of
Quantum Lattice Gas Automata (QLGA)
was established in a series of articles by
Meyer
\cite{meyer1996quantum, meyer1996unitarity, meyer1997quantuma, meyer1997quantumb, meyer1998quantum} and Boghosian \cite{boghosian1997quantum, boghosian1998simulating},
while Succi \cite{succi1993lattice, succi1996numerical}
independently established the link between quantum mechanics and kinetic theory.\footnote{Quantum Cellular Automata (QCA) and QLGA  are distinct classes of algorithms \cite{shakeel2013quantum}. For a review of QCA, research, we refer the reader to the review of \citet{farrelly2020review}.}
Shortly following this initial wave of research, a series of works by Yepez
introduced QLGA implementations for distributed
quantum computer networks \cite{yepez1998quantum, yepez2001quantumdiffusion},
showing their link to the LBM \cite{yepez2001quantum},
as well as  their suitability for solving the
Burgers equation \cite{yepez2002quantum} and the many-body
Schr\"odinger equation \cite{yepez2002efficient}.
Following this wave of research around the turn of the millennium, 
QLGA research remained stagnant for nearly two decades.

More recently, the interest in QLGA has surged in tandem with its QLBM counterpart.
Unlike the LBM, the boolean nature of LGA particles and the exclusionary principle
of velocity channel occupancy make linear operations sufficient to express the evolution
of the model.
In addition, the ensemble averaging that was classically conducted
to combat the statistical noise inherent to coarse LGA simulations
can be naturally mitigated through the superposition
that quantum registers afford.
Recent developments of in the QLGA space include the work of \citet{love2019quantum},
who introduced quantum circuits for streaming and collision
for a two-dimensional lattice discretization, and showed the
invariants that emerge as a consequence of the quantum encoding.
The encoding used by \citet{love2019quantum} had already been established
more than two decades earlier by Boghosian
\cite{boghosian1997quantum, boghosian1998simulating}, who outlined
that QLGA can model the Schr\"{o}dinger equation through
propagation and collision rules.
\citet{fonio2023quantum} provide additional quantum circuits for collision
and measurement in an encoding that shares the same foundational 
principles as that of Yepez \cite{yepez2001quantumdiffusion}.
\citet{schalkers2024importance} introduce the novel Space-Time
data encoding and include circuits for streaming and superposed collision
for a two-dimensional lattice.
\citet{kocherla2024fully} extend the encoding of \citet{love2019quantum}
with a phase estimation protocol to reduce the measurement overhead.
\citet{zamora2025efficient} adapt the LGA loop to an encoding that
affords exponential memory compression of the grid at the cost
of requiring measurement and reinitialization after every time step.
\citet{zamora2025float} develop a quantum variation of Integer LGA (ILGA) 
developed by \citet{blommel2018integer} to model fractional collisions
between lattice sites that model more expressive
integer (as opposed to boolean) occupancy states.
\citet{fonio2025adaptive} introduce an alternative extension to ILGA
by designing a novel collision operator that aims to recover LBM equilibria,
implemented through the LCU \cite{childs2012hamiltonian} technique.

The approaches detailed in the most recent QLGA developments by \citet{fonio2025adaptive}
and \citet{zamora2025efficient, zamora2025float} all require
reinitialization at each time step.
Similar to the LCU-based QLBM approaches,
this severely hinders the performance of the algorithms
by requiring quantum state tomography and state
preparation to occur at every iteration.
\citet{schalkers2024importance} and \citet{fonio2023quantum} showed
that such limitations are inherent to the
encodings commonly pursued in the QLGA and QLBM literature,
and efficient reinitialization remains an open challenge for such approaches.
However, encodings where reinitialization is not a necessity \emph{do} exist.
Boghosian \cite{boghosian1997quantum, boghosian1998simulating} and \citet{love2019quantum} introduced and detailed such an encoding, which utilizes
a qubit for each velocity channel in the system.
The Space-Time data encoding introduced by \citet{schalkers2024importance}
is an expanded computational basis state encoding, with 
additional qubit subregisters that allow for information to propagate from neighboring
lattice sites.
The number of qubits required for information propagation scales
with the number of time steps to be simulated, with the upper bound
of this scaling reduces to exactly the same encoding used by Boghosian \cite{boghosian1997quantum, boghosian1998simulating} and \citet{love2019quantum}.
Though this encoding does not allow for any memory compression in
the limit of the number of time steps \cite{fonio2023quantum, zamora2025float},
it does allow for the simultaneous simulation of
exponentially many lattice configurations in parallel by means of quantum superposition.

Though recent advances have increased the capabilities of QLGA algorithms,
several challenges persist.
In this work, we specifically target three key limitations of current QLGA methods
based on computational basis state encodings.
First, the literature seldom addresses specific methods for initialization
and boundary condition treatment, which are prerequesites for
solving real-world applications.
Second, the QLGA collision operators proposed in current algorithms
are discretization-specific and often operate on heavily restricted
state-equivalence criteria.
Third, measurement techniques for computing quantities
of interest are either formulated for different encodings \cite{schalkers2024momentum}
or utilize costly phase estimation procedures \cite{fonio2023quantum}.
In this work, we introduce novel quantum circuits building blocks
that cater to each step of the QLGA loop and address these three obstacles.
We offer different variations of implementations,
that trade off expressiveness for efficiency,
and detail their capabilities and limitations.
We analyze the complexity of each building block in terms of 
one- and two-qubit gates and accompany theoretical developments
with practical examples for small-scale LGA systems.

\begin{figure}
	\centering
\begin{tikzpicture}[node distance=2cm]

\tikzstyle{box} = [rectangle, rounded corners, minimum width=3cm, minimum height=1cm, text centered, draw=black, fill=gray!20]
\tikzstyle{arrow} = [thick,->,>=stealth]
    \node (start) [box] {Initialization};
    \draw node[above of=start,yshift=-1.25cm] {\Cref{sec:sp-2-1-initial-conditions}};
    \node (stream) [box, right of=start, xshift=2cm,  text width=4cm,] {Streaming \\ Boundary Conditions};
    \draw node[above of=stream,yshift=-1.25cm] {\Cref{sec:sp-2-2-streaming-boundary}};
    \node (collision) [box, right of=stream, xshift=3.25cm] {Collision};
    \draw node[above of=collision,yshift=-1.25cm] {\Cref{sec:sp-2-3-collision}};
    \node (measure) [box, right of=collision, xshift=2.25cm] {Measurement};
    \draw node[above of=measure,yshift=-1.25cm] {\Cref{sec:sp-2-4-measurement}};
    
    % Arrows
    \draw [arrow] (start) -- (stream);
    \draw [arrow] (stream) -- node[above] {$t \leftarrow t+1$} (collision);
    \draw [arrow] (collision) -- (9.25, -1) -- node[above] {$t<N_t$} (4, -1) -- (stream);
    \draw [arrow] (collision) -- node[above] {$t = N_t$} (measure);
\end{tikzpicture}
	\caption{QLGA algorithm overview for $N_t$ time steps.\label{fig:sp-1-1-qlga-overview}}
\end{figure}
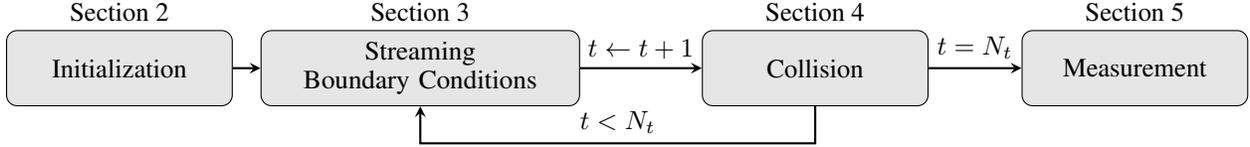

The structure of the paper follows the layout of the QLGA loop
depicted in \Cref{fig:sp-1-1-qlga-overview}.
\Cref{sec:sp-2-1-initial-conditions} introduces circuits for initialization followed by
streaming and boundary conditions in \Cref{sec:sp-2-2-streaming-boundary},
collision in \Cref{sec:sp-2-3-collision}, and measurement with respect
to quantities of interest and forces in \Cref{sec:sp-2-4-measurement}.
\Cref{sec:sp-4-results} shows the end-to-end
application of the introduced
building blocks in simulating one- and two-dimensional systems,
and \Cref{sec:sp-4-conclusion} concludes the paper.
The remainder of this section restates the basic properties of the
encoding we use throughout,
which we additionally generalize to different lattice discretizations
and compare to competing methods in the literature.

\subsection{The Space-Time Data Encoding and Running Examples\label{sec:sp-1-2-intro-examples}}

Classical LGA systems that model $N_g$ gridpoints with $q$ velocity channels
each require $qN_g$ bits to simulate.
The QLGA encoding introduced by Boghosian \cite{boghosian1997quantum, boghosian1998simulating}
and later utilized by  \citet{love2019quantum} and \citet{kocherla2024fully}
maps bits and qubits
in a one-to-one fashion and as such requires exactly $N_gq$ qubits.
By contrast, the sublinear encoding detailed by, \ie,  \citet{fonio2023quantum}
exponentially compresses the grid into $\lceil \log_2 N_g \rceil$ qubits,
that are entangled to a separate register of size $q$, at the cost
of reinitializing the quantum state after each time step.
The Space-Time data encoding introduced by \citet{schalkers2024importance}
can be seen as a compromise between the two previous examples.
All three variations have in common that the size of the velocity register
is \emph{not} compressed, which allows for the representation of all
$2^q$ local velocity configurations simultaneously.

The versatility of the Space-Time encoding comes from the fact that, unlike
the sublinear encoding,
the size of the velocity register is a parameter that can be adjusted
to allow for the efficient simulation of multiple time steps,
at the cost of additional qubits.
As the size of the velocity register grows,
its upper bound is naturally given by $qN_g$,
which is the same as that
of Boghosian \cite{boghosian1997quantum, boghosian1998simulating}.
The fundamental feature of all three encodings is that they 
encode velocities using the computational basis state encoding, \ie, they do \emph{not}
compress the $q$ velocity channel information into $\lceil \log_2 q \rceil$
qubits, which implies that building blocks can be compatible between algorithms.
Since the Space-Time data encoding is formulated in the most general
manner out of the three, we introduce all building blocks
in this work with respect to this encoding as a baseline.
However, since all encodings rely on the same interpretation
of the velocity register, the methods we introduce can
be adapted to the other two encodings with minimal modifications.

The remainder of this section
provides an overview of the properties and the rationale behind
the Space-Time data encoding, and introduces two running examples
used throughout the remainder of this work.
%Following a resurgence in interest surrounding quantum algorithms
%for discrete-velocity methods at the beginning of the decade, several works
%including \todo{add citations} focused on designing quantum
%primitives that perform computations modelled by the Lattice Boltzmann Method.
%Algorithms such as those of \citet{todorova2020quantum} and \citet{schalkers2024efficient}
%implement streaming 
The Space-Time data encoding can be naturally seen
as an extension of the computational basis state encoding that
circumvents the inherent non-unitarity of streaming in this setting \cite{schalkers2024importance}.
This is achieved by entangling a grid (or positional) register
to a velocity register that grows with the number of time steps
to accommodate the information
of neighboring gridpoints, thus fundamentally
altering the nature of streaming compared to other encodings.
The size of the velocity register depends on two factors --
the lattice structure and the number of time steps to be simulated.

The first factor is the specific lattice discretization
that the simulation is carried under.
Particularly, the dimensionality of the system and
the number of channels connecting lattice sites determines
the so-called \emph{neighborhood structure} of the lattice.
This neighborhood structure dictates both
the trajectory that discrete particles can travel in the physical grid,
as well as the required size and structure of the velocity register.
Throughout this work, we distinguish between
discretizations using the commonplace \dq{d}{q} taxonomy, indicating
a $d$-dimensional system with $q$ discrete velocity channels.
The second factor is the number of time steps to be simulated.
The Space-Time data encoding requires that,
for a given number of time steps to be simulated $N_t$,
all information (\ie, particle occupancy of neighboring gridpoints)
that can reach a gridpoint within the $N_t$ time steps
is contained within the velocity register.
In general, however, each gridpoint relies on the information of its
$\mathcal{O}(q)$ neighbors for $1$ time step, and expanding
the neighborhood in the $d$-dimensional space across $N_t$ time steps
envelops $\mathcal{O}(N_t^d)$ gridpoints
with $q$ discrete velocity channels each \citep{schalkers2024importance}.
Naturally, the maximum number of qubits required to perform the simulation
for an arbitrary number of time steps is bounded by $N_g$,
the number of lattice sites of the entire grid.
In what follows, we provide two concrete instances of qubit registers that
encode the information affiliated with the \dq{1}{2} and \dq{2}{4} discretizations,
as well as the exact scaling for the \dq{3}{6} stencil.

\begin{figure}
    \centering
    \hfill
    \subcaptionbox{\dq{2}{4} Space-Time stencil for the simulation of up to $3$ time steps. \label{fig:sp-2-1-examples-diag-sp-stencil-d1q2-flat}}{\centering
	\begin{tikzpicture}[scale=0.3]
    % Grid lines
    \draw[
      help lines,
      line width=0.3pt,
      color=gray!30,
      dashed
    ] (-12, -2) grid[step={($(2, 2) - (0, 0)$)}] (12, 2);
    \draw[
      help lines,
      line width=0.4pt,
      color=black!80,
      dashed
    ] (-14, -2) grid[step={($(4, 4) - (0, 0)$)}] (14, 2);

        % Distance 0
        \doneqtwo{(0,0)}{0.1cm}{2}{0}{1.5}{black}
        
        \draw[] node at (1.25, 0.75) {$\mathrm{q_0}$};
        \draw[] node at (-1.25, 0.75) {$\mathrm{q_1}$};

        % Distance 1
        \doneqtwolabel{(4,0)}{0.1cm}{2}{1}{1.5}{gray}{\tiny{1}}
        \doneqtwolabel{(-4,0)}{0.1cm}{2}{2}{1.5}{gray}{\tiny{1}}

        % Distance 2
        \doneqtwolabel{(8,0)}{0.1cm}{2}{1}{1.5}{gray!20}{\tiny{2}}
        \doneqtwolabel{(-8,0)}{0.1cm}{2}{2}{1.5}{gray!20}{\tiny{2}}

        % Distance 3
        \doneqtwolabel{(12,0)}{0.1cm}{2}{1}{1.5}{gray!10}{\tiny{3}}
        \doneqtwolabel{(-12,0)}{0.1cm}{2}{2}{1.5}{gray!10}{\tiny{3}}
\end{tikzpicture}}%
    \hfill
    \subcaptionbox{\dq{1}{2} register for $3$ time steps. \label{fig:sp-2-1-examples-diag-sp-register-d1q2}}{\input{circuits/circ-sp-register-d1q2-3-timesteps}}%
    \caption{Space-Time data structure and quantum register for the \dq{1}{2} discretization. \label{fig:diag-sp-d1q2-stencil-and-register}}    
\end{figure}

\begin{figure}
    \centering
    \hfill
    \subcaptionbox{\dq{2}{4} Space-Time stencil for the simulation of up to $4$ time steps. \label{fig:sp-2-1-examples-diag-sp-stencil-d2q4-flat}}{	\centering
\begin{tikzpicture}[scale=0.28]

    % Grid lines
    \draw[
      help lines,
      line width=0.3pt,
      color=gray!30,
      dashed
    ] (-16, -16) grid[step={($(2, 2) - (0, 0)$)}] (16, 16);
    \draw[
      help lines,
      line width=0.4pt,
      color=black!80,
      dashed
    ] (-18, -18) grid[step={($(4, 4) - (0, 0)$)}] (18, 18);
        
    % Nodes
    % Distance 0
    \dtwoqfour{(0,0)}{0.1cm}{2}{0}{1.5}{black}
    
	\draw[] node at (1.25, 0.55) {\tiny{$\mathrm{q_0}$}};
	\draw[] node at (0.75, 1.5) {\tiny{$\mathrm{q_1}$}};
    \draw[] node at (-1.25, 0.55) {\tiny{$\mathrm{q_2}$}};
    \draw[] node at (0.75, -1.5) {\tiny{$\mathrm{q_3}$}};

    % Distance 1
    \dtwoqfourlabel{(4,0)}{0.1cm}{2}{1}{1.5}{black!60}{\tiny{1}}
    \dtwoqfourlabel{(-4,0)}{0.1cm}{2}{2}{1.5}{black!60}{\tiny{1}}
    \dtwoqfourlabel{(0,4)}{0.1cm}{2}{3}{1.5}{black!60}{\tiny{1}}
    \dtwoqfourlabel{(0,-4)}{0.1cm}{2}{4}{1.5}{black!60}{\tiny{1}}

    % Distance 2
    \foreach \x in {0,4} {
    		\dtwoqfourlabel{(8-\x,\x)}{0.1cm}{2}{5}{1.5}{gray!30}{\tiny{2}}
    		\dtwoqfourlabel{(-\x,8-\x)}{0.1cm}{2}{5}{1.5}{gray!30}{\tiny{2}}
    		\dtwoqfourlabel{(-8+\x,-\x)}{0.1cm}{2}{5}{1.5}{gray!30}{\tiny{2}}
    		\dtwoqfourlabel{(\x,-8+\x)}{0.1cm}{2}{5}{1.5}{gray!30}{\tiny{2}}
    }
    	
    	% Distance 3
    \foreach \x in {0,4,8} {
    		\dtwoqfourlabel{(12-\x,\x)}{0.1cm}{2}{5}{1.5}{gray!30}{\tiny{3}}
    		\dtwoqfourlabel{(-\x,12-\x)}{0.1cm}{2}{5}{1.5}{gray!30}{\tiny{3}}
    		\dtwoqfourlabel{(-12+\x,-\x)}{0.1cm}{2}{5}{1.5}{gray!30}{\tiny{3}}
    		\dtwoqfourlabel{(\x,-12+\x)}{0.1cm}{2}{5}{1.5}{gray!30}{\tiny{3}}
    }
    
    	% Distance 4
    \foreach \x in {0,4,8,12} {
    		\dtwoqfourlabel{(16-\x,\x)}{0.1cm}{2}{5}{1.5}{gray!00}{\tiny{4}}
    		\dtwoqfourlabel{(-\x,16-\x)}{0.1cm}{2}{5}{1.5}{gray!10}{\tiny{4}}
    		\dtwoqfourlabel{(-16+\x,-\x)}{0.1cm}{2}{5}{1.5}{gray!10}{\tiny{4}}
    		\dtwoqfourlabel{(\x,-16+\x)}{0.1cm}{2}{5}{1.5}{gray!10}{\tiny{4}}
    }
    
\end{tikzpicture}}%
    \hfill
    \subcaptionbox{\dq{2}{4} velocity register for $4$ time steps. \label{fig:sp-2-1-examples-diag-sp-register-d2q4}}{\input{circuits/circ-sp-register-d2q4-1-timestep}}%
    \hfill
    \caption{Space-Time data structure and quantum register for the \dq{2}{4} discretization. \label{fig:diag-sp-d2q4-stencil-and-register}}%
\end{figure}

\paragraph{Example -- \dq{1}{2}, \dq{2}{4}, and \dq{3}{6}.}
We introduce three examples showing how the Space-Time data encoding generalizes
to multiple time steps across different discretizations, beginning
from the simplest instance of \dq{1}{2} depicted in
\Cref{fig:diag-sp-d1q2-stencil-and-register}.
The \dq{1}{2} discretization only distinguishes two channels that
particles can travel across in the $1$ dimension of the system,
to either the left or the right neighboring gridpoint.
Though physically trivial, this scenario
is small enough to demonstrate the core mechanisms and challenges of the Space-Time
encoding, and in this case allows for the simulation of multiple time steps without
reinitialization on quantum simulators for commercial classical hardware available today.

The cornerstone data structure of the Space-Time encoding, which we
refer to as the \emph{Space-Time stencil}, is shown in
\Cref{fig:sp-2-1-examples-diag-sp-stencil-d1q2-flat}, and allows for the simulation
of up to $N_t = 3$ time steps.
The information contained within the stencil includes the velocity profile
of the gridpoint residing at the center of the stencil,
which we refer to as the \emph{physical origin},
as well as the velocity profile of all neighboring gridpoints up to a distance of $N_t$.
We label each gridpoint encoded in the stencil
with its distance from the physical origin,
as this information determines under which circumstances the girdpoint is subject
to computation depending on the time step under simulation.
The number of gridpoints in the stencil is $2N_t + 1$,
and the number of discrete velocity channel occupancies to
track is $4N_t + 2$.
The corresponding quantum register is depicted in \Cref{fig:sp-2-1-examples-diag-sp-register-d1q2}.
Each velocity channel is assigned a qubit labeled $v^{d_o}_{j}$ with $d_o$
its distance from the physical origin,
and $j \in \{0, 1\}$ the velocity channel it corresponds to
($0$ for the positive direction and $1$ for the negative direction).

The extension of this encoding to \dq{2}{4} and $4$ time steps is depicted
in \Cref{fig:diag-sp-d2q4-stencil-and-register}, and in general,
shares the same fundamental features as von Neumann cellular automata \citep{toffoli1987cellular}.
The number of gridpoints from which information can propagate to the physical
origin in $N_t$ time steps scales quadratically with

\begin{equation}
\mathcal{N}_{g}^{\mathrm{D}_2\mathrm{Q}_4}(N_t)=\sum_{t=1}^{N_t} 4t = 2N_t^2 + 2N_t.
\end{equation}

Accounting for the $4$ discrete velocity channels and the physical origin
itself, the number of particles (and therefore qubits) required
to model the system is $4\left(\mathcal{N}_{g}^{\mathrm{D}_2\mathrm{Q}_4}(N_t) + 1\right)=8N_t^2 + 8N_t + 4$.
\Cref{fig:sp-2-1-examples-diag-sp-stencil-d2q4-flat} displays the neighborhood
of $41$ gridpoints, while \Cref{fig:sp-2-1-examples-diag-sp-register-d2q4}
shows the quantum register that encodes the
data required for the simulation of $4$ time steps.
Throughout this paper, we order the qubits to align with the typical
convention of $\mathrm{q_0}$ signifying the positive direction of the $x$-axis,
and all other indices following a counterclockwise order
sorted first by direction (positive or negative) and second by dimension,
as shown on the labels of the velocity origins corresponding to the physical origin in 
\Cref{fig:sp-2-1-examples-diag-sp-stencil-d1q2-flat} and \Cref{fig:sp-2-1-examples-diag-sp-stencil-d2q4-flat}.
In the computational basis state encoding of the \dq{1}{2} discretization,
the state $\ket{01}$ thus encodes a single particle traveling to the right,
while a the state $1/\sqrt{2}(\ket{1010} + \ket{0101})$ of a \dq{2}{4}
stencil represents a superposition of basis states, where particles are 
occupying both velocity channels of the $x$- and $y$-axes, respectively.

Adding a third dimension with two discrete and opposing velocity channels, one obtains
the \dq{3}{6} discretization, which we use as a running example for our collision methods.
For this $3$-dimensional discretization, the number of gridpoints in the Space-Time
stencil is equal to the $N_t^{\text{th}}$ \emph{Ha\"{u}y octahedral number} \cite{deza2012figurate}
and is given by

\begin{equation}
\mathcal{N}_{g}^{\mathrm{D}_3\mathrm{Q}_6}(N_t)=\frac{(2N_t+1)(2N_t^2+2N_t+3)}{3}.
\end{equation}

Since each gridpoint has $6$ velocity channels, the
entire discretization in turn requires $6 \cdot \mathcal{N}_{g}^{\mathrm{D}_3\mathrm{Q}_6}(N_t)=8N_t^3 + 12N_t^2 + 16N_t + 6$ qubits
to model $N_t$ time steps.
It is worth noting that in addition to the velocity register,
and unless the register reaches its bound of $qN_g$ qubits,
the Space-Time data encoding requires additional qubits to entangle the velocity register.
We refer to these additional qubits as \emph{grid} or \emph{positional} qubits.
Irrespective of the $N_t$, the number of grid qubits required is

\begin{equation}
	n_g = \sum_{d} \lceil\log_2  N_{g_d}\rceil,
\end{equation}

with $N_{g_d}$ the number
of lattice sites for each dimension of the system.

\paragraph{Complexity analysis assumptions.} To
accurately capture the computational requirements
of the Space-Time QLGA algorithm, we describe the complexity
of each introduced buidling block in terms of
\emph{native quantum gates}.
By native quantum gates, we generally refer to
single- and two-qubit gate sets that are known to be part of universal gate sets \citep{nielsen2010quantum}.
Where possible, we derive the complexity in terms of commonplace,
non parameterized gates such as the $\X, \mathrm{H}$, and $\CX$ gates.
We assume the number of dimensions $d$,
the number of gridpoints in each dimension (\ie $N_{g_x}$),
the number of discrete velocity channels $q$,
and the number of time steps to be simulated $N_t$ are all variables. 
For readability, we assume grid operations that act on one dimension of the grid
affect $\mathcal{O}(n_g = \sum_{d} \lceil N_{g_d} \rceil)$ qubits,
which is an overestimation in $d>1$ dimensions. 
Where no efficient decomposition is known, we assume the
decomposition of the unitary matrix requires a number of native
quantum gates that grows exponentially with
size of the register it is applied on, as first shown by \citet{barenco1995elementary}.
We express the complexity of each operation
with respect to \emph{one time step}, as the complexity
throughout the runtime of the algorithm scales linearly with $N_t$.
Finally, for a unitary matrix $U$, we use the notation
$\mathrm{C}^pU$ to indicate that the operation is controlled on $p$ qubits.
%This assignment corresponds exactly to the classical counterpart described previously,
%and as such it offers no memory compression over classical simulations.

\section{Initial Conditions\label{sec:sp-2-1-initial-conditions}}

The first step in the LGA pipeline is
initializing the flow field by setting the occupancy
of each lattice according to the simulation specification.
The quantum circuit that performs
this function is tasked with evolving 
the initial $\ket{0}^{\otimes N}$ state into the state
that encodes the classically prescribed flow field.
Such state preparation procedures are notoriously
challenging in quantum computing and often require exponentially complex quantum
circuits or approximations to realize.
Within LGA in particular, complex initial conditions are paramount
for realizing realistic simulations.

As with all building blocks described in this work,
the circuits implementing initial conditions should meet two
criteria to be of any use.
First, they should be expressive enough to encode initial
conditions that are adequate for real-world problems and benchmarks.
Second, they should be efficient enough both
to be constructed using the classical hardware available today,
and to justify a potential advantage to executing the
quantum algorithms over their classical counterparts.
In this section, we propose two methods that
favor expressivity and efficiency, respectively.
We begin by introducing the expressive \emph{pointwise} method,
which allows the assignment of arbitrary velocity profiles to each gridpoint.

\subsection{Pointwise Initialization\label{subsec:sp-2-1-initial-conditions-pw}}

The initial conditions of complex flow fields
may necessitate finely detailed velocity profiles that
differ even within small neighborhoods of gridpoints.
For this purpose, we introduce a quantum primitive for pointwise initialization,
that addresses the grid qubit register in a way that
caters individually to each gridpoint on the lattice.

Let us assume a \dq{2}{4} lattice discretization and consider the situation
in which we want to simulate a system where all particles
initially reside at one lattice site.
We refer to this lattice site as the physical origin.
To assign the velocity profile to the one particular gridpoint without
affecting any of the other lattice sites, we first apply a layer of $\X$ gates
to the grid register that transforms the $\ket{x}\ket{y}$ state to the
$\ket{1}^{\otimes n_{g_x}}\ket{1}^{\otimes n_{g_y}}$ state.
Then, several layers of $\CPX{p}$ gates
act upon the appropriate velocity register to
flip the values of the appropriate qubits to $\ket{1}$,
where $\CPX{p}$ is a $p$-controlled $\X$ gate.
Finally, another layer of $\X$ gates resets
the grid register to its previous state.
To appropriately address the locality encoded in the Space-Time stencil,
this procedure is repeated for each lattice site
where information from the grid point can propagate to the physical origin
within the number of time steps to be simulated.
That is, controlled on the position of all neighboring
gridpoints that contain the information of the physical origin,
the qubits corresponding to the relative position of the physical
origin within the neighbor's data
structure  must be initialized with the same particle occupancy.

\begin{figure}
	\input{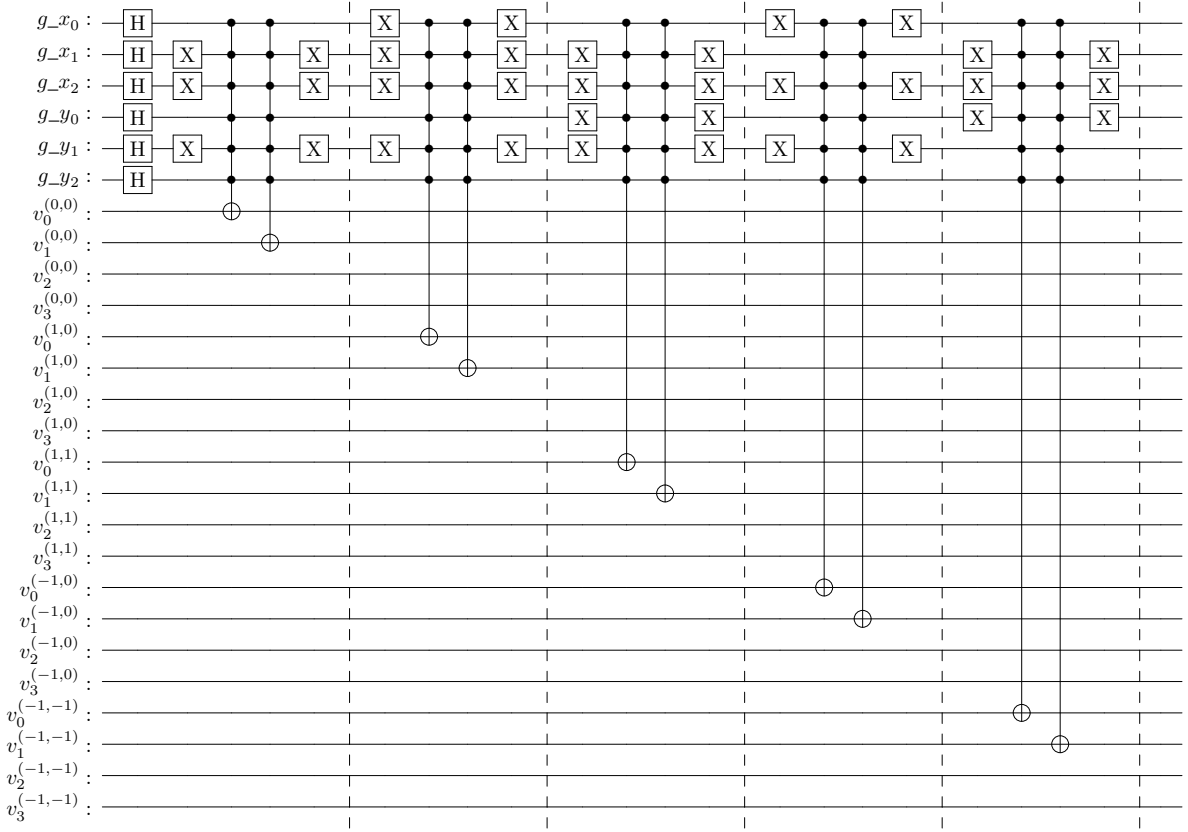}
	\caption{\dq{2}{4} pointwise initialization of the gridpoint at location $(1, 5)$ and velocity profile $\ket{1100}$ for one time step. \label{fig:meth-init-circuits-circ-sp-initial_conditions_d2q4_1_timestep_pw}}
\end{figure}

\paragraph{Example -- \dq{2}{4} pointwise initialization.}
Let us consider a brief example of pointwise initialization that
demonstrates the application of the method to an $8\times 8$ \dq{2}{4} lattice with $N_t = 1$.
We aim to initialize a single gridpoint at 
location $(1, 5)$ with velocity profile $\ket{1100}$,
corresponding to a particle moving in the positive $x$ direction and one
moving in the positive $y$ direction.
\Cref{fig:meth-init-circuits-circ-sp-initial_conditions_d2q4_1_timestep_pw}
depicts the quantum circuits that creates such a state.
The qubit register spans $6$ positional qubits
and $20$ velocity qubits that encode the information
that can travel to the relative origin in one time step.
The circuit begins with a layer of $\mathrm{H}$ gates
that prime the positional qubits in a unifrom superposition of $(\ket{0} + \ket{1})^{\otimes n_g}$.
Following this step, there are $5$ layers of operations
that perform the same procedure
on basis states that correspond to different
physical lattice sites.
All layers permute the basis states of the grid register
such that a particular target state is converted to the  $\ket{1}^{\otimes 6}$ state,
before setting the appropriate velocity qubits and undoing to first operation.
The first layer corresponds to the physical origin and
targets the $\ket{1}\ket{5} \equiv \ket{001}\ket{101}$ state, whereas all
other operations address the neighbors one distance away in each direction.

\paragraph{Complexity Analysis.} Though expressive, the pointwise method 
poses significant drawbacks in terms of efficiency.
Each individual operation by itself is efficient, only requiring $\mathcal{O}(n_g)$
$\X$ gates to set and unset the state of the grid qubits, as well as 
$\mathcal{O}(q)\ \CPX{n_g}$ gates,
which can be decomposed into $O(qn_g^2)\ \CX$ gates \cite{barenco1995elementary}. 
As most building blocks introduced in this paper, this initialization
technique traverses the stencil of the physical origin and initializes
the state of each neighbor, thus repeating each operation $\mathcal{O}(N_t^d)$ times
to initialize a single grid point.
To realize more intricate initial conditions, this operation must be repeated for each
individual gridpoint, which therefore results in an
overall complexity of $\mathcal{O}(N_gN_t^d(n_g + qn_g^2))$.
Under this complexity, any potential advantage obtained by compressing grid information
is lost, as the exponential overhead of addressing each basis state sequantially
leads to performance comparable to classical serial computers.
To address this pitfall of pointwise methods,
we introduce a more efficient,
though less expressive alternative -- \emph{volumetric} initialization.

\subsection{Volumetric Initialization\label{subsec:sp-2-1-initial-conditions-vol}}

The key drawback of the pointwise method is that it does
not take advantage of the superposition of the grid qubits and therefore
incurs an overhead that is linear the number of physical lattice sites.
In what follows, we introduce a more efficient approach that does
exploit the superposition of the positional qubits to
efficiently perform one operation over arbitrarily many lattice sites.
This operation adapts the previous boundary condition implementation by
\citet{schalkers2024efficient} and extends it to the application of initial conditions
and the corresponding edge cases that come with the Space-Time encoding.
The core idea of volumetric operations is to make use of
\emph{quantum comparator circuits} to isolate a desired \emph{volume} 
of physical space where one  operation can be uniformly applied.
Quantum comparator circuits make use of the periodicity of binary additions
to entangle states that adhere to prescribed bounds to the $\ket{1}$ state
of an ancillary qubit.
To realize these circuits, we make use of Draper adders based on the
Quantum Fourier Transform (QFT),
described in \cite{draper2000addition}, and more broadly surveyed in \cite{ruiz2017quantum}.
For a detailed description of how the
Draper adder can be utilized as a comparator,
we refer the reader to \cite{schalkers2024efficient}.
Within the context of initial conditions,
this would be akin to initializing all qubits within a certain
volume with the same velocity profile.
In this section, we describe how cuboid-shaped volumes
can be efficiently implemented using a small number of ancilla qubits.

To model a cuboid-shaped volume, we only require information
about its bounds in each dimension.
Implementing this in a quantum circuit is straight-forward
by utilizing one ancilla qubit per bound and per dimensions, for a total
of $2d$ qubits.
The purpose of each qubit is to entangle with the positional qubits and
encode the information of whether the site lies within ($\ket{1}$)
the target volume or not ($\ket{0}$).
Then, controlled on the appropriate ancilla qubit(s),
we apply the same series of multi-controlled $\X$ gates
that we used to assign the prescribed velocity profile to a single girdpoint in the pointwise case.
The circuit which creates this entanglement is identical to the 
quantum comparator implementation described by \citet{schalkers2024efficient}
-- where our extension differs is in its application to the
Space-Time data structure and its edge cases.
To illustrate the application of these circuits, let us consider a simple \dq{1}{2} example.

\begin{figure}
	\input{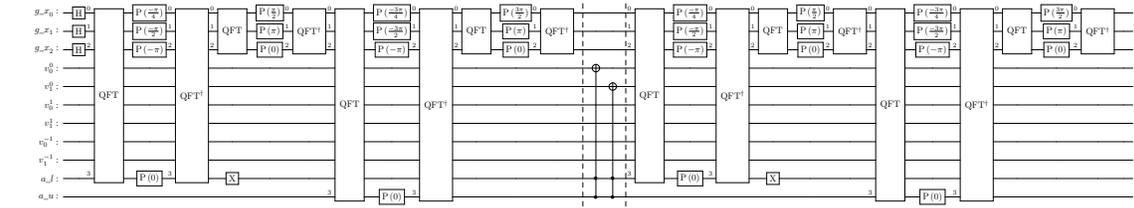}
	\caption{\dq{1}{2} volumetric initialization of the interval $[2, 5]$ and velocity profile $\ket{11}$ for part of one time step. \label{fig:sp-2-1-init-circuits-circ-sp-initial_conditions_d2q4_1_timestep_vol}}
\end{figure}

\paragraph{Example -- \dq{1}{2} volumetric initialization.}
Let us consider the application of volumetric initial conditions to a 
\dq{1}{2} lattice with $8$ lattice sites that we aim to simulate for $3$ time steps.
The initial conditions we aim to initialize are such that 
gridpoints $2, 3, 4$, and $5$ all share with the velocity profile $\ket{11}$.
The circuit implementing this procedure is given in
\Cref{fig:sp-2-1-init-circuits-circ-sp-initial_conditions_d2q4_1_timestep_vol}.
To make use of volumetric operations, $2$ additional qubits $\mathrm{a_l}$ and $\mathrm{a_u}$
are appended to the end of the register, to determine whether gridpoints
lie within the bounds of the interval.
As in the pointwise case, the circuits first
primes the grid register to the $\ket{+}^{\otimes 3}$
state such that all gridpoints are encoded, which leads to the state

\begin{equation}
\ket{\uppsi_1} = \frac{1}{2^{10}}\sum_{k=0}^7 \ket{k}\ket{0}^{\otimes 6}\ket{0}\ket{0}.
\end{equation}

Following this, two Quantum Fourier Transform (QFT)
based comparator circuits set $\mathrm{a_l}$ and $\mathrm{a_u}$
to states indicating whether the points belong in the target volume as

\begin{align}
\begin{split}
\ket{\uppsi_2} & = \frac{1}{2^{10}}(\left(\ket{0} + \ket{1}\right)\ket{0}^{\otimes 6}\ket{0}\ket{1}\\
			   & + \left(\ket{2} + \ket{3} + \ket{4} + \ket{5}\right)\ket{0}^{\otimes 6}\ket{1}\ket{1}\\
			   & + \left(\ket{6} + \ket{7}\right)\ket{0}^{\otimes 6}\ket{1}\ket{0}).
\end{split}
\end{align}

This reflects the fact that (i) the gridpoints at locations $0$ and $1$
are not greater than the lower bound, but are lower than the upper bound ($\ket{a_la_u} = \ket{01}$),
(ii) gridpoints between $2$ and $5$ obey both bounds ($\ket{a_la_u} = \ket{11}$),
and (iii) gridpoints $6$ and $7$ are over the lower bound, but not under the upper bound
($\ket{a_la_u} = \ket{10}$).
Subsequently, the application of the doubly-controlled $\X$ gates
sets the two velocity qubits that correspond to the origin of the stencil
of all gridpoints within the interval to $\ket{11}$, resulting in the state

\begin{align}
\begin{split}
\ket{\uppsi_3} & = \frac{1}{2^{10}}(\left(\ket{0} + \ket{1}\right)\ket{0}^{\otimes 6}\ket{0}\ket{1}\\
			   & + \left(\ket{2} + \ket{3} + \ket{4} + \ket{5}\right)\ket{1}\ket{1}\ket{0}^{\otimes 4}\ket{1}\ket{1}\\
			   & + \left(\ket{6} + \ket{7}\right)\ket{0}^{\otimes 6}\ket{1}\ket{0}).
\end{split}
\end{align}

The final step consists of reversing the setting of the ancilla qubits by
means of the mirrored comparator circuits, which evolves the state to

\begin{align}
\begin{split}
\ket{\uppsi_4} & = \frac{1}{2^{10}}(\left(\ket{0} + \ket{1}\right)\ket{0}^{\otimes 8}\\
			   & + \left(\ket{2} + \ket{3} + \ket{4} + \ket{5}\right)\ket{1}\ket{1}\ket{0}^{\otimes 6}\\
			   & + \left(\ket{6} + \ket{7}\right)\ket{0}^{\otimes 8}).
\end{split}
\end{align}

In this final configuration $\ket{\uppsi_4}$ encodes the desired
velocity profile for the physical origin corresponding to the gridpoints
in the target interval, without affecting any other basis states.
To consistently set the state throughout the entire grid, the circuit shown in
\Cref{fig:sp-2-1-init-circuits-circ-sp-initial_conditions_d2q4_1_timestep_vol}
should be adjusted and repeated for all gridpoints encoded in the Space-Time stencil.
The only differences in the application of the circuit are that (i) 
the phases of the $\mathrm{P}$ gates are adjusted
to fit the different regions of space the gridpoints correspond to
and that (ii) the targets of the $\CPX{2}$ gates change accordingly.
The advantage of the volumetric initialization method
lies in that it can address arbitrarily many physical gridpoints simultaneously.
As a consequence of this method, however, several edge cases arise that
were otherwise trivial in the pointwise counterpart.

\paragraph{Edge cases -- \dq{2}{4} initialization.}
Due to its treatment of contiguous regions of space simultaneously,
the volumetric initialization technique requires careful analysis
in situations in which information can propagate through periodic
boundary conditions.
We illustrate this challenge in a \dq{2}{4} scenario,
and provide a general construction mechanism that works across $1$-$3$D spaces.
To illustrate the challenge, we use a $5 \times 5$ lattice in which
we select the gridpoints in the square domain bound by
$(\mathrm{x_l, x_u})=(1, 3)$ and $(\mathrm{y_l, y_u})=(1, 3)$ with
an arbitrary velocity profile, which is not relevant for this analysis.
We further assume that we are interested in simulating this system
for $N_t=4$ time steps.
\Cref{fig:sp-2-1-init-diag-sp-grid-init-edge-cases} depicts this system,
and the three kinds of edge cases that emerge in this scenario.

\begin{figure}
	\centering
	\hfill
	\subcaptionbox{Nominal case for the application of volumetric initial conditions. \label{fig:sp-2-1-init-diag-sp-grid-init-edge-cases-0}}{\centering
\begin{minipage}{0.55\textwidth}
\begin{tikzpicture}[scale=0.3]
    % Grid lines
    \draw[
      help lines,
      line width=0.3pt,
      color=gray!30,
      dashed
    ] (-8, -8) grid[step={($(2, 2) - (0, 0)$)}] (8, 8);
    \draw[
      help lines,
      line width=0.4pt,
      color=black!80,
      dashed
    ] (-10, -10) grid[step={($(4, 4) - (0, 0)$)}] (10, 10);
        
    % Nodes
    % Distance 0
    \dtwoqfour{(0,0)}{0.1cm}{2}{0}{1.5}{gray}

    % Distance 1
    \dtwoqfour{(4,0)}{0.1cm}{2}{1}{1.5}{gray}
    \dtwoqfour{(-4,0)}{0.1cm}{2}{2}{1.5}{gray}
    \dtwoqfour{(0,4)}{0.1cm}{2}{3}{1.5}{gray}
    \dtwoqfour{(0,-4)}{0.1cm}{2}{4}{1.5}{gray}

    % Distance 2
    \dtwoqfour{(8,0)}{0.1cm}{2}{5}{1.5}{gray}
    \dtwoqfour{(-8,0)}{0.1cm}{2}{5}{1.5}{gray}
    \dtwoqfour{(4,4)}{0.1cm}{2}{5}{1.5}{gray}
    \dtwoqfour{(-4,4)}{0.1cm}{2}{5}{1.5}{gray}
    \dtwoqfour{(-4,-4)}{0.1cm}{2}{5}{1.5}{gray}
    \dtwoqfour{(4,-4)}{0.1cm}{2}{5}{1.5}{gray}
    \dtwoqfour{(0,-8)}{0.1cm}{2}{5}{1.5}{gray}
    \dtwoqfour{(0,8)}{0.1cm}{2}{5}{1.5}{gray}

    \dtwoqfour{(4,8)}{0.1cm}{2}{5}{1.5}{gray}
    \dtwoqfour{(8,4)}{0.1cm}{2}{5}{1.5}{gray}
    \dtwoqfour{(8,8)}{0.1cm}{2}{5}{1.5}{gray}
    \dtwoqfour{(-4,8)}{0.1cm}{2}{5}{1.5}{gray}
    \dtwoqfour{(-8,4)}{0.1cm}{2}{5}{1.5}{gray}
    \dtwoqfour{(-8,8)}{0.1cm}{2}{5}{1.5}{gray}
    \dtwoqfour{(4,-8)}{0.1cm}{2}{5}{1.5}{gray}
    \dtwoqfour{(8,-4)}{0.1cm}{2}{5}{1.5}{gray}
    \dtwoqfour{(8,-8)}{0.1cm}{2}{5}{1.5}{gray}
    \dtwoqfour{(-4,-8)}{0.1cm}{2}{5}{1.5}{gray}
    \dtwoqfour{(-8,-4)}{0.1cm}{2}{5}{1.5}{gray}
    \dtwoqfour{(-8,-8)}{0.1cm}{2}{5}{1.5}{gray}
    \draw[rounded corners, ultra thick] (-6,-6) rectangle (6, 6);
    
    \node[] at (-4, -11) {\small$\mathbf{x_l}$};
    \node[] at (4, -11) {\small$\mathbf{x_u}$};

    \node[] at (-11, -4) {\small$\mathbf{y_l}$};
    \node[] at (-11, 4) {\small$\mathbf{y_u}$};
    
    \dtwoqfour{(0, -4)}{0.1cm}{2}{5}{1.5}{black}
    \dtwoqfour{(0, 0)}{0.1cm}{2}{5}{1.5}{black}
    \dtwoqfour{(0, 4)}{0.1cm}{2}{5}{1.5}{black}

    \dtwoqfour{(-4, -4)}{0.1cm}{2}{5}{1.5}{black}
    \dtwoqfour{(-4, 0)}{0.1cm}{2}{5}{1.5}{black}
    \dtwoqfour{(-4, 4)}{0.1cm}{2}{5}{1.5}{black}

    \dtwoqfour{(4, -4)}{0.1cm}{2}{5}{1.5}{black}
    \dtwoqfour{(4, 0)}{0.1cm}{2}{5}{1.5}{black}
    \dtwoqfour{(4, 4)}{0.1cm}{2}{5}{1.5}{black}
    %\draw[-,solid, ultra thick, color=black, rounded corners] (6,-6) -- (6,6) -- (-6,6) -- (-6,-6) -- (6,-6) -- (6,-5);
\end{tikzpicture}
\end{minipage}
\hfill
\begin{minipage}{0.4\textwidth}
\begin{table}[H]
\centering
\begin{tabular}{c|c}
\toprule
\textbf{Condition} & \textbf{Domain} \\
\midrule
${x_l \leq x}$ & \checkmark \\
${x_u \geq x}$ & \checkmark \\
${y_l \leq y}$ & \checkmark \\
${y_u \geq y}$ & \checkmark \\
\bottomrule
\end{tabular}
\caption{Conditions determining the domain. \label{tab:sp-2-1-init-diag-sp-grid-init-edge-cases-0}}
\end{table}
\end{minipage}}%
	\hfill%
	\\
	\hfill
	\subcaptionbox{Edge case for the application of boundary conditions with periodic overflow in $1$ dimension. \label{fig:sp-2-1-init-diag-sp-grid-init-edge-cases-1}}{\centering
\begin{minipage}{0.55\textwidth}
\begin{tikzpicture}[scale=0.3]
    % Grid lines
    \draw[
      help lines,
      line width=0.3pt,
      color=gray!30,
      dashed
    ] (-8, -8) grid[step={($(2, 2) - (0, 0)$)}] (8, 8);
    \draw[
      help lines,
      line width=0.4pt,
      color=black!80,
      dashed
    ] (-10, -10) grid[step={($(4, 4) - (0, 0)$)}] (10, 10);
        
    % Nodes
    % Distance 0
    \dtwoqfour{(0,0)}{0.1cm}{2}{0}{1.5}{gray}

    % Distance 1
    \dtwoqfour{(4,0)}{0.1cm}{2}{1}{1.5}{gray}
    \dtwoqfour{(-4,0)}{0.1cm}{2}{2}{1.5}{gray}
    \dtwoqfour{(0,4)}{0.1cm}{2}{3}{1.5}{gray}
    \dtwoqfour{(0,-4)}{0.1cm}{2}{4}{1.5}{gray}

    % Distance 2
    \dtwoqfour{(8,0)}{0.1cm}{2}{5}{1.5}{gray}
    \dtwoqfour{(-8,0)}{0.1cm}{2}{5}{1.5}{gray}
    \dtwoqfour{(4,4)}{0.1cm}{2}{5}{1.5}{gray}
    \dtwoqfour{(-4,4)}{0.1cm}{2}{5}{1.5}{gray}
    \dtwoqfour{(-4,-4)}{0.1cm}{2}{5}{1.5}{gray}
    \dtwoqfour{(4,-4)}{0.1cm}{2}{5}{1.5}{gray}
    \dtwoqfour{(0,-8)}{0.1cm}{2}{5}{1.5}{gray}
    \dtwoqfour{(0,8)}{0.1cm}{2}{5}{1.5}{gray}

    \dtwoqfour{(4,8)}{0.1cm}{2}{5}{1.5}{gray}
    \dtwoqfour{(8,4)}{0.1cm}{2}{5}{1.5}{gray}
    \dtwoqfour{(8,8)}{0.1cm}{2}{5}{1.5}{gray}
    \dtwoqfour{(-4,8)}{0.1cm}{2}{5}{1.5}{gray}
    \dtwoqfour{(-8,4)}{0.1cm}{2}{5}{1.5}{gray}
    \dtwoqfour{(-8,8)}{0.1cm}{2}{5}{1.5}{gray}
    \dtwoqfour{(4,-8)}{0.1cm}{2}{5}{1.5}{gray}
    \dtwoqfour{(8,-4)}{0.1cm}{2}{5}{1.5}{gray}
    \dtwoqfour{(8,-8)}{0.1cm}{2}{5}{1.5}{gray}
    \dtwoqfour{(-4,-8)}{0.1cm}{2}{5}{1.5}{gray}
    \dtwoqfour{(-8,-4)}{0.1cm}{2}{5}{1.5}{gray}
    \dtwoqfour{(-8,-8)}{0.1cm}{2}{5}{1.5}{gray}
    
    \node[] at (4, -11) {\small$\mathbf{x_l}$};
    \node[] at (-8, -11) {\small$\mathbf{x_u}$};

    \node[] at (-11, -4) {\small$\mathbf{y_l}$};
    \node[] at (-11, 4) {\small$\mathbf{y_u}$};
    
	\dtwoqfour{(8, -4)}{0.1cm}{2}{5}{1.5}{black}
    \dtwoqfour{(8, 0)}{0.1cm}{2}{5}{1.5}{black}
    \dtwoqfour{(8, 4)}{0.1cm}{2}{5}{1.5}{black}

    \dtwoqfour{(4, -4)}{0.1cm}{2}{5}{1.5}{black}
    \dtwoqfour{(4, 0)}{0.1cm}{2}{5}{1.5}{black}
    \dtwoqfour{(4, 4)}{0.1cm}{2}{5}{1.5}{black}
    
    \dtwoqfour{(-8, -4)}{0.1cm}{2}{5}{1.5}{black}
    \dtwoqfour{(-8, 0)}{0.1cm}{2}{5}{1.5}{black}
    \dtwoqfour{(-8, 4)}{0.1cm}{2}{5}{1.5}{black}
    
    \draw[-,solid, ultra thick, color=black, rounded corners, style=densely dashdotdotted] (-10,-6) -- (-6,-6) -- (-6, 6) -- (-10, 6);
    \draw[-,solid, ultra thick, color=black, rounded corners] (10,-6) -- (2,-6) -- (2, 6) -- (10, 6);
\end{tikzpicture}
\end{minipage}
\hfill
\begin{minipage}{0.4\textwidth}
\begin{table}[H]
\centering
\begin{tabular}{c|c|c}
\toprule
\multirow{2}{*}{\textbf{Condition}} & \multicolumn{2}{c}{\textbf{Domain}} \\
\cmidrule(l){2-3}
& \tikz[baseline]{\draw[] (0.5, 0.5) -- (0, 0.5) -- (0, 0) -- (0.5, 0);} & \tikz[baseline]{\draw[densely dashdotdotted] (0, 0) -- (0.5, 0) -- (0.5, 0.5) -- (0, 0.5);} \\
\midrule
${x_l \leq x}$ & \checkmark & \\
${x_u \geq x}$ & & \checkmark \\
${y_l \leq y}$ & \checkmark & \checkmark \\
${y_u \geq y}$ & \checkmark & \checkmark \\
\bottomrule
\end{tabular}
\caption{Conditions determining the domains. \label{tab:sp-2-1-init-diag-sp-grid-init-edge-cases-1}}
\end{table}
\end{minipage}}%
	\hfill%
	\\
	\hfill
	\subcaptionbox{Edge case for the application of boundary conditions with periodic overflow in $2$ dimensions. \label{fig:sp-2-1-init-diag-sp-grid-init-edge-cases-2}}{\centering
\begin{minipage}{0.55\textwidth}
\begin{tikzpicture}[scale=0.3]
    % Grid lines
    \draw[
      help lines,
      line width=0.3pt,
      color=gray!30,
      dashed
    ] (-8, -8) grid[step={($(2, 2) - (0, 0)$)}] (8, 8);
    \draw[
      help lines,
      line width=0.4pt,
      color=black!80,
      dashed
    ] (-10, -10) grid[step={($(4, 4) - (0, 0)$)}] (10, 10);
        
    % Nodes
    % Distance 0
    \dtwoqfour{(0,0)}{0.1cm}{2}{0}{1.5}{gray}

    % Distance 1
    \dtwoqfour{(4,0)}{0.1cm}{2}{1}{1.5}{gray}
    \dtwoqfour{(-4,0)}{0.1cm}{2}{2}{1.5}{gray}
    \dtwoqfour{(0,4)}{0.1cm}{2}{3}{1.5}{gray}
    \dtwoqfour{(0,-4)}{0.1cm}{2}{4}{1.5}{gray}

    % Distance 2
    \dtwoqfour{(8,0)}{0.1cm}{2}{5}{1.5}{gray}
    \dtwoqfour{(-8,0)}{0.1cm}{2}{5}{1.5}{gray}
    \dtwoqfour{(4,4)}{0.1cm}{2}{5}{1.5}{gray}
    \dtwoqfour{(-4,4)}{0.1cm}{2}{5}{1.5}{gray}
    \dtwoqfour{(-4,-4)}{0.1cm}{2}{5}{1.5}{gray}
    \dtwoqfour{(4,-4)}{0.1cm}{2}{5}{1.5}{gray}
    \dtwoqfour{(0,-8)}{0.1cm}{2}{5}{1.5}{gray}
    \dtwoqfour{(0,8)}{0.1cm}{2}{5}{1.5}{gray}

    \dtwoqfour{(4,8)}{0.1cm}{2}{5}{1.5}{gray}
    \dtwoqfour{(8,4)}{0.1cm}{2}{5}{1.5}{gray}
    \dtwoqfour{(8,8)}{0.1cm}{2}{5}{1.5}{gray}
    \dtwoqfour{(-4,8)}{0.1cm}{2}{5}{1.5}{gray}
    \dtwoqfour{(-8,4)}{0.1cm}{2}{5}{1.5}{gray}
    \dtwoqfour{(-8,8)}{0.1cm}{2}{5}{1.5}{gray}
    \dtwoqfour{(4,-8)}{0.1cm}{2}{5}{1.5}{gray}
    \dtwoqfour{(8,-4)}{0.1cm}{2}{5}{1.5}{gray}
    \dtwoqfour{(8,-8)}{0.1cm}{2}{5}{1.5}{gray}
    \dtwoqfour{(-4,-8)}{0.1cm}{2}{5}{1.5}{gray}
    \dtwoqfour{(-8,-4)}{0.1cm}{2}{5}{1.5}{gray}
    \dtwoqfour{(-8,-8)}{0.1cm}{2}{5}{1.5}{gray}
    \node[] at (4, -11) {\small$\mathbf{x_0}$};
    \node[] at (-8, -11) {\small$\mathbf{x_1}$};

    \node[] at (-11, -4) {\small$\mathbf{y_u}$};
    \node[] at (-11, 4) {\small$\mathbf{y_l}$};

    \dtwoqfour{(8, 8)}{0.1cm}{2}{5}{1.5}{black}
    \dtwoqfour{(8, 4)}{0.1cm}{2}{5}{1.5}{black}
    \dtwoqfour{(4, 8)}{0.1cm}{2}{5}{1.5}{black}
    \dtwoqfour{(4, 4)}{0.1cm}{2}{5}{1.5}{black}

    \dtwoqfour{(-8, 8)}{0.1cm}{2}{5}{1.5}{black}
    \dtwoqfour{(-8, 4)}{0.1cm}{2}{5}{1.5}{black}

    \dtwoqfour{(-8, -8)}{0.1cm}{2}{5}{1.5}{black}

    \dtwoqfour{(8, -8)}{0.1cm}{2}{5}{1.5}{black}
    \dtwoqfour{(4, -8)}{0.1cm}{2}{5}{1.5}{black}
    
    \draw[-,solid, ultra thick, color=black, rounded corners] (10, 2) -- (2, 2) -- (2, 10);
    \draw[-,solid, ultra thick, color=black, style=densely dotted, rounded corners] (10, -6) -- (2, -6) -- (2, -10);
    
    \draw[-,solid, ultra thick, color=black, style=densely dashdotdotted, rounded corners] (-10, 2) -- (-6, 2) -- (-6, 10);
    \draw[-,solid, ultra thick, color=black, style=densely dashed, rounded corners] (-10, -6) -- (-6, -6) -- (-6, -10);
\end{tikzpicture}
\end{minipage}
\hfill
\begin{minipage}{0.4\textwidth}
\begin{table}[H]
\centering
\begin{tabular}{c|c|c|c|c}
\toprule
\multirow{2}{*}{\textbf{Condition}} & \multicolumn{4}{c}{\textbf{Domain}} \\
\cmidrule(l){2-5}
 & \tikz[baseline]{\draw[] (0, 0.5) -- (0, 0) -- (0.5, 0);} & \tikz[baseline]{\draw[densely dotted] (0.5, 0.5) -- (0, 0.5) -- (0, 0);} & \tikz[baseline]{\draw[densely dashed] (0, 0.5) -- (0.5, 0.5) -- (0.5, 0);} & \tikz[baseline]{\draw[densely dashdotdotted] (0, 0) -- (0.5, 0) -- (0.5, 0.5);}\\
\midrule
${x_l \leq x}$ & \checkmark & \checkmark & & \\
${x_u \geq x}$ & & & \checkmark & \checkmark \\
${y_l \leq y}$ & \checkmark & & & \checkmark \\
${y_u \geq y}$ & & \checkmark & \checkmark & \\
\bottomrule
\end{tabular}
\caption{Conditions determining the domains. \label{tab:sp-2-1-init-diag-sp-grid-init-edge-cases-2}}
\end{table}
\end{minipage}}%
	\hfill%
	\caption{Comparison of edge cases for \dq{2}{4} volumetric initialization. \label{fig:sp-2-1-init-diag-sp-grid-init-edge-cases}}    
\end{figure}

\Cref{fig:sp-2-1-init-diag-sp-grid-init-edge-cases}a depicts the nominal initialization scenario,
in which the gridpoints that the volumetric operation primes
all belong to one contiguous region, and the periodic boundary conditions do not affect initialization.
In this instance, the $4$ ancilla qubits encoding conditions listed in 
\Cref{tab:sp-2-1-init-diag-sp-grid-init-edge-cases-0} act as controls
for the corresponding $\mathrm{CX}$ gate(s) simultaneously.
A more complex scenario occurs when initializing the volume of gridpoints
that are at a distance of $(+2, 0)$ from the physical origin, as shown in
\Cref{fig:sp-2-1-init-diag-sp-grid-init-edge-cases}b.
Under these conditions, the gridpoints belonging to the volume are split
by the $x$-boundary of the domain, and the $4$ conditions describing the
square are no longer sound, as now $x_l > x_u$.

Notably, the same encoding over the $4$ ancilla qubits,
together with minimal changes to the quantum circuit
are sufficient to fit such a scenario.
The two volumes can now be addressed separately, by using
a subset of the $4$ conditions for each of them, as listed in 
\Cref{tab:sp-2-1-init-diag-sp-grid-init-edge-cases-1}.
The $\mathrm{CX}$ gate(s) that instantiate the desired velocity
profile onto the part of the volume delimited by the solid line
are controlled on all qubits except for the one encoding $x_u \geq x$,
as the upper bound is now the boundary of the domain in the $x$ dimension,
and therefore all gridpoints (including the ones in the target domain)
implicitly adhere to this condition.
Symmetrically, the remainder part of the domain no longer requires the 
$x_l \leq x$ condition to hold, and therefore controls are only placed on both
bounds of the non-overflowing dimension and on
the upper bound of the overflowing dimension, $x_u$.
Using this method, the comparator circuits used for the scenario of
\Cref{tab:sp-2-1-init-diag-sp-grid-init-edge-cases-0}
are structurally similar,
except for the positional shift,
and the only difference in complexity stems from the different
utilization of the $\mathrm{CX}$ gate(s).

The same straightforward reasoning about the boundaries of the domain
enable the efficient initialization of the flow field
in the most complex scenario that can occur in two dimensions,
described in \Cref{fig:sp-2-1-init-diag-sp-grid-init-edge-cases}c
and \Cref{tab:sp-2-1-init-diag-sp-grid-init-edge-cases-2}.
This instance corresponds to initializing the grid qubits
distanced $(+2, +2)$ from the physical origin, which is the farthest
away that information can propagate within $4$ time steps.
Under these circumstances, the contiguous volume is split into
$4$ segments, each of which can be characterized by two conditions.
The realizations of this circuit only requires $4$ consecutive applications
of $\mathrm{C^2X}$ gates per segment.
In general, the number of overflows that can occur is at most $2^d$, as
the domain can overflow in each dimension independently,
and is therefore not linked to the velocity discretization.
For a domain that has overflown in $0 \leq o \leq d$ dimensions,
the velocity profile can be instantiated onto the volume
by means of $2^{o + 1}$ multi-controlled $\X$, each controlled on $2d-o$ qubits.

\paragraph{Complexity analysis.}
Generalizing the previous example to an arbitrary
\dq{d}{q} discretization, the operation of setting one relative
neighbor to the appropriate velocity profile requires $2d$
comparator operations, each consisting of one application of the
$\qft$ and one application of the $\qft^\dagger$.
Both of these operations are applied to $n_g + 1$ qubits
and can therefore be decomposed into $\mathcal{O}(n_g^2)$
Hadamard and controlled phase shift gates \cite{musk2020comparison}.
In addition to the $\qft$ layers, the comparators require an additional
$\mathcal{O}(n_g)$ phase gates, which does not affect the order of the scaling.
There are at most $2^d$ controlled $\X$ gates that set the velocity
profile of each gridpoint, which require $O(d)$ controls each.
Therefore, the $\mathcal{O}(2^d)\ \CPX{d}$ gates of this step can be decomposed into
$\mathcal{O}(2^dd^2)\ \CX$ gates each,
following the decomposition of \citet{barenco1995elementary}.
To obtain a consistent quantum state, the application should be
repeated for all gridpoints in the space-time data structure
from which information can reach the origin, incurring
a cost of $N_t^d$, much like the pointwise method.
The key difference lies in that, unlike the pointwise method,
the volumetric initialization technique addresses
all gridpoints in the target volume simultaneously, requiring only one
traversal of the stencil instead of the $N_g$.
Thus, the cumulative complexity of the volumetric initialization
technique requires $\mathcal{O}(N_t^d(n_g^2 + 2^dd^2))$ native quantum gates.
Finally, we note that the lower computational
cost of the volumetric application of initial conditions
comes at the expense of expressiveness.
For volumetric operations to be applicable, the entire volume
must be initialized to exactly the same velocity configuration for all gridpoints
contained within.

\section{Streaming and Boundary Conditions\label{sec:sp-2-2-streaming-boundary}}

In this section, we address the streaming and boundary condition
steps of the LGA loop, as they both account for the
movement of particles across the discrete velocity channels.
Streaming is particularly efficient in the Space-Time encoding,
as the extended computational basis state allows
for particle occupancy sites to propagate
by means of swap gates, as described in
Section 4 of the original paper \cite{schalkers2024importance}.

The core concept behind the generalization of the streaming procedure
is that of a so-called \emph{streaming line}.
A streaming line in the Space-Time stencil
conists of the qubits that encode the same velocity channel $c$
of neighboring gridpoints linked by $c$ and its opposing velocity channel $\bar{c}$.
The \dq{1}{2} discretization contains a single streaming line,
while \dq{2}{4} includes $4(N_t - 1) + 2$ streaming lines, as each new
time step introduces $4$ additional streaming lines and increases the size
of each existing streaming line by $2$.
Generally \dq{d}{q} discretizations require that each gridpoint is traversed by
$\lfloor q/2 \rfloor$ streaming lines per gridpoint due to the potential inclusion of rest particles.
Since particles only travel along discrete velocity channels once
per time step, each additional time step of the simulation requires $\mathcal{O}(q)$
additional swaps per streaming line, for each of the $\mathcal{O}(N_t^d)$
gridpoints that reside at the edges of their respective streaming lines,
leading to a total number of swap gates that scales with $\mathcal{O}(qN_t^d)$.
Since streaming lines are independent, the same arrangement of 
swap gates described in the original paper for \dq{2}{4} \cite{schalkers2024importance}
can be applied per each direction of the
streaming line, leading to a logarithmic circuit depth.

Conceptually, the application of boundary conditions in LGA
is closely related to streaming,
as particles are simply transported in physical space along velocity channels.
Unlike in streaming, however, the precise trajectory particles
traverse during their interaction with solids
depends on the shape of the geometry,
and the particular kind of boundary condition that is prescribed
(\ie bounce-back or specular reflection).
Moreover, the implementation of this subroutine must
preserve the locality and reversibility
that the Space-Time encoding relies on.
In the remainder of this section, we address how bounce-back and specular reflection
boundary conditions around rigid bodies
can be implemented in the Space-Time encoding,
and introduce both pointwise and volumetric building blocks.

\subsection{Pointwise Imposition of Boundary Conditions\label{subsec:sp-2-2-boundary-pw}}

The application of boundary conditions using the pointwise method
follows the same rationale detailed in \Cref{subsec:sp-2-1-initial-conditions-pw},
where each gridpoint is addressed sequentially, and each relative
position of the Space-Time stencil is treated individually.
The implementation differs, however, in the operation that takes place at each gridpoint.
Where initial conditions simply require that the state
of the qubit relating to one particular velocity channel is conditionally flipped,
boundary conditions necessitate that the information is transferred from one gridpoint to another.
Implementing such an operation is equivalent to
realizing a controlled streaming step with subtly different semantics.
The circuit that performs such an operation can be realized
by means of a multi-controlled swap
operation for pre-determined velocity channels.

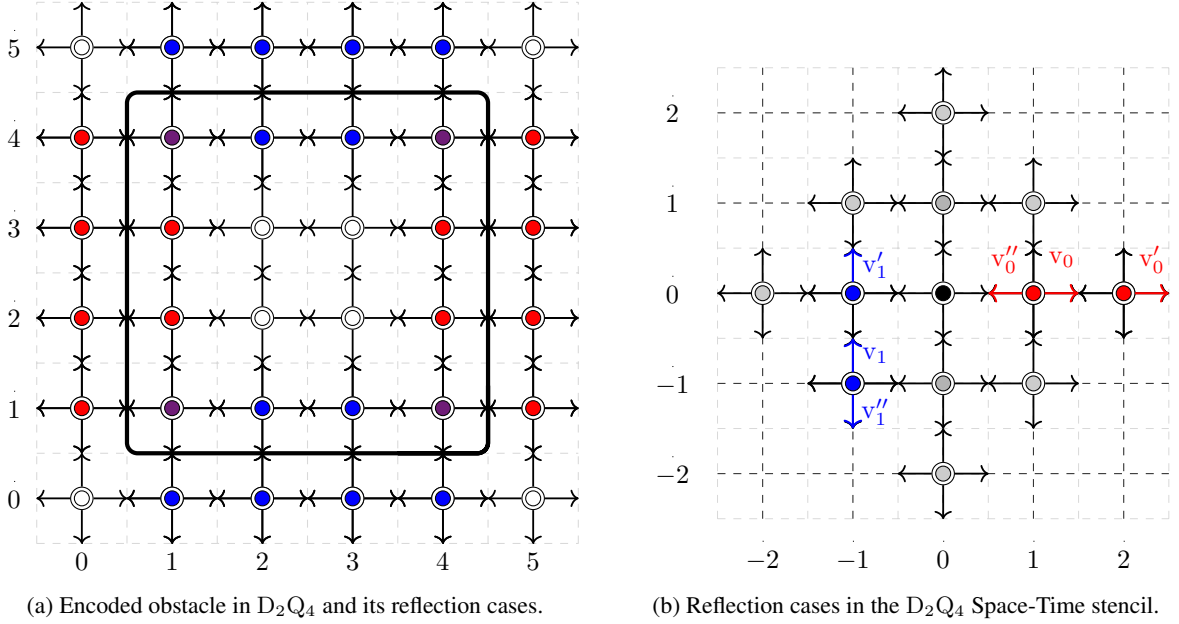
\begin{figure}
    \centering
    \hfill
    \subcaptionbox{Encoded obstacle in \dq{2}{4} and its reflection cases. \label{fig:sp-2-3-diag-sp-boundary-square-pointwise-grid}}{\centering
\begin{tikzpicture}[scale=0.3]
    % Grid lines

    \draw[
      help lines,
      line width=0.4pt,
      color=gray!30,
      dashed
    ] (-12, -12) grid[step={($(4, 4) - (0, 0)$)}] (12, 12);

    \foreach \x in {-10,-6,-2,2,6,10} {
        \foreach \y in {-10,-6,-2,2,6,10} {
            \dtwoqfour{(\x,\y)}{0.1cm}{2}{0}{1.5}{white}
        }
    }

    % Axes labels
    \foreach \x/\xtext in {0,...,5} {
        \draw (\x*4-10,-12) -- (\x*4-10,-12) node[anchor=north] {$\xtext$};
    }

    \foreach \x/\xtext in {0,...,5} {
        \draw (-13,\x*4-9.25) -- (-13,\x*4-9.25) node[anchor=north] {$\xtext$};
    }
    
    \dtwoqfour{(-6,-2)}{0.1cm}{2}{0}{1.5}{red}
    \dtwoqfour{(-6,2)}{0.1cm}{2}{0}{1.5}{red}
    \dtwoqfour{(6,2)}{0.1cm}{2}{0}{1.5}{red}
    \dtwoqfour{(6,-2)}{0.1cm}{2}{0}{1.5}{red}
    
    \dtwoqfour{(-2,6)}{0.1cm}{2}{0}{1.5}{blue}
    \dtwoqfour{(2,6)}{0.1cm}{2}{0}{1.5}{blue}
    \dtwoqfour{(-2,-6)}{0.1cm}{2}{0}{1.5}{blue}
    \dtwoqfour{(2,-6)}{0.1cm}{2}{0}{1.5}{blue}
    
    \dtwoqfour{(6,6)}{0.1cm}{2}{0}{1.5}{tud purple}
    \dtwoqfour{(-6,-6)}{0.1cm}{2}{0}{1.5}{tud purple}
    \dtwoqfour{(-6,6)}{0.1cm}{2}{0}{1.5}{tud purple}
    \dtwoqfour{(6,-6)}{0.1cm}{2}{0}{1.5}{tud purple}
    
    \foreach \x in {-10, 10} {
        \foreach \y in {-6,-2,2,6} {
            \dtwoqfour{(\x,\y)}{0.1cm}{2}{0}{1.5}{red}
        }
    }

    \foreach \x in {-10, 10} {
        \foreach \y in {-6,-2,2,6} {
            \dtwoqfour{(\y,\x)}{0.1cm}{2}{0}{1.5}{blue}
        }
    }

\draw[-,solid, ultra thick, black, rounded corners] (4, -8) -- (8,-8) -- (8,8) -- (-8,8) -- (-8, -8) -- (8, -8) -- (8, -5);
\end{tikzpicture}}%
    \hfill
    \subcaptionbox{Reflection cases in the \dq{2}{4} Space-Time stencil. \label{fig:sp-2-3-diag-sp-boundary-square-pointwise-stencil}}{\centering
\begin{tikzpicture}[scale=0.3]
    % Grid lines
    \draw[
      help lines,
      line width=0.3pt,
      color=gray!30,
      dashed
    ] (-10, -10) grid[step={($(2, 2) - (0, 0)$)}] (10, 10);
    \draw[
      help lines,
      line width=0.4pt,
      color=black!80,
      dashed
    ] (-10, -10) grid[step={($(4, 4) - (0, 0)$)}] (10, 10);
        
    % Nodes
    % Distance 0
    \dtwoqfour{(0,0)}{0.1cm}{2}{0}{1.5}{black}

    % Distance 1
    \dtwoqfour{(4,0)}{0.1cm}{2}{1}{1.5}{gray!60}
    \dtwoqfour{(-4,0)}{0.1cm}{2}{2}{1.5}{gray!60}
    \dtwoqfour{(0,4)}{0.1cm}{2}{3}{1.5}{gray!60}
    \dtwoqfour{(0,-4)}{0.1cm}{2}{4}{1.5}{gray!60}

    % Distance 2
    \dtwoqfour{(8,0)}{0.1cm}{2}{5}{1.5}{gray!40}
    \dtwoqfour{(-8,0)}{0.1cm}{2}{5}{1.5}{gray!40}
    \dtwoqfour{(4,4)}{0.1cm}{2}{5}{1.5}{gray!40}
    \dtwoqfour{(-4,4)}{0.1cm}{2}{5}{1.5}{gray!40}
    \dtwoqfour{(-4,-4)}{0.1cm}{2}{5}{1.5}{gray!40}
    \dtwoqfour{(4,-4)}{0.1cm}{2}{5}{1.5}{gray!40}
    \dtwoqfour{(0,-8)}{0.1cm}{2}{5}{1.5}{gray!40}
    \dtwoqfour{(0,8)}{0.1cm}{2}{5}{1.5}{gray!40}
    
    % Axes labels
    \foreach \x/\xtext in {-2,...,2} {
        \draw (\x*4,-11) -- (\x*4,-11) node[anchor=north] {$\xtext$};
    }

    \foreach \x/\xtext in {-2,...,2} {
        \draw (-12,\x*4+0.75) -- (-12,\x*4+0.75) node[anchor=north] {$\xtext$};
    }
    
    \dtwoqfour{(4,0)}{0.1cm}{2}{5}{1.5}{red}
    \dtwoqfour{(8,0)}{0.1cm}{2}{5}{1.5}{red}
	\draw[->, thick, red] (4.5, 0) -- (6, 0) node at (5.25, 1.5) {$\mathrm{v_0}$};
	\draw[->, thick, red] (8.5, 0) -- (10, 0) node at (9.25, 1.5) {$\mathrm{v_0'}$};
	\draw[->, thick, red] (3.5, 0) -- (2, 0) node at (2.75, 1.5) {$\mathrm{v_0''}$};
	
	\dtwoqfour{(-4,0)}{0.1cm}{2}{5}{1.5}{blue}
    \dtwoqfour{(-4,-4)}{0.1cm}{2}{5}{1.5}{blue}
    \draw[->, thick, blue] (-4, -3.5) -- (-4, -2) node at (-3, -2.75) {$\mathrm{v_1}$};
    \draw[->, thick, blue] (-4, 0.5) -- (-4, 2) node at (-3, 1.25) {$\mathrm{v_1'}$};
    \draw[->, thick, blue] (-4, -4.5) -- (-4, -6) node at (-3, -5.25) {$\mathrm{v_1''}$};
\end{tikzpicture}}%
    \hfill%
    \caption{Reflection cases of a square spanning $(1, 4) \times (1, 4)$ on a $6\times 6$ \dq{2}{4} grid in $2$-time step Space-Time stencil. \label{fig:sp-2-3-diag-sp-boundary-square-pointwise}}    
\end{figure}

\paragraph{Example -- \dq{2}{4} square.}
\Cref{fig:sp-2-3-diag-sp-boundary-square-pointwise} depicts
the application on boundary conditions on a $6 \times 6$ \dq{2}{4} grid
where particles reflect off a solid square obstacle spanning $(1, 4) \times (1, 4)$.
The physical grid is shown in
\Cref{fig:sp-2-3-diag-sp-boundary-square-pointwise-grid}, where
red gridpoints exchange particle
occupancy information across the $y$-axis
channels, while gridpoints colored blue do so along the $x$-axis.
Purple gridpoints are at the corners of the object,
and exchange information through both axes.
For instance, the red gridpoint at $(1, 2)$ only reflects the particle that
travelled inside the object through $(0, 2)$, while the blue gridpoint at position
$(2, 1)$ exchanges information with its neighbor at $(0, 1)$.

\Cref{fig:sp-2-3-diag-sp-boundary-square-pointwise-stencil} tracks
this interaction in the Space-Time stencil, where $\mathrm{v_0}$ and $\mathrm{v_1}$ are
the two particles described in the previous scenario \emph{before streaming}.
Following streaming, $\mathrm{v}_j$ is swapped onto the qubit
coresponding to its neighbor, $\mathrm{v}_j'$,
which in this case is positioned inside the solid obstacle.
The boundary condition circuit then performs a second swap between
$\mathrm{v}_j'$ and $\mathrm{v}_j''$, placing the particle onto the opposite direction of the
same streaming line of the neighbor it originated from.
Since there are no particles inside the object before streaming and streaming
only happens across streaming lines, the qubit encoding
$\mathrm{v}_j''$ is always in the state $\ket{0}$ before this operation takes place,
which guarantees the reversibility of the method.

\begin{figure}
	\input{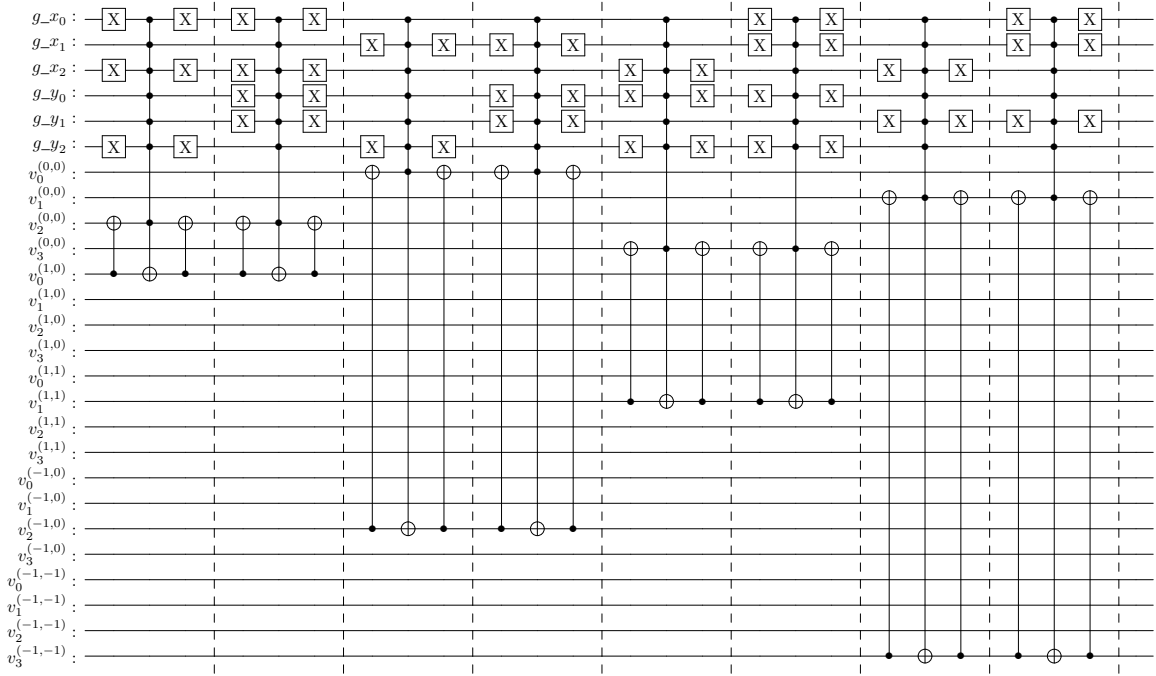}
	\caption{\dq{2}{4} pointwise reflection for a square spanning $(1, 4) \times (1, 4)$ for one time step. \label{fig:sp-2-3-reflection-circuits-circ-sp-refleciton-d2q4-1-timestep-pw}}
\end{figure}

The circuit implementing pointwise boundary conditions for a more concise
instance of a $(1, 4) \times (1 , 4)$ square obstacle for one time step is
shown in \Cref{fig:sp-2-3-reflection-circuits-circ-sp-refleciton-d2q4-1-timestep-pw}.
The structure of the circuit is the same as the initial conditions example,
except for the multi-controlled swap operation,
which in this case we decompose into one multi-controlled $\X$ gate
and two $\CX$ gates using the method described by \citet{heese2022representation}.

\paragraph{The feasibility of Specular Reflection.}
The application of post-streaming swaps between fixed points in the Space-Time stencil
can be tweaked to implement various kinds of LGA boundary conditions.
Bounce-back boundary conditions are applicable in the Space-Time encoding
in all common \dq{d}{q} discretizations, as following
reflection against the wall through channel $c$,
particles occupy channel $\bar{c}$ of the same gridpoint,
which guarantees that information is contained within
the locality of the stencil.
Specular reflection boundary conditions, on the other hand,
require that the trajectory of particles
is reversed only in the dimension(s) in which the
particle made contact with the solid surface
(\ie, the direction(s) normal to the surface of contact).
Whether this velocity channel belongs to a gridpoint
already encoded in the stencil is dependent on the discretization.
While commonplace discretizations like \dq{2}{9} are fully suitable for specular
reflection in the Space-Time encoding,
Stencils that contain diagonal channels that
connect gridpoints in such a way that the gridpoint
at which the particle arrives post-collision is not
connected directly to the origin by another channel are less suitable.
The \dq{3}{8} model propsed by \cite{nor2007three} is one such discretization.
To accomodate models with these properties, the notion of locality in the stencil has
to be extended such that all gridpoints affected by streaming and boundary conditions
within $N_t$ time steps are reachable.
This does not, however, affect the Space-Time encoding in the limit,
where all velocity channels in the lattice are allotted a qubit.
In this instance, sound reflection semantics of many kinds
are efficiently implemented by means of regular (non-controlled) $\mathrm{SWAP}$
gates, irrespective of the discretization. 

\paragraph{Complexity Analysis.}
As is the case of initial conditions, pointwise boundary conditions can be used to implement arbitrary
geometrical shapes under arbitrary discretizations, at the expense of performance.
To reflect $1$ particle between two positions
in the Space-Time stencil, the circuit requires $\mathcal{O}(n_g)$ $\X$ gates
twice to prepare and revert the $\ket{1}^{\otimes n_g}\ket{v}$ state, as well as $2$ $\CX$
gates and one $\CPX{n_g+1}$ gate to decompose the $\mathrm{C}^{n_g}\mathrm{SWAP}$
gate, as described in \citep{heese2022representation}, for a total
of $\mathcal{O}(n_g^2)$ native gates.
Each particle reflection must take place in each stencil where the pair
of points that exchange information are both present.
For \dq{2}{4}, the number of pairs is

\begin{align}
\begin{split}
2N_t + 4\sum_{t=1}^{N_t}t = 2N_t^2 + 2N_t - 2,
\end{split}
\end{align}

as each additional time step to be simulated
adds $2$ additional pairs of points for the previous layer of the stencil,
as well as $2$ additional pairs at the extremities.
In general this number scales again with $\mathcal{O}(qN_t^d)$.
This procedure should in general be repeated for the $\mathcal{O}(q)$
velocity channels that particles travel on, as well as for each
grid point at the perimeter of the solid domain.
Assuming there are $N_p$ such perimeter points, and
$N_p \leq N_g$, the overall complexity of the boundary condition step
is $\mathcal{O}(N_pN_t^dq(n_g + n_g^2))$.

The locality of the Space-Time encoding affords several techniques
that in practice can diminish the complexity of the
circuits that implement boundary conditions.
First, any pair of points in the Space-Time stencil that
would be subject to boundary condition treatment, but which are entangled
to gridpoints inside of the solid domain can be neglected.
Since information is always local to the stencil,
the state of the velocity qubits at that grid location
remains $\ket{0}^{\otimes\mathcal{O}(N_t^d)}$, and the swaps are therefore pointless.
The same argument can be used to reduce the complexity of the initial conditions described in
the previous section, as any velocity information
entangled to positional qubits that encode positions inside the solid domain can be skipped.
Finally, there are instances in which the application of the pointwise method
requires $\mathcal{O}(N_g)$ repetitions.
Two such instances include (i) situations in which the space-time
stencil is large enough to approximate a sizeable
proportion of the entire physical domain, and (ii) cases in which the
surface of the solid geometry covers a large area of the domain.
In cases where the stencils are relatively small and the geometry is enclosed to a small volume
of the domain, pointwise boundary conditions may prove practically feasible.
For instances where this is not the case, we next introduce the volumetric
alternative to boundary conditions, which significantly improves scaling.

\subsection{Volumetric Imposition of Boundary Conditions\label{subsec:sp-2-2-boundary-vol}}

The application of volumetric operations to
boundary conditions follows the same rationale as in \Cref{subsec:sp-2-1-initial-conditions-vol} --
each volume of gridpoints where the same
operation is simultaneously addressed after isolating it
by means of comparator operations.
Similar to initial conditions, this operation must be performed
for each volume where the information is present in the Space-Time stencil.
Unlike initial conditions, however, the properties of this volume
are dependent on the shape of the solid domain and the lattice discretization.
In general, the volumes affected by boundary conditions are only concerned
with the gridpoints that lie at the interface between the solid and the fluid domains.
As a consequence, this volume is at most $(d-1)$-dimensional,
which decreases both the number of qubits and the complexity
of the circuits that implement these operations.

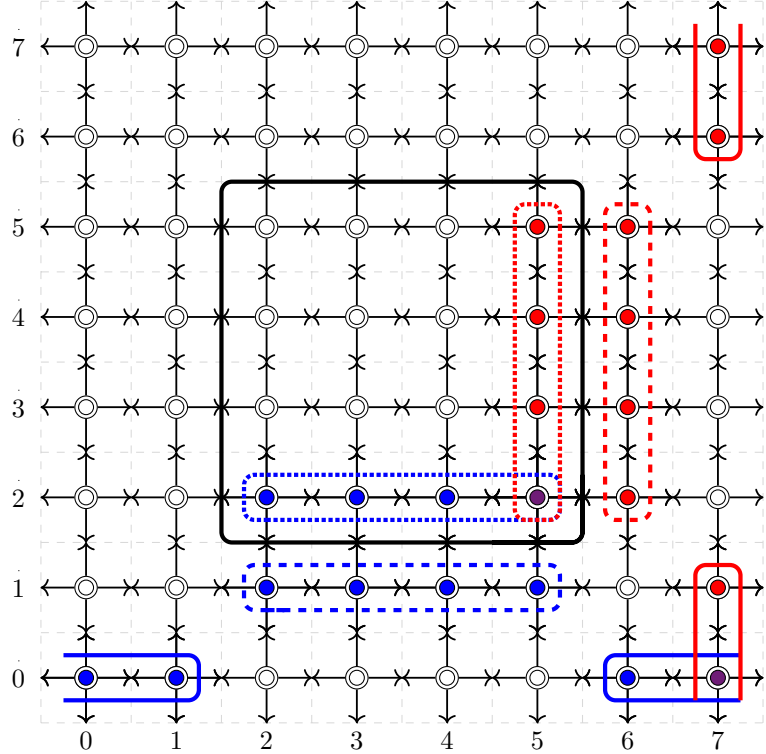
\begin{figure}
	\centering
\begin{tikzpicture}[scale=0.3]
    % Grid lines

    \draw[
      help lines,
      line width=0.4pt,
      color=gray!30,
      dashed
    ] (-16, -16) grid[step={($(4, 4) - (0, 0)$)}] (16, 16);

    \foreach \x in {-14,-10,-6,-2,2,6,10,14} {
        \foreach \y in {-14,-10,-6,-2,2,6,10,14} {
            \dtwoqfour{(\x,\y)}{0.1cm}{2}{0}{1.5}{white}
        }
    }

    % Axes labels
    \foreach \x/\xtext in {0,...,7} {
        \draw (\x*4-14,-16) -- (\x*4-14,-16) node[anchor=north] {$\xtext$};
    }

    \foreach \x/\xtext in {0,...,7} {
        \draw (-17,\x*4-13.25) -- (-17,\x*4-13.25) node[anchor=north] {$\xtext$};
    }

     \foreach \y in {-6,-2,2,6} {
            \dtwoqfour{(6,\y)}{0.1cm}{2}{0}{1.5}{red}
            \dtwoqfour{(10,\y)}{0.1cm}{2}{0}{1.5}{red}
            \dtwoqfour{(\y, -6)}{0.1cm}{2}{0}{1.5}{blue}
            \dtwoqfour{(\y, -10)}{0.1cm}{2}{0}{1.5}{blue}
     }
     
     \dtwoqfour{(6, -6)}{0.1cm}{2}{0}{1.5}{tud purple}
     \dtwoqfour{(14, -14)}{0.1cm}{2}{0}{1.5}{tud purple}
     \dtwoqfour{(14, -10)}{0.1cm}{2}{0}{1.5}{red}
     \dtwoqfour{(14, 10)}{0.1cm}{2}{0}{1.5}{red}
     \dtwoqfour{(14, 14)}{0.1cm}{2}{0}{1.5}{red}
     \dtwoqfour{(10, -14)}{0.1cm}{2}{0}{1.5}{blue}
     \dtwoqfour{(-10, -14)}{0.1cm}{2}{0}{1.5}{blue}
     \dtwoqfour{(-14, -14)}{0.1cm}{2}{0}{1.5}{blue}

\draw[-,solid, ultra thick, black, rounded corners] (4, -8) -- (8,-8) -- (8,8) -- (-8,8) -- (-8, -8) -- (8, -8) -- (8, -5);

\draw[-,densely dotted, ultra thick, blue, rounded corners] (-6, -7) -- (7,-7) -- (7, -5) -- (-7, -5) -- (-7, -7) -- (-6, -7);
\draw[-,dashed, ultra thick, blue, rounded corners] (-6, -11) -- (7,-11) -- (7, -9) -- (-7, -9) -- (-7, -11) -- (-5, -11);
\draw[-,solid, ultra thick, blue, rounded corners] (15,-15) -- (9, -15) -- (9, -13) -- (15, -13);
\draw[-,solid, ultra thick, blue, rounded corners] (-15,-15) -- (-9, -15) -- (-9, -13) -- (-15, -13);

\draw[-,densely dotted, ultra thick, red, rounded corners] (7, 6) -- (7,-7) -- (5,-7) -- (5, 7) -- (7, 7) -- (7, 6);
\draw[-,dashed, ultra thick, red, rounded corners] (11, 6) -- (11,-7) -- (9,-7) -- (9, 7) -- (11, 7) -- (11, 6);
\draw[-,solid, ultra thick, red, rounded corners] (15, 15) -- (15,9) -- (13,9) -- (13, 15);
\draw[-,solid, ultra thick, red, rounded corners] (15, -15) -- (15,-9) -- (13,-9) -- (13, -15);
\end{tikzpicture}
	\caption{\dq{2}{4} volumetric reflection cases for a square spanning $(2, 5) \times (2, 5)$ for up to $5$ time steps. \label{fig:sp-2-3-diag-sp-boundary-square-vol}}
\end{figure}

\Cref{fig:sp-2-3-diag-sp-boundary-square-vol} shows the edge cases volumetric operations
must account for in axis-aligned geometry for a \dq{2}{4} discretization.
As before, gridpoints highlighted in red exchange information along the $x$-axis,
while blue gridpoints do so in the $y$-axis.
The gridpoints surrounded by the dotted lines highlight points inside the
solid domain that particles stream into and therefore require boundary treatment.
Importantly, since these points belong to a line on the $y$-axis ($x$-axis), all 
relative neighbors in the space-time stencil that they exchange information
with also belong to a similar line.
Therefore, the use of volumetric operations described in \Cref{subsec:sp-2-1-initial-conditions-vol}
generalizes to boundary conditions in a straightforward way,
as the same techniques can be applied to isolate the gridpoints where
boundary interactions occur.
Unlike initial conditions, however, boundary conditions address
volumes that are $(d-1)$-dimensional, and can overflow therefore occurs
in at most $(d-1)$ (as opposed to $d$) dimensions.
In \Cref{fig:sp-2-3-diag-sp-boundary-square-vol}, gridpoints
surrounded by the dashed outline show contiguous volumes of space in the nominal case,
while gridpoints surrounded by solid lines (corresponding to distances $(+2, +4)$ and $(-2, -4)$, respecitvely)
highlight how overflow can occur on the physical grid.

\begin{figure}
	\input{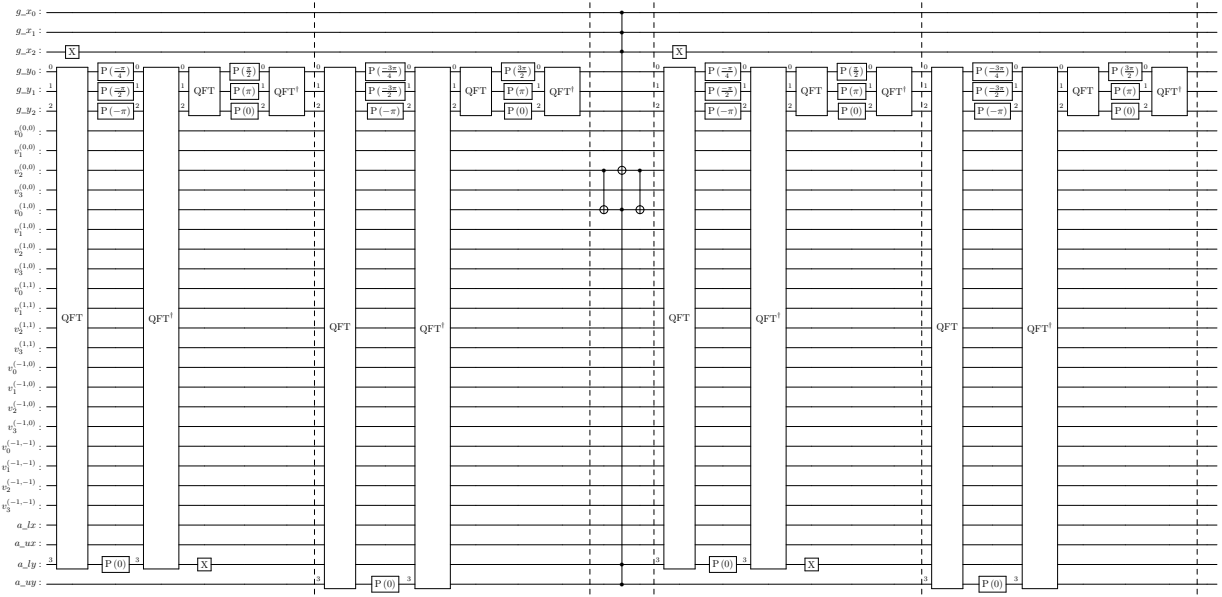}
	\caption{\dq{2}{4} volumetric reflection for the wall of a square spanning $(2, 5) \times (2, 5)$. \label{fig:sp-2-3-circ-sp-refleciton-d2q4-1-timestep-vol}}
\end{figure}

\Cref{fig:sp-2-3-circ-sp-refleciton-d2q4-1-timestep-vol} shows the quantum circuit that
performs the volumetric imposition of boundary conditions for the gridpoints
surrounded by the red dashed line in \Cref{fig:sp-2-3-diag-sp-boundary-square-vol}.
This volume spans the interval $[2, 5]$ on the $y$-axis and has a fixed position of $6$ on the $x$-axis.
Up to the second barrier,
the circuit sets the $\mathrm{a_{l_y}}$ and $\mathrm{a_{u_y}}$
ancillae to the appropriate state
by means of the $\qft$-based Draper comparator, while the $\X$ gate on the $\mathrm{g\_x_2}$
converts the state of the $x$ grid qubits from $\ket{6} \equiv \ket{110}$ to $\ket{111}$
in preparation for the multi-controlled operation.
Following this preparation step, the same multi-controlled swap decomposition
outlined in \Cref{subsec:sp-2-2-boundary-pw} performs the
swap operation that places the particles back in the fluid domain.
Finally, the preceding operations are reversed.
The edge cases outlined by the solid demarcations
are addressed identically as described in
\Cref{fig:sp-2-1-init-diag-sp-grid-init-edge-cases} \ie,
the same ancilla qubits act as controls in multiple subsequent swaps.
Following the same line of reasoning, the volumetric imposition of boundary
conditions require $\mathcal{O}(N_sN_t^dq(n_g^2+2^{d-1}(d-1)^2))$
one- and two-qubit gates,
where $N_s$ is the number of axis-aligned segments that are subject to boundary treatment.
The additional $q$ term accounts for the fact that, depending on the stencil,
multiple operations may be required per segment
to traverse the corresponding physical space for each reflected channel.

While the volumetric methods introduced thus far decrease
the complexity of initialization and boundary treatment,
they only allow for these operations to be applied in volumes
bound by axis-aligned segments.
In what follows, we introduce extensions that broaden volumetric operations to more complex shapes.

\subsection{Staircase approximations \label{sec:staircase_approx}}

To overcome the limitations of axis-aligned volumetric operations,
we require quantum circuits that are capable of efficiently describing
the components of more intricate shapes.
In this section, we introduce primitives that isolate diagonal segments
on a $2$-dimensional grid, providing similar scaling to the axis-aligned counterparts.
To demonstrate the practical utility of these primitives, we consider an instance
of a circle discretization, which shows how
the combination of axis-aligned and diagonal segments can
describe more complex shapes.

\begin{figure}
	\centering 
 \begin{tikzpicture}[scale=0.3]
    % Grid lines
    \draw[
      help lines,
      line width=0.3pt,
      color=gray!30,
      dashed
    ] (-16, -16) grid[step={($(2, 2) - (0, 0)$)}] (20, 20);
        
    \foreach \x in {-14,-10,-6,-2,2,6,10,14,18} {
        \foreach \y in {-14,-10,-6,-2,2,6,10,14,18} {
            \dtwoqfour{(\x,\y)}{0.1cm}{2}{0}{1.5}{white}
        }
    }

    % Axes labels
    \foreach \x/\xtext in {0,...,8} {
        \draw (\x*4-14,-16) -- (\x*4-14,-16) node[anchor=north] {$\xtext$};
    }

    \foreach \x/\xtext in {0,...,8} {
        \draw (-17,\x*4-13) -- (-17,\x*4-13) node[anchor=north] {$\xtext$};
    }

    \draw[thick, dotted] (2, 2) circle (14);
    \draw[ultra thick, rounded corners] (-12, 7) -- (-12, -4) -- (-8, -4) -- (-8, -8) -- (-4, -8) -- (-4, -12)
    -- (8, -12) -- (8, -8) -- (12, -8) -- (12, -4) -- (16, -4)
    -- (16, 8) -- (12, 8) -- (12, 12) -- (8, 12) -- (8, 16)
    -- (-4, 16) -- (-4, 12) -- (-8, 12) -- (-8, 8) -- (-12, 8) -- (-12, 6);
    
        % Top part
        \dtwoqfour{(2, 14)}{0.1cm}{2}{0}{1.5}{blue}
        \dtwoqfour{(6, 14)}{0.1cm}{2}{0}{1.5}{tud purple}
        \dtwoqfour{(-2, 14)}{0.1cm}{2}{0}{1.5}{tud purple}

        % Bottom part
        \dtwoqfour{(2, -10)}{0.1cm}{2}{0}{1.5}{blue}
        \dtwoqfour{(6, -10)}{0.1cm}{2}{0}{1.5}{tud purple}
        \dtwoqfour{(-2, -10)}{0.1cm}{2}{0}{1.5}{tud purple}

        % Left
         \dtwoqfour{(-10, -2)}{0.1cm}{2}{0}{1.5}{tud purple}
         \dtwoqfour{(-10, 2)}{0.1cm}{2}{0}{1.5}{red}
         \dtwoqfour{(-10, 6)}{0.1cm}{2}{0}{1.5}{tud purple}

        % % Right
         \dtwoqfour{(14, -2)}{0.1cm}{2}{0}{1.5}{tud purple}
         \dtwoqfour{(14, 2)}{0.1cm}{2}{0}{1.5}{red}
         \dtwoqfour{(14, 6)}{0.1cm}{2}{0}{1.5}{tud purple}

         \dtwoqfour{(10, -6)}{0.1cm}{2}{0}{1.5}{tud purple}
         \dtwoqfour{(10, 10)}{0.1cm}{2}{0}{1.5}{tud purple}
         \dtwoqfour{(-6, 10)}{0.1cm}{2}{0}{1.5}{tud purple}
         \dtwoqfour{(-6, -6)}{0.1cm}{2}{0}{1.5}{tud purple}
    
    \draw[->, thick, red] (-13.5, 2) -- (-12, 2) node at (-12.75, 3) {$\mathrm{v_0}$};
    \draw[->, thick, red] (-9.5, 2) -- (-8, 2) node at (-8.75, 3) {$\mathrm{v_0'}$};
    \draw[->, thick, red] (-14.5, 2) -- (-16, 2) node at (-15.25, 3) {$\mathrm{v_0''}$};
    
    \draw[->, thick, red] (-9.5, -6) -- (-8, -6) node at (-8.75, -5) {$\mathrm{v_1}$};
    \draw[->, thick, red] (-5.5, -6) -- (-4, -6) node at (-4.75, -5) {$\mathrm{v_1'}$};
    \draw[->, thick, red] (-10.5, -6) -- (-12, -6) node at (-11.25, -5) {$\mathrm{v_1''}$};
    
    \draw[->, thick, blue] (-6, -9.5) -- (-6, -8) node at (-7, -8.75) {$\mathrm{v_2}$};
    \draw[->, thick, blue] (-6, -5.5) -- (-6, -4) node at (-7, -4.75) {$\mathrm{v_2'}$};
    \draw[->, thick, blue] (-6, -10.5) -- (-6, -12) node at (-7, -11.25) {$\mathrm{v_2''}$};
    
    \draw[thick, rounded corners, tud purple] (7, -11) -- (16, -2) -- (14, 0) -- (4, -10) -- (6, -12) -- (12, -6);
    
    \end{tikzpicture}
	\caption{Discretized approximation of a circle centered at $(4, 4)$ with a radius of $3.5$ gridpoints on a $9 \times 9$ \dq{2}{4} lattice.}
	\label{fig:sp-2-3-diag-refleciton-d2q4-circle}
\end{figure}
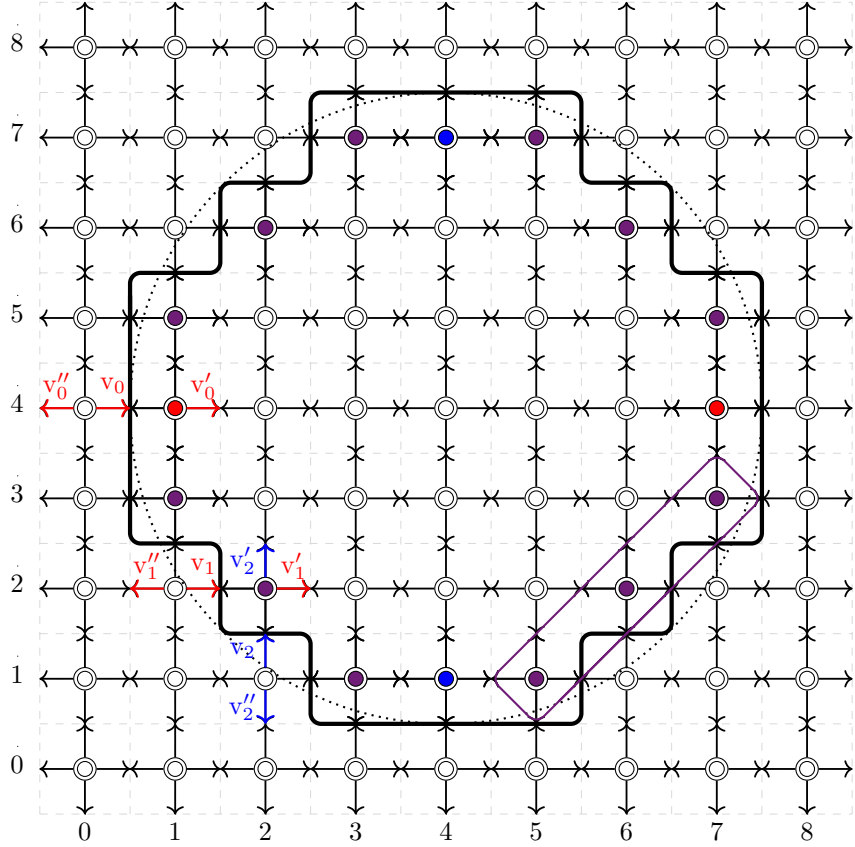

Consider the instance displayed in \Cref{fig:sp-2-3-diag-refleciton-d2q4-circle},
where we are concerned with reflecting particles off a circle centered at $(4, 4)$ on a $9\times 9$ grid.
The dotted line displays the continuous shape of the circle (with a radius of $3.5 \Delta x$),
while the solid line traces its discretized approximation.
We refer to this method of encoding solid objects with smooth
boundaries into the \dq{2}{4} grid as a \emph{staircase approximation}.
As before, the points highlighted in red are
subject to reflection across the horizontal channel,
whereas gridpoints in blue reflect only the vertical channel.
Purple gridpoints reflect both.
Of note in this staircase approximation of a circle is that points
belong to only one of three categories.
The first two categories are the axis-aligned segments described in previous sections,
which include only the $4$ red and blue points.
The third category includes segments of purple points which, importantly,
are always situated on diagonal segments.

This observation calls for the design of a primitive that,
as with axis-aligned segments for cuboid objects,
can efficiently separate diagonal points from non-diagonal
points irrespective of the segment's size.
To design such a primitive, we note that
a diagonal segment is described by 3 characteristics --
a lower bound, an upper bound, and an increment.
The increment describes the discrete
change in physical space that when iteratively applied to the
lower bound, yields all points of the segment, up to the upper bound.
Assuming an explicit ordering of the bounds, such as clockwise,
can make the increment implicit.
Consider the diagonal highlighted in 
purple in \Cref{fig:sp-2-3-diag-refleciton-d2q4-circle}.
This segment would be described by
the lower bound $(5, 1)$, the upper bound $(7, 3)$,
and the increment $(+1, +1)$.
In what follows, we describe a quantum circuit implementation
that isolates the qubits of this segment by setting ancilla qubit $\mathrm{a_d}$ to $\ket{1}$.

\begin{figure}
	\input{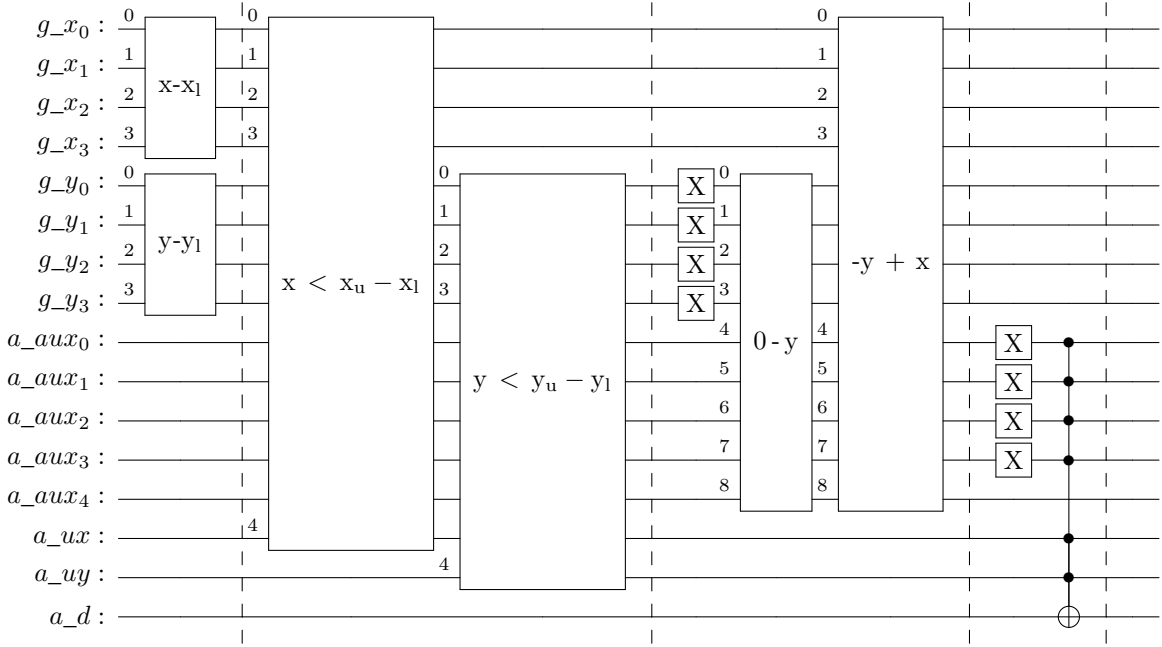}
	\caption{Volumetric identification operation of the a diagonal segment.}
	\label{fig:sp-2-3-circ-sp-reflection-circle-diagonal-vol}
\end{figure}

\Cref{fig:sp-2-3-circ-sp-reflection-circle-diagonal-vol} shows the quantum circuit.
For brevity, we omit the velocity qubits
and the exact gates that implement the addition
and subtractions, as their structure
is identical to the Draper adders shown in previous examples.
To ensure the lower bounds are adhered to, the circuit begins by subtracting $x_l$ and $y_l$
from their respective registers, effectively offseting the diagonal
such that it starts at the origin.
This is an alternative way of implementing the lower bounds, which uses a similar number
of gates to the comparator, but does not require any ancilla qubits.
Next, two comparator operations set the state
of the upper bound ancilla qubits to
$\ket{1}$ in the same way as for the axis-aligned case,
except  that upper bounds must be adjusted as \ie, $x_u-x_l$
to account for the lower bound subtraction.
To isolate gridpoints where the condition $x=y$ holds,
$2$ adders perform the addition of $x$ and $-y$
onto a third ancilla register $\mathrm{a_{aux}}$,
which we use to encode the number $\ket{x-y}$.
Points belong to the diagonal we are targetting
only if $x=y$, and therefore $\ket{a_{aux}}=\ket{0}^{\otimes 5}$.
Finally, we invert the state of the auxiliary register
such that the state $\ket{1}^{\otimes 4}$ is entangled with the
gridpoints belonging to the diagonal, and perform a
$\CPX{6}$ operation controlled on the auxiliary register and the upper bound qubits
and targetting the $\mathrm{a_d}$ qubit.
Under this state, the same controlled swap operations of previous sections
can be employed to realize the boundary conditions
on all points of the diagonal simultaneously,
before undoing each addition to reset the grid state.

\paragraph{Complexity Analysis.} The complexity of this single operation
is dominated by the $2$ Draper adders that compute the
$x-y$ state on the auxiliary register.
Assuming the same $\qft$-based implementation is employed, each adder requires
$\mathcal{O}(4n_g^2)$ gates, as the size of the auxiliary register
is one qubit larger than the largest physical register, to account for potential overflow.
Extending this circuit to the cover the entire circle,
we obtain a bound of $\mathcal{O}(N_sN_t^dq(4n_g^2+2^{d-1}(d-1)^2))$,
which is similar to the axis-parallel segments.

\section{Collision\label{sec:sp-2-3-collision}}

Collision operators in LGA redistribute particles located
at the same lattice site inkeeping with mass and momentum
conservation \cite{wolf2004lattice}.
The exact rules that govern collision are specific
to the velocity discretization and, classically,
the redistribution of particles can either be deterministic or stochastic.
Importantly, the boolean nature of particle discretizations in LGA
makes it such that collision operators are generally linear, and therefore
good candidates for a quantum circuit implementations.

\begin{figure}
	\centering
	\subcaptionbox{One-to-one collision and streaming in the \dq{2}{4} discretization.\label{fig:sp-2-4-collision-determinism-models-deterministic}}{\begin{tikzpicture}[scale=0.25,shift={(8,0)}]
	% Grid lines
	\draw[
	help lines,
	line width=0.4pt,
	color=gray!30,
	dashed
	] (-8, -8) grid[step={($(4, 4) - (0, 0)$)}] (4, 4);

	\foreach \x in {-6,-2,2} {
		\foreach \y in {-6,-2,2} {
			\dtwoqfour{(\x,\y)}{0.1cm}{2}{0}{1.5}{white}
		}
	}
	
	\dtwoqfourcolor{(-2,2)}{0.1cm}{2}{0}{1.5}{white}{red}{black}{red}{black}
	\dtwoqfourcolor{(-2,-2)}{0.1cm}{2}{0}{1.5}{white}{black}{red}{black}{red}
	\dtwoqfourcolor{(2,-6)}{0.1cm}{2}{0}{1.5}{white}{red}{black}{black}{black}
	
	\draw[->, ultra thick, black] (5, -2) -- (13, -2) node at (9, -1) {Collision};
	
	% Picture 2
	\begin{scope}[shift={(22,0)}]
		
		% Grid lines
		\draw[
		help lines,
		line width=0.4pt,
		color=gray!30,
		dashed
		] (-8, -8) grid[step={($(4, 4) - (0, 0)$)}] (4, 4);

		\foreach \x in {-6,-2,2} {
			\foreach \y in {-6,-2,2} {
				\dtwoqfour{(\x,\y)}{0.1cm}{2}{0}{1.5}{white}
			}
		}
		
	\dtwoqfourcolor{(-2,2)}{0.1cm}{2}{0}{1.5}{white}{black}{red}{black}{red}
	\dtwoqfourcolor{(-2,-2)}{0.1cm}{2}{0}{1.5}{white}{red}{black}{red}{black}
	
	\dtwoqfourcolor{(2,-6)}{0.1cm}{2}{0}{1.5}{white}{red}{black}{black}{black}
		
		\draw[->, ultra thick, black] (5, -2) -- (13, -2) node at (9, -1) {Streaming};
	\end{scope}

	% Picture 3
	\begin{scope}[shift={(44,0)}]
		
		% Grid lines
		\draw[
		help lines,
		line width=0.4pt,
		color=gray!30,
		dashed
		] (-8, -8) grid[step={($(4, 4) - (0, 0)$)}] (4, 4);

		\foreach \x in {-6,-2,2} {
			\foreach \y in {-6,-2,2} {
				\dtwoqfour{(\x,\y)}{0.1cm}{2}{0}{1.5}{white}
			}
		}
		
	\dtwoqfourcolor{(-2,-2)}{0.1cm}{2}{0}{1.5}{white}{black}{black}{black}{red}
	\dtwoqfourcolor{(2,-2)}{0.1cm}{2}{0}{1.5}{white}{red}{black}{black}{black}
	\dtwoqfourcolor{(-6,-2)}{0.1cm}{2}{0}{1.5}{white}{black}{black}{red}{black}
	\dtwoqfourcolor{(-2,-6)}{0.1cm}{2}{0}{1.5}{white}{black}{red}{black}{black}
	\dtwoqfourcolor{(-6,-6)}{0.1cm}{2}{0}{1.5}{white}{red}{black}{black}{black}
		
	\end{scope}
\end{tikzpicture}}\\
	\vspace{0.5cm}
	\centering
	\subcaptionbox{Superposed collision through uniform superposition followed by streaming in the \dq{2}{4} discretization.\label{fig:sp-2-4-collision-determinism-models-nondeterministic}}{\input{diagrams/diag-sp-collision-nondeterministic}}
	\caption{Comparison of one-to-one and superposed collision models in the \dq{2}{4} discretization.\label{fig:sp-2-4-collision-determinism-models}}
\end{figure}

In this section, we consider two ways of designing and
implementing LGA collision operators in the Space-Time
encoding -- \emph{one-to-one} and \emph{superposed}.
\Cref{fig:sp-2-4-collision-determinism-models}
shows the difference between the two models in a \dq{2}{4} setting.
One-to-one collision models act as depicted in
\Cref{fig:sp-2-4-collision-determinism-models-deterministic}, where the state
of each lattice site is permuted onto a pre-determined outcome such that
macroscopic quantities (\ie mass, momentum) are conserved.
This was the approach proposed by the earliest LGA models, and was later superseded by
ensemble averaging over stochastic models, with the aim of reducing noise.

\Cref{fig:sp-2-4-collision-determinism-models-nondeterministic}
depicts the superposed collision operator in the \dq{2}{4} discretization.
The difference between the two models is that the superposed approach
introduces a stochastic component to the permutation of states.
In practice, classical implementations of LGA would sample random numbers that would
then map each state onto a discrete outcome from a set of states with
equivalent macroscopic quantities.
In designing the quantum circuit for this operator, we need not select
one sample out of the set of possible outcomes, but can instead
represent all feasible states simultaneously.
This in turn allows for the simultaneous simulation
of exponentially many independent lattice configurations,
which is equivalent to a lattice-level ensemble in the taxonomy
proposed by \citet{wang2025quantum}.
It is this representational power of the extended computational basis state
encoding that circumvents the need for ensemble averaging in classical LGA.
Though both one-to-one and superposed QLGA collision operators have already
been proposed in the literature \cite{love2019quantum, schalkers2024importance, fonio2023quantum, zamora2025efficient}, they are all designed
for specific discretizations or follow a rigid definition of state equivalency.
In the remainder of this section, we
provide both concrete quantum circuits and a generic formulation
that apply superposed collision for arbitrary \dq{d}{q} stencils.

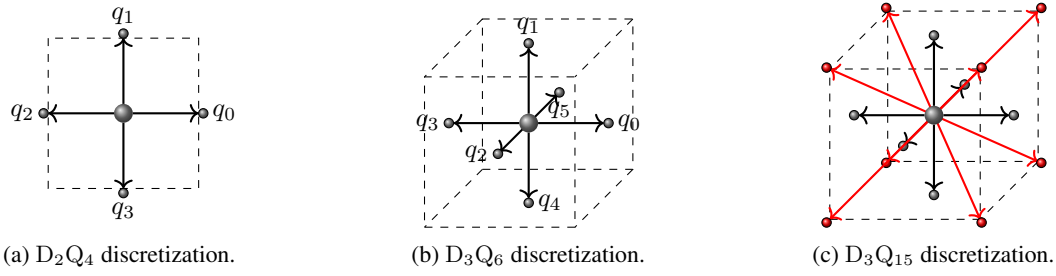
\begin{figure}
\centering
\hfill
\subcaptionbox{\dq{2}{4} discretization. \label{fig:sp-2-4-stencil-d2q4}}[0.3\textwidth]
{\begin{tikzpicture}
    % Define the cube size and sphere radius
    \def\cubesize{2}
    \def\radius{0.125}
	\draw[dashed,black] (-1,1) -- (-1,-1) -- (1,-1) -- (1,1) -- (-1,1);

    	 \draw[->,thick] (-1*\radius,0) -- (-1,0) ;
    	 \draw[->,thick] (1*\radius,0) -- (1,0);
    	 \draw[->,thick] (0,-1*\radius) -- (0,-1);
    	 \draw[->,thick] (0,1*\radius) -- (0,1);
    	 %\draw[->,thick] (0,\x*\radius) -- (0,\x);
    
    \draw[ball color=gray] (1.06125,0,0) circle[radius=\radius/2] node [right,align=center] {$q_0$};
    \draw[ball color=gray] (-1.06125,0,0) circle[radius=\radius/2] node [left,align=center] {$q_2$};
    \draw[ball color=gray] (0,-1.06125,0) circle[radius=\radius/2] node [below,align=center] {$q_3$};
    \draw[ball color=gray] (0,1.06125,0) circle[radius=\radius/2] node [above,align=center] {$q_1$};
    
    % Draw the sphere
    \shade[ball color=gray] (0,0,0) circle[radius=\radius];
\end{tikzpicture}}
\centering
\hfill
\subcaptionbox{\dq{3}{6} discretization.\label{fig:sp-2-4-stencil-d3q6}}[0.3\textwidth]
{\begin{tikzpicture}
    % Define the cube size and sphere radius
    \def\cubesize{2}
    \def\radius{0.125}
    \draw[dashed,black] (-1,-1,-1) -- (1,-1,-1) -- (1,1,-1) -- (-1,1,-1) -- cycle; % Bottom face
    \draw[dashed,black] (-1,-1,1) -- (1,-1,1) -- (1,1,1) -- (-1,1,1) -- cycle;   % Top face
    \draw[dashed,black] (-1,-1,-1) -- (-1,-1,1);
    \draw[dashed,black] (1,-1,-1) -- (1,-1,1);
    \draw[dashed,black] (1,1,-1) -- (1,1,1);
    \draw[dashed,black] (-1,1,-1) -- (-1,1,1);
    
    \foreach \x in {-1,1} {
    		% Arrows to faces
        \draw[->,thick] (\x*\radius,0,0) -- (\x,0,0);
        \draw[->,thick] (0,\x*\radius,0) -- (0,\x,0);
        \draw[->,thick] (0,0,\x*\radius) -- (0,0,\x);
        
    }
    
    \draw[ball color=gray] (1.06125,0,0) circle[radius=\radius/2] node [right,align=center] {$q_0$};
    \draw[ball color=gray] (-1.06125,0,0) circle[radius=\radius/2] node [left,align=center] {$q_3$};
    \draw[ball color=gray] (0,-1.06125,0) circle[radius=\radius/2] node [right,align=center] {$q_4$};
    \draw[ball color=gray] (0,1.06125,0) circle[radius=\radius/2] node [above,align=center] {$q_1$};
    \draw[ball color=gray] (0,0,-1.06125) circle[radius=\radius/2] node [below,align=center] {$q_5$};
    \draw[ball color=gray] (0,0,1.06125) circle[radius=\radius/2] node [left,align=center] {$q_2$};
    
    % Draw the sphere
    \shade[ball color=gray] (0,0,0) circle[radius=\radius];
\end{tikzpicture}}
\centering
\hfill
\subcaptionbox{\dq{3}{15} discretization.\label{fig:sp-2-4-stencil-d3q15}}[0.3\textwidth]
{\begin{tikzpicture}
    % Define the cube size and sphere radius
    \def\cubesize{2}
    \def\radius{0.125}
    \draw[dashed,black] (-1,-1,-1) -- (1,-1,-1) -- (1,1,-1) -- (-1,1,-1) -- cycle; % Bottom face
    \draw[dashed,black] (-1,-1,1) -- (1,-1,1) -- (1,1,1) -- (-1,1,1) -- cycle;   % Top face
    \draw[dashed,black] (-1,-1,-1) -- (-1,-1,1);
    \draw[dashed,black] (1,-1,-1) -- (1,-1,1);
    \draw[dashed,black] (1,1,-1) -- (1,1,1);
    \draw[dashed,black] (-1,1,-1) -- (-1,1,1);
    
    \foreach \x in {-1,1} {
    		% Arrows to faces
        \draw[->,thick] (\x*\radius,0,0) -- (\x,0,0);
        \draw[->,thick] (0,\x*\radius,0) -- (0,\x,0);
        \draw[->,thick] (0,0,\x*\radius) -- (0,0,\x);
        
    }
    
    \draw[ball color=gray] (1.06125,0,0) circle[radius=\radius/2];
    \draw[ball color=gray] (-1.06125,0,0) circle[radius=\radius/2];
    \draw[ball color=gray] (0,-1.06125,0) circle[radius=\radius/2];
    \draw[ball color=gray] (0,1.06125,0) circle[radius=\radius/2];
    \draw[ball color=gray] (0,0,-1.06125) circle[radius=\radius/2];
    \draw[ball color=gray] (0,0,1.06125) circle[radius=\radius/2];
    
    \foreach \x in {-1,1} {
    		\foreach \y in {-1,1} {
    			\foreach \z in {-1,1} {
    			   \draw[->,thick, red] (0,0,0) -- (\x,\y,\z);
    			   \draw[ball color=red] (\x+\x*0.030625,\y+\y*0.030625,\z+\z*0.030625) circle[radius=\radius/2];
    			}
    		}

    }
    
    % Draw the sphere
    \shade[ball color=gray] (0,0,0) circle[radius=\radius];
\end{tikzpicture}}
\hfill\strut
\caption{Example velocity discretizations.\label{fig:sp-2-4-stencils}}
\end{figure}

Throughout this section, we use the \dq{2}{4}, \dq{3}{6}, and \dq{3}{15} velocity discretizations
as depicted in \Cref{fig:sp-2-4-stencils}
to illustrate the challenges different configurations
give rise to, and how our methods address them.
We define collision operators that perform computations in terms of \emph{equivalence classes}.
To build up the set of all equivalence classes attached to velocity discretization,
we first define the set all possible momentum values of a \dq{d}{q} discretization as

\begin{equation}
	\mathbb{M}_{\mathrm{D}_d\mathrm{Q}_q} = \left\{ \sum_{j=0}^{q-1}k_j\boldsymbol{e}_j~|~k \in \{0, 1\}^q \right\},
\end{equation}

with the $\boldsymbol{e}_j$ vectors defining the momentum contribution of each
velocity channel according to the discretization,
and $k_j$ the $j^{\text{th}}$ bit in the binary representation of $k$.
The exact values of these vectors are given by \citet{kruger2017lattice}
for several common discretizations.
Let $E_{(\mathrm{D}_d\mathrm{Q}_q, m, \boldsymbol{\mu} )} \subseteq \{0, 1\}^q$  be a set
of local velocity configurations that have mass $m$ and
momentum $\boldsymbol{\mu} \in \mathbb{M}_{\mathrm{D}_d\mathrm{Q}_q} $
under a given velocity discretization \dq{d}{q}, where each velocity
is represented as in the computational basis state encoding.
Furthermore, let $\| E_{(\mathrm{D}_d\mathrm{Q}_q, m, \boldsymbol{\mu} )} \|$
be the number of velocity profiles in the class and let

\begin{equation}
	\mathbb{E}_{\mathrm{D}_d\mathrm{Q}_q} = \left\{E_{(\mathrm{D}_d\mathrm{Q}_q, m, \boldsymbol{\mu} )}~|~m\in \mathbb{N}, \boldsymbol{\mu}  \in \mathbb{M}_{\mathrm{D}_d\mathrm{Q}_q} \right\}
\end{equation}

be the set of all equivalence classes of the velocity discretization.
We note that our definition of equivalence classes leads to larger sets of equivalent
lattice configurations, as we do not constrain them based on specific 
rotation angles as in \citep{fonio2023quantum} and \citep{zamora2025float}.
This increases the expressiveness of the model, as more novel states can emerge
from the same configuration.
When interpreting LGA as a random walk, this model change inherently
increases the exploration space.

So far as designing collision operators is concerned, we restrict ourselves to
transformations acting on configurations with $m \geq 2$ as in simpler
equivalence classes the collision is trivial.
For the \dq{2}{4} discretization depicted in \Cref{fig:sp-2-4-stencil-d2q4}, there
is a single set non-trivial equivalence class $E_{(\mathrm{D_2Q_4}, 2, (0, 0)^\mathrm{T})}=\{ 1010, 0101 \}$,
that contains two particles moving in opposing directions across the same streaming line \cite{schalkers2024importance}.
The $8$ relevant equivalence classes of \dq{3}{6} (\Cref{fig:sp-2-4-stencil-d3q6}) are:

\begin{subequations}
\begin{align}
	& E_{(\mathrm{D}_3\mathrm{Q}_6, 2, (0, 0, 0)^\mathrm{T})} & = \{ 100100, 010010, 001001 \},\label{eq:sp-2-4-d3q6-0}\\
	& E_{(\mathrm{D}_3\mathrm{Q}_6, 4, (0, 0, 0)^\mathrm{T})} & = \{ 110110, 101101, 011011 \},\\
	& E_{(\mathrm{D}_3\mathrm{Q}_6, 3, (1, 0, 0)^\mathrm{T})} & = \{ 110010, 101001 \},\\
	& E_{(\mathrm{D}_3\mathrm{Q}_6, 3, (-1, 0, 0)^\mathrm{T})} & = \{ 010110, 001101 \},\\
	& E_{(\mathrm{D}_3\mathrm{Q}_6, 3, (0, 1, 0)^\mathrm{T})} & = \{ 110100, 011001 \},\\
	& E_{(\mathrm{D}_3\mathrm{Q}_6, 3, (0, -1, 0)^\mathrm{T})} & = \{ 100110, 001011 \},\\
	& E_{(\mathrm{D}_3\mathrm{Q}_6, 3, (0, 0, 1)^\mathrm{T})} & = \{ 101100, 011010 \},\\
	& E_{(\mathrm{D}_3\mathrm{Q}_6, 3, (0, 0, -1)^\mathrm{T})} & = \{ 100101, 010011 \},
\end{align}
\end{subequations}

while there are a total of $2832$ equivalence classes for \dq{3}{15}.\footnote{The code we used to generate all equivalence classes 
	according to our definition is available with the replication package \cite{georgescu2025qlgabuildingblocksreplication}.}
In the following sections, we describe the process behind
implementing quantum circuits for arbitrary equivalence classes,
as well as how these circuits can be concatenated to obtain generic complete
collision operators for a given discretization.

\subsection{Rationale and Small Collision Circuits}

\begin{figure}
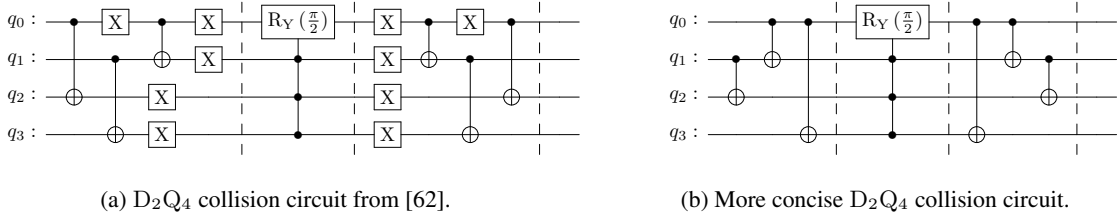

	\centering
	\hfill
	\subcaptionbox{\dq{2}{4} collision circuit from \cite{schalkers2024importance}.\label{fig:sp-2-4-collision-d2q4-old}}{\input{circuits/circ-sp-collision-d2q4-old}}
	\centering
	\hfill
	\subcaptionbox{More concise \dq{2}{4} collision circuit.\label{fig:sp-2-4-collision-d2q4-new}}{\input{circuits/circ-sp-collision-d2q4-new}}
	\hfill\strut
	\caption{\dq{2}{4} collision circuits.\label{fig:sp-2-4-collision-d2q4}}
\end{figure}

In previous work, \citet{schalkers2024importance} implemented a
collision operator for the Space-Time encoding of the \dq{2}{4}
discretization as depicted in \Cref{fig:sp-2-4-collision-d2q4-old}.
Conceptually, the circuit can be broken down into $3$ logical steps
that transform the quantum state encoding information
at a lattice site as

\begin{equation}
	\ket{\psi}_{\mathrm{C}} \leftarrow \mathrm{U_P}^\dagger \mathrm{U_R} \mathrm{U_P} \ket{\psi}.
\label{eq:sp-2-4-collision-order}
\end{equation}

The three steps procede as follows. The local lattice configuration
first undegoes \emph{permutation} by a unitary matrix $U_\mathrm{P}$ that,
for an equivalence class $E$,  maps relevant states to a superposition of basis states
$\sum_{k=0}^{\| E \|-1}\ket{k}\ket{1}^{\otimes (q - \lceil \log_2 \| E \| \rceil)}$.
For \dq{2}{4}, these states would a linear combination of $\ket{0111}$ and $\ket{1111}$.
One possible and simple implementation of this circuit is by the first
partition of the circuit shown in \Cref{fig:sp-2-4-collision-d2q4-old}.
The second step then \emph{redistributes} the amplitude of the states that belong to the
equivalence class in a uniform-magnitude superposition
on the first qubit, controlled on the $3$ other qubits.
The third step is a reversal of the first step, which permutes the basis states back
to their original configuration.
This circuit effectively models an evenly superposed LGA
collision, as it evenly distributes the probability
mass between the two states of the equivalence class.
Each step in this \emph{\underline{p}ermute-\underline{r}edistribute-un\underline{p}ermute} (PRP)
sequence requires careful considerations and is feasible only under certain constraints,
which we delve into in the following paragraphs.

%\begin{algorithm}
%\caption{Collision circuit construction algorithm}\label{alg:sp-2-4-collision-algorithm}
%\begin{algorithmic}[1]
%\Statex \textbf{Input:} Discretization $\mathrm{D}_d\mathrm{Q}_q$, Equivalence Class $E_{(\mathrm{D}_d\mathrm{Q}_q, m, \boldsymbol{\mu} )}$, Local Velocity Quantum State $\ket{\psi}$
%\Statex \textbf{Output:} Post-Collision Quantum State $\ket{\psi_PCP}$
%\State $\ket{\psi_P} \gets \textsc{Permute}(\ket{\psi})$ \algorithmiccomment{Map relevant states to $\ket{k}\ket{1}^{\otimes E_C- \lceil \log_2 E_C \rceil}$}
%\State $\ket{\psi_{PC}} \gets \textsc{Transform}(\ket{\psi_P})$ \Comment{Apply transformation according to the model}
%\State $\ket{\psi_{PCP}} \gets  \textsc{Permute}^{-1}(\ket{\psi_{PC}})$ \Comment{Undo initial permutation}
%\end{algorithmic}
%\end{algorithm}

\paragraph{Permutations.} Permutations are only necessary to map particular basis states onto
pre-determined outcomes. For an arbitrary equivalence class
$E_{(\mathrm{D}_d\mathrm{Q}_q, m, \boldsymbol{\mu} )}$ with
$E_C = \| E_{(\mathrm{D}_d\mathrm{Q}_q, m, \boldsymbol{\mu} )}  \|$
elements, the permutation
should map the members of the equivalence class onto the
$\sum_{k=0}^{E_C - 1}\ket{k}\ket{1}^{\otimes (q - \lceil \log_2 E_C \rceil)}$ states,
where (without loss of generality) $0 \leq k < E_C$ and $\ket{k}$ is encoded in the
first $\lceil \log_2 E_C \rceil$ qubits of the $E_C$-wide register.
The states that velocity configurations not belonging to $E_{(\mathrm{D}_d\mathrm{Q}_q, m, \boldsymbol{\mu} )}$
are mapped to is not relevant under these conditions, as we explore in the next paragraph.
In practice, this means there are multiple feasible permutations
that the quantum circuit can perform to model the same physical behavior.
\Cref{fig:sp-2-4-collision-d2q4-new} gives an example of such an alternative
for the \dq{2}{4} discretization,
where the equivalence class members are mapped to the appropriate
domain using fewer gates.

\paragraph{Particle Redistribution.} Much like the permutation step,
the redistribution step can take several forms,
all of which act on the first $\lceil \log_2 E_C \rceil$ qubits of the register.
The $\RY{\uppi/2}$ gate that induces the superposition on the
states of interest can thus be substituted for a different (controlled)
gate depending on the physical behavior that the model attempts to capture.
Permutations applied to the first $\lceil \log_2 E_C \rceil$ qubits
of the register (a simple $\CPX{3}$ gate for \dq{2}{4})
model one-to-one collisions, whereas operations that introduce
superpositions simultaneously model all possible outcomes of collisions.
Crucially, whichever model the circuit attempts to implement,
the redistribution step must not introduce any basis state $\ket{x}$ that falls
in the range $E_C < x < 2^{\lceil \log_2 E_C \rceil}$,
as (under the assumptions established in the previous paragraph)
that would encode a velocity configuration that does not belong to the equivalence class.
Uncomputing the permutation of such a state would result
in behavior that is non-physical.
This is not an issue under the \dq{2}{4} discretization, but becomes challenging
with stencils such as \dq{3}{6}, as we explore next.

\begin{figure}
	\input{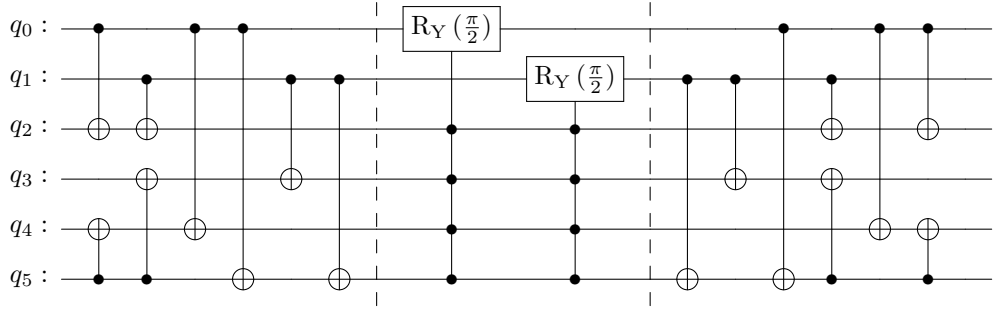}
	\caption{Inexact collision circuit for $E_{(\mathrm{D}_3\mathrm{Q}_6, 2, (0, 0, 0)^\mathrm{T})}$. \label{fig:sp-2-4-circ-sp-collision-d3q6-wrong}}
\end{figure}

\paragraph{Example -- \dq{3}{6} collision.} Let us consider how one could apply \Cref{eq:sp-2-4-collision-order}
to $E_{(\mathrm{D}_3\mathrm{Q}_6, 2, (0, 0, 0)^\mathrm{T})}$ described in \Cref{eq:sp-2-4-d3q6-0}.
\Cref{fig:sp-2-4-circ-sp-collision-d3q6-wrong} shows one such circuit
that follows the PRP principle
acting on the $6$ qubits requried to encode
the \dq{3}{6} velocity set.
The first permutation step is comprised of $8\ \CX$ gates
and has a depth of $4$, while
$2$ $\RY{\uppi/2}$ gates
create a superposition of the first 
$\lceil \log_2 (\| E_{(\mathrm{D}_3\mathrm{Q}_6, 2, (0, 0, 0)^\mathrm{T})} \|) \rceil = 2$ qubits.
This means that the states affected by the rotation matrices
are $\ket{001111}, \ket{011111}, \ket{101111}$, and $\ket{111111}$.
While the permutation correctly maps the states of $E_{(\mathrm{D}_3\mathrm{Q}_6, 2, (0, 0, 0)^\mathrm{T})}$
onto the first three of these states, it
also leaves the $\ket{111111}$ state unchanged.
This in turn means that applying the circuit in \Cref{fig:sp-2-4-circ-sp-collision-d3q6-wrong}
to, for instance, the state $\ket{100100}$ results in

\begin{align}
\begin{split}
	\mathrm{U}^\dagger_\mathrm{P}\mathrm{U}_\mathrm{R}\mathrm{U}_\mathrm{P}\ket{100100} &= \mathrm{U}^\dagger_\mathrm{P}\mathrm{U}_\mathrm{R}\ket{101111}\\
	& = \mathrm{U}^\dagger_\mathrm{P}\frac{1}{2}\left(-\ket{001111} - \ket{011111} + \ket{101111} + \ket{111111} \right)\\
	& = \frac{1}{2}\left(\ket{001001} - \ket{010010} + \ket{100100} - \ket{111111} \right)
\end{split}
\end{align}

which in addition to the three constituents of $E_{(\mathrm{D}_3\mathrm{Q}_6, 2, (0, 0, 0)^\mathrm{T})}$
also encodes the state where all streaming channels of the lattice site are occupied.
Clearly, this violates the mass conservation requirement of the collision
operator and compromises the validity of the simulation.

This incompatibility is not tied to any particular equivalence class but
generally occurs when the cardinality of an equivalence class is not exactly
a power of $2$.
While redistribution operators such as the $\RY{\cdot}$ gate
are concise and exact for stencils such as \dq{2}{4}, their utility is limited
on more complex discretizations.
In the following section,
we describe a generic unitary redistribution operator
that can act on arbitrary stencil discretizations that addresses this issue.

\subsection{Arbitrary and Complete Collision Circuits}

To build up the collision operators for arbitrary discretizations,
we start from a discrete Fourier transform ($\dft$) matrix.
For the purposes of this section we define this matrix as

\begin{align}
\begin{split}
	\dft(N) & = \frac{1}{\sqrt{N}}\left( e^{\frac{-2jk\pi i}{N}} \right)_{j,k=0,\cdots,N-1}\\
	& = 
	\frac{1}{\sqrt{N}}
	\begin{pmatrix}
	1 & 1 & 1 & 1 & \cdots & 1\\
	1 & e^{\frac{-2\pi i}{N}} & e^{\frac{-4\pi i}{N}} & e^{\frac{-6\pi i}{N}} & \cdots & e^{\frac{-2(N-1)\pi i}{N}}\\
	1 & e^{\frac{-4\pi i}{N}} & e^{\frac{-8\pi i}{N}} & e^{\frac{-12\pi i}{N}} & \cdots & e^{\frac{-2\cdot 2(N-1)\pi i}{N}}\\
	1 & e^{\frac{-6\pi i}{N}} & e^{\frac{-12\pi i}{N}} & e^{\frac{-18\pi i}{N}} & \cdots & e^{\frac{-2\cdot 3(N-1)\pi i}{N}}\\
	\vdots & \vdots & \vdots & \vdots & \ddots & \vdots\\
	1 & e^{\frac{-2\cdot(N-1)\pi i}{N}} & e^{\frac{-2\cdot 2(N-1)\pi i}{N}} & e^{\frac{-2 \cdot 3\cdot(N-1)\pi i}{N}} & \cdots & e^{\frac{-2\cdot(N-1)(N-1)\pi i}{N}}
	\end{pmatrix},
\end{split}
\end{align}

with $N$ the size of the of the matrix.
We use this to define the larger collision operator $\mathrm{Coll}(k, N)$ as

\begin{align}
	\mathrm{Coll}(k, N) &= 
	\begin{pmatrix}
		\dft(k) & \mathbb{O}_{k\times(N-k)}\\
		\mathbb{O}_{(N-k)\times k} & \mathbb{I}_{N-k}
	\end{pmatrix},
\label{eq:sp-2-4-coll-matrix}
\end{align}

with $N$ the size of the matrix, and $k \leq N$ the size of the $\dft$ sublock.
To realize collision in the Space-Time encoding
for a discretization $\mathrm{D}_d\mathrm{Q}_q$, and a velocity class $E$, we
set $k=\| E \|$ and $N=2^{\lceil \log_2 \| E \| \rceil}$.
When applied to a $q$-qubit basis state $\ket{\psi}$, performs the following transformation

\begin{equation}
	\mathrm{Coll}(\| E \|, 2^{\lceil \log_2 \| E \| \rceil}) \ket{\psi} = 
	\begin{cases}
		\frac{1}{\sqrt{\| E \|}} \sum_{k=0}^{\| E \| - 1} e^{\frac{-2k\pi i}{\| E \|}}\ket{k} & \ket{\psi} \in \{ \ket{0}, \cdots, \ket{\| E \| - 1}\}\\
		\ket{\psi} & \text{otherwise},
	\end{cases}
\end{equation}

which can be understood as creating a superposition of the first $k$ basis states 
when applied to one of them while
leaving the remainder $N-k$ basis states unaffected.
The amplitudes of the basis states in superposition are all of the form
$e^{\frac{-2\cdot k\pi i}{N}}/\sqrt{N}$, which when squared leads
a probability of $1/N$ and therefore a uniform-magnitude
superposition of all $N$ basis states of interest.\footnote{
The unitarity of the $\mathrm{Coll}(k, N)$ matrix follows directly from the unitarity of the $\dft$ matrix as

\begin{align*}
	\begin{split}
		\mathrm{Coll}(k, N)\mathrm{Coll}(k, N)^\dagger & =
		\begin{pmatrix}
			\dft(k) & \mathbb{O}_{k\times(N-k)}\\
			\mathbb{O}_{(N-k)\times k} & \mathbb{I}_{N-k}
		\end{pmatrix}
		\begin{pmatrix}
			\dft(k) & \mathbb{O}_{k\times(N-k)}\\
			\mathbb{O}_{(N-k)\times k} & \mathbb{I}_{N-k}
		\end{pmatrix}^\dagger \\
		& = \begin{pmatrix}
			\dft(k) & \mathbb{O}_{k\times(N-k)}\\
			\mathbb{O}_{(N-k)\times k} & \mathbb{I}_{N-k}
		\end{pmatrix}
		\begin{pmatrix}
			\dft(k)^\dagger & \mathbb{O}_{k\times(N-k)}\\
			\mathbb{O}_{(N-k)\times k} & \mathbb{I}_{N-k}
		\end{pmatrix}\\
		& = \begin{pmatrix}
			\dft(k)\dft(k)^\dagger + \mathbb{O}_{k\times(N-k)}\mathbb{O}_{(N-k)\times k} & \dft(k)\mathbb{O}_{k\times(N-k)} + \mathbb{O}_{k\times(N-k)}\mathbb{I}_{N-k}\\
			\mathbb{O}_{(N-k)\times k}\dft(k) + \mathbb{I}_{N-k}\mathbb{O}_{k\times(N-k)}& \mathbb{O}_{(N-k)\times k}\mathbb{O}_{k\times(N-k)} + \mathbb{I}_{N-k}\mathbb{I}_{N-k}
		\end{pmatrix}\\
		& = \begin{pmatrix}
			\mathbb{I}_k & \mathbb{O}_{k\times (N-k)}\\
			\mathbb{O}_{(N-k)\times k} & \mathbb{I}_{N-k}
		\end{pmatrix} = \mathbb{I}_{N}.
	\end{split}
\end{align*}
}

It is important to note that $q$ is fixed according to the choice of
lattice discretization, and size of
the $\dft$ block depends on the equivalence class of the discretization.
Because of these properties, we  
can determine the complexity of the circuit that implements $\mathrm{Coll}(k, N)$
exactly for commonplace discretizations,
as well as reason about the complexity of the operator in general.
Before obtaining such a general expression of complexity, 
we first have to analyze how the PRP
procedure can be expanded to cover all equivalence
classes of a particular discretization.

\begin{figure}
\input{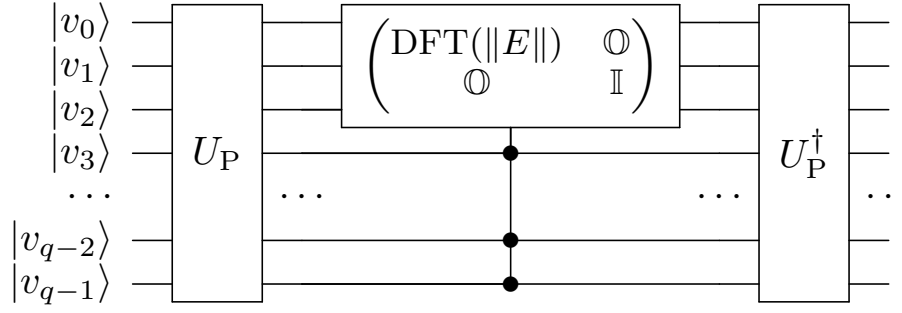}
\caption{Schematic of the generic collision operator for an equivalence class $E$ with $4 < \|E\| < 8$. \label{fig:circ-sp-2-4-genericollision-schematic}}
\end{figure}

Let us consider a system of $n_g$ grid qubits,
and analyze the circuit that performs collision for one time step,
while assuming that $N_t$ time  steps are still to be simulated.
Under an arbitrary $\mathrm{D}_d\mathrm{Q}_q$ discretization
with equivalence classes in $\mathbb{E}_{\mathrm{D}_d\mathrm{Q}_q}$ , this requires
the customary $\mathcal{O}(qN_t^d)$ neighboring qubits.
For any individual equivalence class $E \in \mathbb{E}_{\mathrm{D}_d\mathrm{Q}_q}$,
the PRP-based collision operator is given by

\begin{equation}
\mathbb{I}^{\otimes n_g} \otimes \left( \left(\mathrm{U_P}(E)^\dagger\right)^{\otimes \mathcal{O}(N_t^d)}\left(\mathrm{C}^{(q-\lceil\log_2\| E\| \rceil)}\mathrm{Coll}(\| E \|, 2^{\lceil \log_2 \| E \| \rceil})\right)^{\otimes \mathcal{O}(N_t^d)} \mathrm{U_P}(E)^{\otimes \mathcal{O}(N_t^d)} \right),
\label{eq:sp-2-4-prp-collision-local}
\end{equation}

with $\mathrm{C}^{(\cdot)}\mathrm{Coll}(\| E \|, q)$ the $q$-qubit matrix obtained by applying the matrix given in
\Cref{eq:sp-2-4-coll-matrix} controlled on the trailing $q - \lceil \log_2 \| E \| \rceil$ qubits.
\Cref{fig:circ-sp-2-4-genericollision-schematic} shows a schematic representation
of the quantum circuit acting only on the velocity register.
Intuitively, this applies the equivalence class-specific
collision operator in parallel $\mathcal{O}(N_t^d)$ times for each $q$-qubit subregister,
while leaving leaving the grid register unaffected.
The general collision operator can then be expressed as 

\begin{equation}
\mathbb{I}^{\otimes n_g} \otimes \prod_{E \in \mathbb{E}_{\mathrm{D}_d\mathrm{Q}_q}} \left( \left(\mathrm{U_P}(E)^\dagger\right) ^{\otimes \mathcal{O}(N_t^d)}\left(\mathrm{C}^{(q-\lceil\log_2\| E\| \rceil)}\mathrm{Coll}(\| E \|, 2^{\lceil \log_2 \| E \| \rceil})\right)^{\otimes \mathcal{O}(N_t^d)} \mathrm{U_P}(E)^{\otimes \mathcal{O}(N_t^d)} \right),
\label{eq:sp-2-4-prp-collision-total}
\end{equation}

that is, each equivalence class can be addressed sequentially
on each $q$-qubit velocity register.
The crucial property that enables
this sequential application is that
no two equivalence class-specific collision operators act
on the same basis states,
as there are no common velocity configurations between equivalence classes.

\paragraph{Complexity Analysis.} \Cref{eq:sp-2-4-prp-collision-total} provides an expression
of collision operators that allows us to easily derive their gate complexity.
Each equivalence class requires two permutation circuits and one redistribution circuit.
Recent results by \citet{herbert2024almost} describe an efficient way to
construct circuits that implement transpositions of $n$-qubit
basis states using $\mathcal{O}(n)$ gates.
Since at this level the collision circuit acts in parallel on $q$-qubit subregisters,
and the permutation step only requires that the $\| E \|$ states
of the equivalence class are mapped to specific states, the permutation
step can be implemented as a series at most $\| E \|$
transpositions, and is, therefore, $\mathcal{O} (\| E \|q)$ per subregister, and
$\mathcal{O} (N_t^d\| E \|q)$ across the entire circuit.
The final step of undoing the permutation can be implemented
by mirroring the permutation circuit and thus the two steps share the same complexity.
The most complex redistribution circuit requires the decomposition of 
the $\mathrm{C}^{(q-\lceil\log_2\| E\| \rceil)}\mathrm{Coll}(\| E \|, 2^{\lceil \log_2 \| E \| \rceil})$ matrix.

In general, the decomposition of any arbitrary unitary matrix onto a simple
universal gate set requires a number of single- and two-qubit
gates that is in general exponential
in the number of qubits the matrix is applied on \citep{barenco1995elementary}.
While this does generally compromise the efficiency of the algorithm
and requires heavy classical processing, there
are several properties of this matrix
that may allow for practical decompositions.
First, the size of the matrix is bound by $q$, the number of discrete velocities
in the discretization.
Second, the bottom $2 ^ {q - \lceil \log_2 \| E \| \rceil}$
entries of the matrix are only used to restrict the application of
the $\mathrm{Coll}(\| E \|, 2^{\lceil \log_2 \| E \| \rceil})$
matrix through controls, and the matrix is therefore sparse and structured.
Third, the denser $\mathrm{Coll}(\| E \|, 2^{\lceil \log_2 \| E \| \rceil})$
submatrix is itself bounded by the size
of the equivalence class with the highest cardinality of the specific discretization,
and the number of $0$ entries in the matrix is equal to $q^2 - \| E \|^2 - q + \| E \|$.
For instance, the most populated equivalence classes of the
\dq{3}{15} discretization are $E_{\mathrm{D}_3\mathrm{Q}_{15}, j, (0, 0, 0)^{\mathrm{T}}}$,
for $j \in \{6, 7, 8, 9\}$, 
which contain $73$ velocity profiles.
The matrix encoding the corresponding collision operator is therefore
$\mathrm{Coll}(73, 128)$, which acts on 7 qubits and is $\approx 67.48\%$ sparse.

Further straight-forward techniques,
such as caching transpilations and re-using one transpiled circuit across all time-steps
and across multiple simulations can further significantly reduce
the classical resources necessary to produce the quantum circuits.
Convenient situations in which equivalence classes can be treated with
sequential applications of (multi-controlled) $\RY{\cdot}$ gates
can also be implemented more efficiently and require less complex decompositions.
The implementation of one-to-one collision rules can be performed by means
of transpositions, which only require $\mathcal{O}(q^2)\ \CX$ gates for the redistribution step.
In general, however, the decomposition of the large collision matrix
clearly dominates the complexity of the algorithm and therefore the
number of native quantum gates required to implement
the collision operator scales with
$\mathcal{O}\left(\| \mathbb{E}_{\mathrm{D}_d\mathrm{Q}_q}\| N_t^d \left( \| E_M \|q + 2^q \right) \right)$,
where $E_M = \max_{E \in \mathbb{E}_{\mathrm{D}_d\mathrm{Q}_q}} \| E \|$.
Finally, we note that the DFT compoenent of our redistribution operator
sometimes appears in the literature as a \emph{reduced order QFT}.
For more details, we refer the interested reader to
the works of \citet{kitaev1995quantum}, and \citet{mosca2004exact},
as well as the overview of \citet{kempe2003quantum}

\section{Measurement\label{sec:sp-2-4-measurement}}

In this section, we detail techniques for extracting
physical information encoded in the quantum state by means of observables
and quantum circuits.
We first introduce observables that can be used to compute the mass, density, and pressure
over arbitrary regions of space, before describing quantum circuits and observables that compute
the forces acting on an object by means of the momentum exchange method.
Throughout this section, we express physical quantities in lattice units.

\subsection{Mass, Density, and Pressure Measurement\label{subsec:sp-2-4-measurement-mass-pressure}}

In lattice gas hydrodynamics, pressure is typically computed as

\begin{equation}
p = c_s^2 \rho,
\end{equation}

with $c_s$ the \emph{speed of sound} in lattice units and $\rho$ is
the local particle density \cite{frisch1987lattice}.
We formulate an observable $P$ such that the expectation
value $\langle P \rangle$ is equal to the (scaled) mass at a fixed lattice site,
before generalizing its expression to arbitrary regions of space.
Density is directly determined by the local mass and the velocity discretization,
and the speed of sound $c_s$ is given by the lattice discretization
and therefore independent of the state of the system.
We therefore express $P$ in such a way that it computes mass, which
in turn enables the calculation
of density and pressure by trivial classical post-processing.
Since LGA particles are boolean by nature, the particle density
at a given lattice site $x$ for a \dq{d}{q}
discretization with equally weighted particle channels is

\begin{equation}
\rho_{x} = \frac{1}{q}\sum_{j=0}^{q-1} f_j(x),
\end{equation}

with $f_j$ the particle occupancy of each of the $q$ channels.
In the computational basis state encoding of the velocity configuration,
this information is encoded in $q$ qubits with $\ket{0}$
encoding an empty channel and $\ket{1}$ an occupied one.
For a particular qubit $0 \leq j \leq q-1$ encoding such a channel,
the observable that extracts the probability of measuring the state $\ket{1}$ is

\begin{equation}
n_j = \mathbb{I}^{\otimes j} \otimes \frac{1}{2}\left( \mathbb{I} - \mathbb{Z} \right) \otimes \mathbb{I}^{\otimes (q - 1 - j)},
\label{eq:sp-2-5-number-matrix}
\end{equation}

where $\mathbb{Z}$ is Pauli $Z$-matrix.
Intuitively, this is a projection of the $\ket{1}$ state of the particular velocity
channel, and the value of $\bra{\psi}n_i\ket{\psi}$ is the
probability of measuring the $j^{\text{th}}$ qubit in state $\ket{1}$.
Extending this to compute the mass encoded in a $q$-qubit velocity register, we get
the observable

\begin{equation}
N = \sum_{j = 0}^{q-1} n_j,
\end{equation}

which encodes the probabiliy of measuring any
velocity channel in the state $\ket{1}$.
Since mass is tied to the amplitude of the $\ket{1}$ state,
of the velocity qubits, the matrix $N$ can be re-written as

\begin{align}
\begin{split}
 N &= \text{diag}\left( \left( \sum_{j=0}^{q-1} k_j \right)_{k=0,\cdots,q-1} \right),
\end{split}
\label{eq:sp-2-5-diagonal-density}
\end{align}

where $k_j$ is the $j^{\text{th}}$ bit in the binary representation of
the $q$-bit number $k$.
Intuitively, this matrix weighs each basis state
by the number of $1$s in its binary encoding.
This is also called the \emph{Hamming weight}
of the bitstring,
which in the computational basis state encoding
corresponds exactly to the number of particles present at the lattice site.
Assuming the quantum state is encoded as $\ket{v}\ket{g}$
\ie the velocity qubits precede the positional qubits,
and that the positional qubits are in a superposition
that includes the target lattice site $x$,
we can isolate this position with the operator

\begin{equation}
N_x = \left( \sum_{j = 0}^{q-1} n_j \right) \otimes \ket{x}\bra{x},
\label{eq:sp-2-5-density-with-grid}
\end{equation}

which projects the particle mass encoded in the velocity register conditioned
on the grid qubits being in state $\ket{x}$.
Finally, the density, and therefore pressure,
over a region $\Gamma$ can be computed by means of the operator

\begin{align}
\begin{split}
P_\Gamma & = \sum_{x\in \Gamma} N_x  = \sum_{x\in \Gamma} \left( \sum_{j = 0}^{q-1} \left( \mathbb{I}^{\otimes j} \otimes \frac{1}{2}\left( \mathbb{I} - \mathbb{Z} \right) \otimes \mathbb{I}^{\otimes (q - 1 - j)}  \right) \right) \ket{x}\bra{x}.
\end{split}
\end{align}

Assuming a one-to-one collision model and
computing the expectation of $P$ with respect to an arbitrary QLGA
quantum state $\ket{\psi}$ yields

\begin{equation}
\bra{\psi}P_\Gamma\ket{\psi} \propto \sum_{x\in \Gamma} \sum_{j=0}^{q-1} f_j(x),
\end{equation}

which in turn leads to the mass over $\Gamma$ being $m_\Gamma = \frac{2^{n_g}}{\| \Gamma \|}\bra{\psi}P_\Gamma\ket{\psi}$
and the density $\rho_\Gamma = \frac{2^{n_g}}{q\| \Gamma \|}\bra{\psi}P_\Gamma\ket{\psi}$.

\paragraph{Example -- \dq{1}{2} mass measurement.}
We begin with the simplest working example that
demonstrates how the operator we described previously
correctly computes the mass of a basic \dq{1}{2} system.
For this instance, consider a lattice with just $2$ gridpoints,
and therefore one positional qubit.
Since there are no non-trivial equivalence classes in the \dq{1}{2} discretization,
there is no superposition over the velocity qubits.
%Finally, assume all streaming steps have been performed, and as such
%only the velocity qubits that correspond to the physical origin
%are of interest for the velocity computation.
Let the state of the system be

\begin{equation}
\ket{\uppsi_1} = \frac{1}{\sqrt{2}} \left( \ket{0}\ket{10} + \ket{1}\ket{11} \right).
\end{equation}

That is, the leftmost gridpoint $\ket{0}$ has a particle on one of its
velocity channels ($\ket{10}$), while the rightmost gridpoint $\ket{1}$
has $2$ particles ($\ket{11}$).
We verify the claim that the application of $P_\Gamma$ derived previously
to $\ket{\uppsi_1}$ correctly computes a scalar that can be post-processed into
the values of mass, density, and pressure, at the gridpoint corresponding to $\ket{0}$.
First, constructing the $N$ matrix as described in \Cref{eq:sp-2-5-diagonal-density}, we obtain

\begin{equation}
\mathrm{N}_{\mathrm{D}_1\mathrm{Q}_2} = \text{diag}((0, 1, 1, 2)),
\end{equation}

which weighs the four $2$-qubit basis states 
by their Hamming weight in the binary encoding.
To additionally project the $\ket{0}$ grid point
that we are interested in, we adapt the form given in \Cref{eq:sp-2-5-density-with-grid}
to obtain

\begin{align}
\begin{split}
\mathrm{N}_{(\mathrm{D}_1\mathrm{Q}_2, \{ 0 \})} & = \left(\ket{0}\bra{0}\right) \otimes \text{diag}((0, 1, 1, 2))\\
& = \begin{pmatrix}
1 & 0\\
0 & 0
\end{pmatrix} \otimes \begin{pmatrix}
0 & 0 & 0 & 0\\
0 & 1 & 0 & 0\\
0 & 0 & 1 & 0\\
0 & 0 & 0 & 2
\end{pmatrix}\\
& = \begin{pmatrix}
\text{diag}((0, 1, 1, 2)) & \mathbb{O}_{4\times 4}\\
\mathbb{O}_{4\times 4}& \mathbb{O}_{4\times 4}
\end{pmatrix}.\\
\end{split}
\end{align}

Since we are only interested in the properties of a single gridpoint,
$\mathrm{N}_{(\mathrm{D}_1\mathrm{Q}_2, \{ 0 \})} = P_\Gamma$ and the expectation value
$\bra{\uppsi_1}P_\Gamma\ket{\uppsi_1} = \frac{1}{2}$ gives the probability
mass corresponding to the states of interest, weighted by their mass.
Multiplying this value by $2^{n_g}$, we obtain $1$, the total mass over the area,
which in turn leads to density and pressure.
The physical interpretation of this procedure is straightforward
for the deterministic \dq{1}{2} model,
but becomes more involved once superposition-driven collision
is applied, as we show next.

\paragraph{Example -- \dq{2}{4} mass measurement.}
To demonstrate the applicability of the mass measurement technique
to a more general case, we use a \dq{2}{4} instance that includes a collision step.
For brevity, we omit the positional qubits and non-local velocity qubits
and instead focus on the $4$ qubits belonging to a physical origin.
Assume the state

\begin{equation}
\ket{\uppsi_1} = \frac{1}{\sqrt{2}} \left( \ket{1000} + \ket{1010} \right)
\label{eq:sp-2-5-measurment-example-pre-collision}
\end{equation}

is obtained as a result of streaming, \ie, from a left neighbor where no
collision took place in the previous time step, and from a
down neighbor where collision did occur.
The $\ket{1010}$ basis state belongs
to $E_{(\mathrm{D_2Q_4}, 2, (0, 0)^\mathrm{T})}=\{ 1010, 0101 \}$,
and as a result, its amplitude is shared with the $\ket{0101}$ state, resulting
in a post-collision configuration of

\begin{equation}
\ket{\uppsi_2} = \frac{1}{\sqrt{2}} \ket{1000} + \frac{1}{2}\ket{1010} + \frac{1}{2}\ket{0101}.
\label{eq:sp-2-5-measurment-example-post-collision}
\end{equation}

To verify the correctness of the mass measurement observable with respect
to the collision operator, we evaluate its expectation with respect to
$\ket{\uppsi_1}$ and $\ket{\uppsi_2}$ separately.
The corresponding observable is

\begin{equation}
\mathrm{N}_{\mathrm{D}_2\mathrm{Q}_4} = \text{diag}((0, 1, 1, 2, 1, 2, 2, 3, 1, 2, 2, 3, 2, 3, 3, 4)).
\end{equation}

Applying this operator to the pre-collision state yields

\begin{equation}
\bra{\uppsi_1}\mathrm{N}_{\mathrm{D}_2\mathrm{Q}_4} \ket{\uppsi_1} = \frac{1}{2}(1 + 2) = \frac{3}{2},
\end{equation}

as the Hamming weight of both basis states in
\Cref{eq:sp-2-5-measurment-example-pre-collision}
is equal, and the sum is therefore equal to the average mass of all states in the superposition.
Applying the same observable to the post-collision of \Cref{eq:sp-2-5-measurment-example-post-collision} configuration results in 

\begin{equation}
\bra{\uppsi_2}\mathrm{N}_{\mathrm{D}_2\mathrm{Q}_4} \ket{\uppsi_2} = \frac{1}{2} 1 + \frac{1}{4}(2 + 2) = \frac{3}{2}.
\end{equation}

The three basis states of $\ket{\uppsi_2}$ retain the same
cumulative contribution to the local mass as the pre-collision configuration.
In general, assuming collision affects a single
basis state $\ket{k}$ with amplitude $\alpha$ that
belongs to equivalence class $E$, the post-collision configuration
divides the amplitude of $\ket{k}$ uniformly along $\| E \|$ basis states
that, by virtue of belonging to the same equivalence class,
all have the same mass.
Assuming the mass of $\ket{k}$ is $m_k = \sum_{j=0}^{q-1}\ket{k_j}$,
since $|\alpha|^2m_k=|\alpha|^2\sum_{j=1}^{\| E \|} \frac{1}{\| E \|} m_k$,
collision preserves the expectation of the observable $N$.

\subsection{Momentum Exchange Method\label{subsec:sp-2-4-measurement-qmem}}

The Momentum Exchange Method (MEM) was developed for
Lattice Boltzmann Methods by Ladd \citep{ladd1994numerical, ladd1994numericalp2}
as a way to compute the forces acting on solid objects suspended in a fluid.
In the context of the continuous probability distributions of the LBM,
the MEM prescribes that the force acting on the solid domain is equal to

\begin{equation}
\mathbf{F}_\Gamma = \sum_{j=0}^{q-1} \left( \mathbf{e}_jf_j(\mathbf{x}_\Gamma) - \mathbf{e}_{\bar{j}}f_{\bar{j}}(\mathbf{x}_\Gamma) \right),
\end{equation}

where $\Gamma$ is the domain that comprises the solid-fluid interface
(under the assumption that the solid is free of fluid particles),
and $\mathbf{e}_j$ are vectors that encode the direction of the each
streaming channel according to the \dq{d}{q} discretization.
Similarly, $\mathbf{e}_{\bar{j}}$ is the directional vector
of the channel that a particle traveling channel $j$ occupies
after the application of boundary conditions.
For simplicity, we assume the simulation models bounce-back boundary conditions,
which simplifies the force calculation to

\begin{equation}
\mathbf{F}_\Gamma = \sum_{j=0}^{q-1} 2\mathbf{e}_jf_j(\mathbf{x}_\Gamma),
\label{eq:2-5-mem}
\end{equation}

as the opposing velocity channels that particles
occupy before and after boundary treatment are
$\mathbf{e}_j$ and $-\mathbf{e}_j$, respectively.
For the exact values of the $\mathbf{e}_j$ vectors corresponding to common \dq{d}{q} discretizations,
we refer the reader to \cite{kruger2017lattice}.

The first Quantum Momentum Exchange Method (QMEM) was introduced by \citet{schalkers2024momentum}
for an algorithm based on the amplitude encoding.
In their work, $2d$ observables each calculate one component
of the force vector, along each direction of each dimension.
However, since in the amplitude encoding,
the amplitude of particular basis state singularly determines
the mass occupying a particular velocity channel,
this method is not directly compatible with 
computational basis state encodings.
In this section, we adapt the quantum circuit implementation of the QMEM
observable described in \cite{schalkers2024momentum}
to the LGA model underlying the Space-Time encoding.

\paragraph{QMEM in the Space-Time Encoding.} The core differences between the LBM and LGA
MEM implementations stem from the discretization
and quantum register structure, respectively.
First, particles in LGA are boolean and therefore the $f_j$ terms in
\Cref{eq:2-5-mem} are either $0$ or $1$.
Second, as previously described in \Cref{subsec:sp-2-4-measurement-mass-pressure},
the weight of each basis state in a superposition at a local gridpoint
must be accounted for proportionately to
the force it exerts on the boundary of the solid object.
Furthermore, the locality of the Space-Time stencil
prevents the straightforward implementation from the original work \cite{schalkers2024momentum}
from measuring the appropriate quantity of interest, as it 
relies on particles having travelled into the boundary layer of the object.
Since information is local to grid positions in the Space-Time encoding,
this scenario never occurs and therefore requires a different approach.

Formulating observables that account for these criteria is
straightforward and can be done following the same steps as in \Cref{subsec:sp-2-4-measurement-mass-pressure}.
For each dimension $k$, we can define a diagonal matrix

\begin{align}
\begin{split}
N_{k} = \sum_{j=0}^{q-1} 2n_j \cdot\mathbf{e}_j(k),
\end{split}
\end{align}

with $\mathbf{e}_j(k)$ the $k^\text{th}$ entry of the directional vector $\mathbf{e}_j$,
and $n_j$ as defined in \Cref{eq:sp-2-5-number-matrix}.
In practice, this eliminates entries of the $N_{k}$ matrix that do not
contribute to the force component corresponding to that dimension.
The matrix could be further split up into two sparser observables
that only account for components in one particular
direction of the force as described in \cite{schalkers2024momentum}.
Unlike in the mass computation example, however, the final observable is no longer constructed with respect
to the qubits corresponding to the physical origin, as that would entail
initializing gridpoints inside the boundary of the obstacle,
which we avoid as detailed in \Cref{subsec:sp-2-2-boundary-pw}.
Instead, we can construct the observable as

\begin{equation}
N_{(x, k)} = \mathbb{I}^{\otimes (q\cdot o)}\left( \sum_{j = 0}^{q-1} n_j \right)\mathbb{I}^{\otimes(q\cdot(n_r-o-1))} \otimes \left( \ket{x'}\bra{x'} \right),
\end{equation}

with $x'$ the position of a gridpoint \emph{within the fluid domain} that contains information
regarding gridpoints that belong to the boundary of the solid domain in dimension $k$,
$n_r$ the number of $q$-qubit velocity registers the simulation contains,
and $o$ an appropriate offset.
Intuitively, this observable is expressed with respect to the velocity
qubits of gridpoints within the solid domain that are entangled with
positional qubits in the fluid domain.
Since the method follows the same rationale and reasoning as outline in \Cref{subsec:sp-2-4-measurement-mass-pressure},
we omit additional examples and instead describe a quantum circuit implementation that amasses
the amplitude of the states of interest onto an ancilla qubit.

\begin{figure}
    \centering
    \hfill
    \subcaptionbox{The gridpoints contributing to the MEM computation. \label{fig:sp-2-5-diag-sp-measurement-d2q4-grid-square}}{\centering
\begin{tikzpicture}[scale=0.3]
    % Grid lines

    \draw[
      help lines,
      line width=0.4pt,
      color=gray!30,
      dashed
    ] (-12, -12) grid[step={($(4, 4) - (0, 0)$)}] (12, 12);

    \foreach \x in {-10,-6,-2,2,6,10} {
        \foreach \y in {-10,-6,-2,2,6,10} {
            \dtwoqfour{(\x,\y)}{0.1cm}{2}{0}{1.5}{white}
        }
    }

    % Axes labels
    \foreach \x/\xtext in {0,...,5} {
        \draw (\x*4-10,-12) -- (\x*4-10,-12) node[anchor=north] {$\xtext$};
    }

    \foreach \x/\xtext in {0,...,5} {
        \draw (-12.75,\x*4-9.25) -- (-12.75,\x*4-9.25) node[anchor=north] {$\xtext$};
    }

     \foreach \y in {-6,-2,2,6} {
            \dtwoqfour{(-6,\y)}{0.1cm}{2}{0}{1.5}{red}
            \dtwoqfour{(-10,\y)}{0.1cm}{2}{0}{1.5}{red}
     }
\draw[-,solid, ultra thick, black, rounded corners] (4, -8) -- (8,-8) -- (8,8) -- (-8,8) -- (-8, -8) -- (8, -8) -- (8, -5);

\draw[-,solid, ultra thick, red, rounded corners] (-11, 6) -- (-11,-7) -- (-9,-7) -- (-9, 7) -- (-11, 7) -- (-11, 6);

\draw[->, thick, red] (-9.5, -2) -- (-8, -2);
\draw[->, thick, red] (-9.5, 2) -- (-8, 2);
\draw[->, thick, red] (-9.5, -6) -- (-8, -6);
\end{tikzpicture}}%
    \hfill%
    \subcaptionbox{The target qubit in the Space-Time stencil. \label{fig:sp-2-5-diag-sp-measurement-d2q4-grid-stencil}}{	\centering
\begin{tikzpicture}[scale=0.28]

    % Grid lines
    \draw[
      help lines,
      line width=0.3pt,
      color=gray!30,
      dashed
    ] (-8, -8) grid[step={($(2, 2) - (0, 0)$)}] (8, 8);
    \draw[
      help lines,
      line width=0.4pt,
      color=black!80,
      dashed
    ] (-10, -10) grid[step={($(4, 4) - (0, 0)$)}] (10, 10);
        
    % Nodes
    % Distance 0
    \dtwoqfour{(0,0)}{0.1cm}{2}{0}{1.5}{black}

    % Distance 1
    \dtwoqfour{(4,0)}{0.1cm}{2}{1}{1.5}{black!60}
    \dtwoqfour{(-4,0)}{0.1cm}{2}{2}{1.5}{black!60}
    \dtwoqfour{(0,4)}{0.1cm}{2}{3}{1.5}{black!60}
    \dtwoqfour{(0,-4)}{0.1cm}{2}{4}{1.5}{black!60}
    
    \draw[->, thick, red] (4.5, 0) -- (6, 0) node at (5.25, 1.25) {$\mathrm{v_1}$};
\end{tikzpicture}}%
    \hfill%
    \caption{Momentum Exchange example on a $6 \times 6$ \dq{2}{4} grid on the left wall of a square obstacle. \label{fig:sp-2-5-diag-measurement-grid-stencil}}
\end{figure}
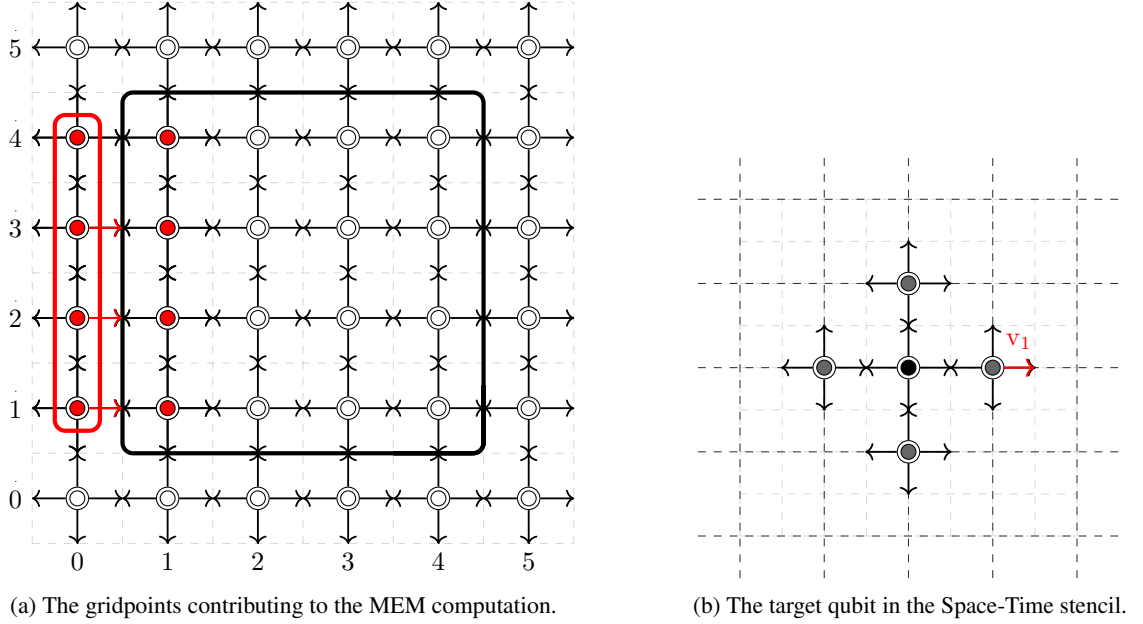

\paragraph{Quantum Circuit Implementation and \dq{2}{4} example.}
Amassing the amplitude of the particles impinging on the object in one particular
dimension can be achieved by utilizing the
techniques described in \Cref{sec:sp-2-1-initial-conditions} and \Cref{sec:sp-2-2-streaming-boundary},
as well as the $2d$ ancillae qubits used for volumetric operations
and $1$ additional ancilla qubit, $\mathrm{a_o}$.
The quantum circuit first isolates the region of space
in the fluid domain that interfaces with the solid domain by means of
the volumetric operations described in previous sections.
Following this step, at most $q2^d$ $\CPX{2d+1}$ gates
flip the state of the $\mathrm{a_o}$ qubit based
on the state of the volumetric ancillae
and the velocity qubit(s) that contribute to the particular
component of the force vector we attempt to measure.

To demonstrate the applicability of this procedure, we consider
a \dq{2}{4} example depicted in \Cref{fig:sp-2-5-diag-measurement-grid-stencil}.
\Cref{fig:sp-2-5-diag-sp-measurement-d2q4-grid-square} shows the system,
in which we are interested in measuring the horizontal component of the force vector
acting in the positive streaming direction on the left wall of the square obstacle.
The simplified state of the system before streaming is

\begin{equation}
\ket{\psi_{\text{pre}}} \propto \ket{a_o}\ket{a_{l,y}}\ket{a_{u,y}}\ket{x}\ket{y}\ket{v^l_0v^l_1v^l_2v^l_3}\ket{v^n_0v^n_1v^n_2v^n_3},
\end{equation}

which inlcudes the measurement and volumetric ancillae for the dimension we
are targetting, as well as the positional $x$ and $y$ qubits,
the $4$ local velocity qubits $v_j^l$,
and the $4$ velocity of the right neighbor of the physical origin $v^n_l$.
For brievity, we omit all other qubits of the register,
as they are not relevant for this computation.
Assume the starting state of the system is

\begin{align}
\begin{split}
\ket{\uppsi_1} & \propto \ket{0}\ket{00} \left( (\ket{0}\ket{1}\ket{1000} + \ket{0}\ket{2}\ket{1100} \right.\\
& + \ket{0}\ket{3}\ket{1101} + \ket{0}\ket{3}\ket{0000})\ket{0000} \\
&\left. + \ket{x}\ket{y}\ket{v^{\text{loc}}_{\text{out}}}\ket{v^{\text{n}}_{\text{out}}} \right).
\label{eq:sp-2-5-mem-pre-streaming}
\end{split}
\end{align}

The state of \Cref{eq:sp-2-5-mem-pre-streaming} assumes the four gridpoints
enclosed in the red rectangle in \Cref{fig:sp-2-5-diag-sp-measurement-d2q4-grid-square}
are in $4$ different states of $\ket{1000}$, $\ket{1100}$,
$\ket{1101}$, and $\ket{0000}$, respectively.
The three velocity channels in state $\ket{1}$ that contribute to the MEM computation
are highlighted in red in \Cref{fig:sp-2-5-diag-sp-measurement-d2q4-grid-square}.
As no particles have streamed into the object yet
and no volumetric operations have been performed,
the ancillae and neighboring velocity qubits are all $\ket{0}$.
All other basis states are only
represented generically as they are not relevant to the MEM.
Following streaming, the updated quantum state becomes

\begin{align}
\begin{split}
\ket{\uppsi_2} & \propto \ket{0}\ket{00} \left( \ket{0}\ket{1}\ket{v^{\text{loc}}_1}\ket{1000} + \ket{0}\ket{2}\ket{v^{\text{loc}}_2}\ket{1000} \right.\\
& + \ket{0}\ket{3}\ket{v^{\text{loc}}_3}\ket{1000} + \ket{0}\ket{4}\ket{v^{\text{loc}}_3}\ket{0000}\\
&\left. + \ket{x}\ket{y}\ket{v^{\text{loc}}_{\text{out}}}\ket{v^{\text{n}}_{\text{out}}} \right).
\end{split}
\label{eq:sp-2-5-mem-post-streaming}
\end{align}

In \Cref{eq:sp-2-5-mem-post-streaming}, $\ket{\uppsi_2}$ encodes that the velocity
occupancy of the first velocity channel
has streamed from the (now no longer of interest) $\ket{v^{\text{loc}}_j}$
qubits of the physical origin
to the qubits representing the right neighbor.
As such, the qubit of interest is now in state $\ket{1}$ for three of the four solid gridpoints.
At this point, the comparator operations can be applied to set the appropriate state onto the second and third ancillae as

\begin{align}
\begin{split}
\ket{\uppsi_3} & \propto \ket{0}\ket{11}\left( \ket{0}\ket{1}\ket{v^{\text{loc}}_1}\ket{1000} + \ket{0}\ket{2}\ket{v^{\text{loc}}_2}\ket{1000} \right)\\
& + \ket{0}\ket{11}\left(\ket{0}\ket{3}\ket{v^{\text{loc}}_3}\ket{1000} + \ket{0}\ket{4}\ket{v^{\text{loc}}_3}\ket{0000}\right)\\
& + \ket{0}\ket{k_1k_2}\ket{x}\ket{y}\ket{v^{\text{loc}}_{\text{out}}}\ket{v^{\text{n}}_{\text{out}}},\\
\end{split}
\end{align}

where $\ket{k_1k_2}$ symbolizes any $2$-qubit quantum state that is not $\ket{11}$.
Up to this point, the quantum circuit is identical to the
volumetric boundary conditions implementation described in \Cref{subsec:sp-2-2-boundary-vol}.
It is for this reason that we express the quantum circuit with respect to the neighboring qubits
of physical origin that belong to the boundary of the fluid domain.
We could alternatively compute the same quantity by measuring the post-boundary condition
velocity qubit in the opposite channel $\bar{c}$ of the physical origin,
or the channel $c$ of the physical origin pre-streaming.
Since no collision occurs between these steps, the same information travels across
these channels, and the quantum circuits therefore compute the same value.
We opt for this manner of implementing as it fits in exactly
with the realization of the volumetric boundary conditions.
Where the circuits differ is that to appropriately amass the amplitude
onto the $\ket{1}$ state of the leading qubit,
a $\CPX{3}$ gate is applied to the $\mathrm{a_o}$ qubit, controlled on the
two comparator qubits and the velocity qubit(s) that contribute to the force component.
In this case, this is the first qubit of the neighboring gridpoint,
visually depicted in \Cref{fig:sp-2-5-diag-sp-measurement-d2q4-grid-stencil}.
The state following this procedure is 

\begin{align}
\begin{split}
\ket{\uppsi_4} & \propto \ket{1}\ket{11} \left( \ket{0}\ket{1}\ket{v^{\text{loc}}_1}\ket{1000} + \ket{0}\ket{2}\ket{v^{\text{loc}}_2}\ket{1000} + \ket{0}\ket{3}\ket{v^{\text{loc}}_3}\ket{1000}\right)\\
& + \ket{0}\ket{11}\ket{0}\ket{4}\ket{v^{\text{loc}}_3}\ket{0000}\\
& + \ket{0}\ket{k_1k_2}\ket{x}\ket{y}\ket{v^{\text{loc}}_{\text{out}}}\ket{v^{\text{n}}_{\text{out}}}.
\end{split}
\end{align}

In $\ket{\uppsi_4}$, the $\mathrm{a_o}$ qubit
can be measured to extract the target amplitude.
Since the operations that isolate the positional qubits
and the $\CPX{3}$ gate do not affect the amplitude of the target
qubits, the same line of reasoning described
in the previous section can be followed to demonstrate
its compatibility with the superposed collision operators.
To generalize this procedure, $\mathcal{O}(q)$ ancilla qubits are
necessary, that can each be set to $\ket{1}$ for each
velocity channel that travels in the direction of the force component we attempt to measure.
\section{Results \label{sec:sp-4-results}}

In this section, we show the behavior of \dq{1}{2} and \dq{2}{4} systems simulated
through the building blocks introduced in this work.
We focus on showing the correct time-evolution of the system,
rather than verifying the numerical validity
of the model for particular target equations.
This is an intentional limitation of the study, as we aim to address
the latter through a broader empirical analysis in future works.
Implementations for all described building blocks are available
in the open-source \qlbm~software
package \cite{georgescu2025qlbm}.
Simulations were carried out using the Qiskit \cite{qiskit2024} \texttt{AerSimulator},
and visualizations are created through Paraview \cite{paraview}.
All scripts used to produce the
results and quantum circuits described in this paper
are available in the replication package
\cite{georgescu2025qlgabuildingblocksreplication}.

\begin{figure}
	\centering
	\hfill
	\subcaptionbox{The system in the $1^{\text{st}}$ time step. \label{fig:sp-3-1-results-d1q2-multistep-16-00-steps}}{\includegraphics[width=0.5\linewidth,trim=1cm 16cm 1cm 1cm,clip]{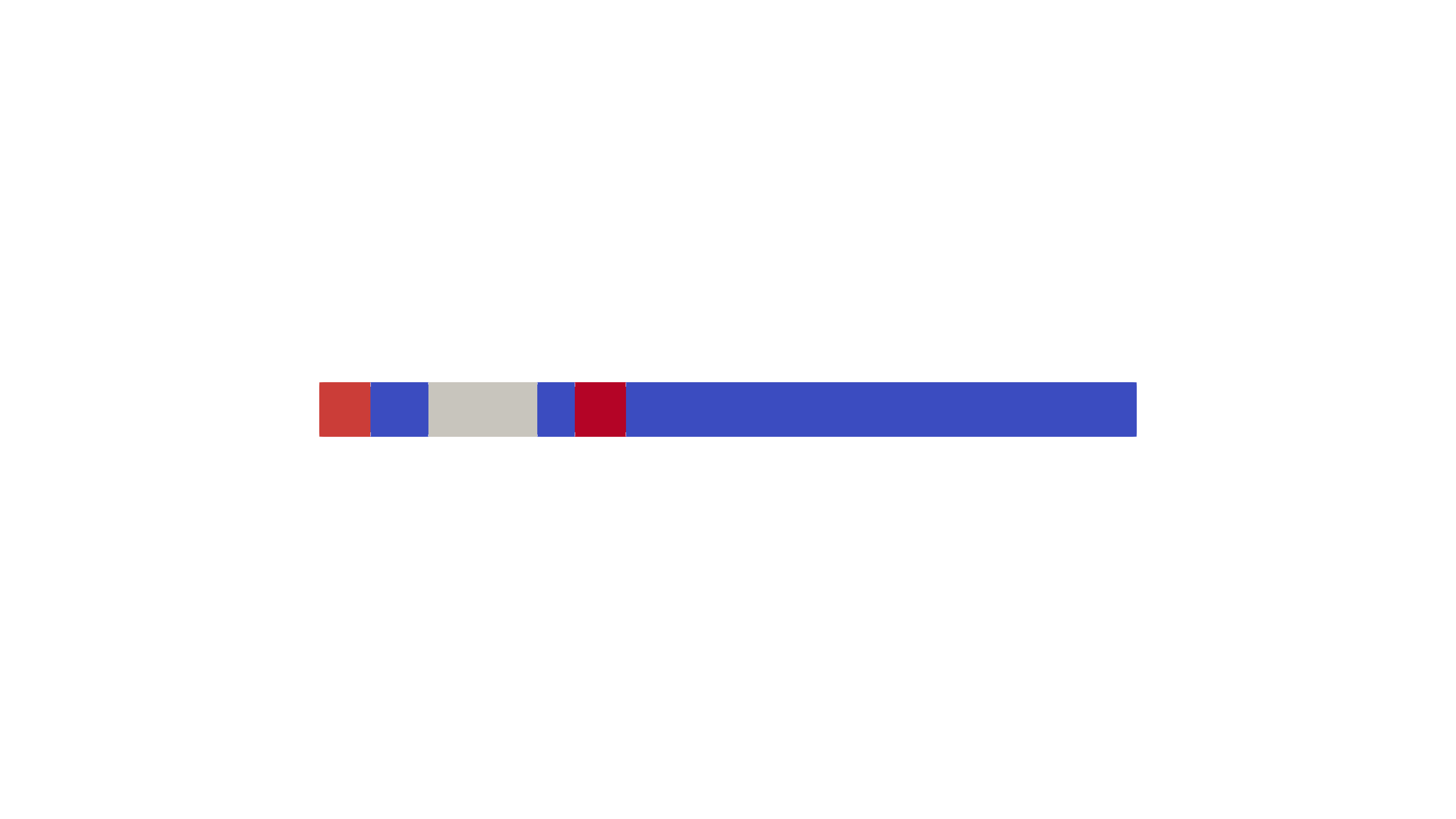}}%
	\hfill
	\subcaptionbox{The system in the $4^{\text{th}}$ time step ($1$ circuit repetition). \label{fig:sp-3-1-results-d1q2-multistep-16-01-steps}}{\includegraphics[width=0.5\linewidth,trim=1cm 16cm 1cm 10cm,clip]{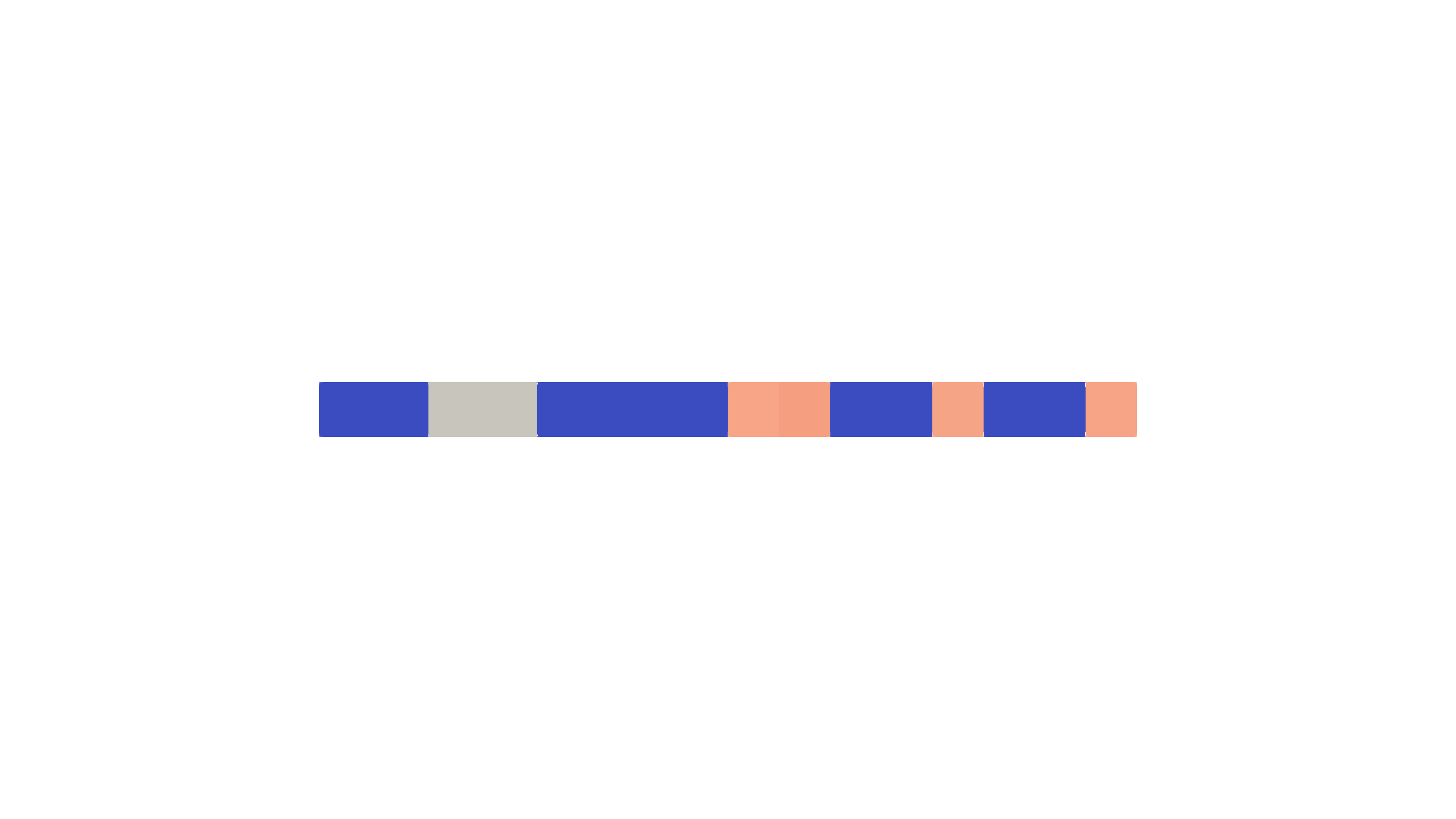}}%
	\hfill%
	\\
	\hfill
	\subcaptionbox{The system in the $8^{\text{th}}$ time step ($2$ circuit repetitions). \label{fig:sp-3-1-results-d1q2-multistep-16-02-steps}}{\includegraphics[width=0.5\linewidth,trim=1cm 12cm 1cm 10cm,clip]{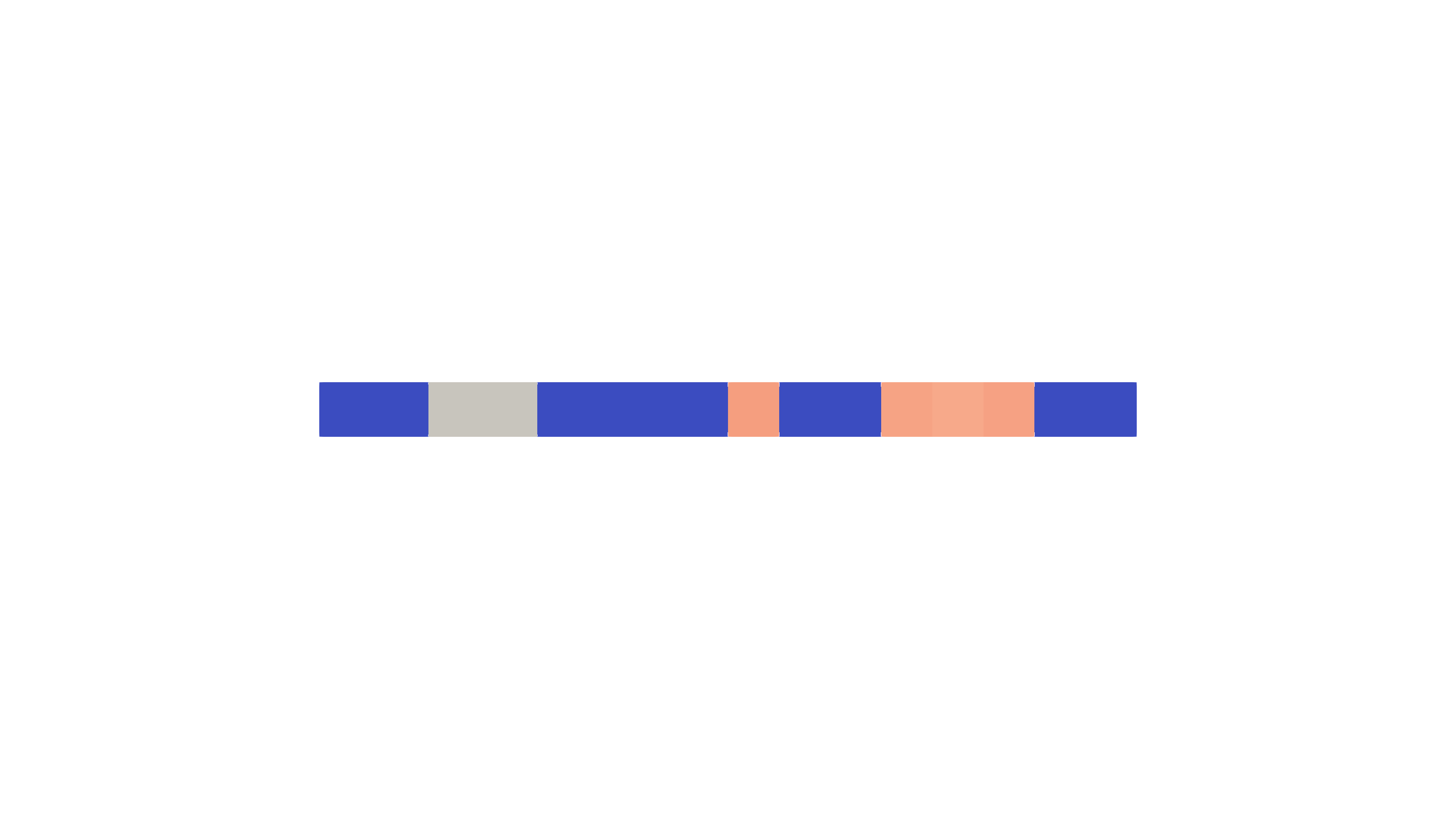}}%
	\hfill
	\subcaptionbox{The system in the $12^{\text{th}}$ time step ($3$ circuit repetitions). \label{fig:sp-3-1-results-d1q2-multistep-16-03-steps}}{\includegraphics[width=0.5\linewidth,trim=1cm 12cm 1cm 10cm,clip]{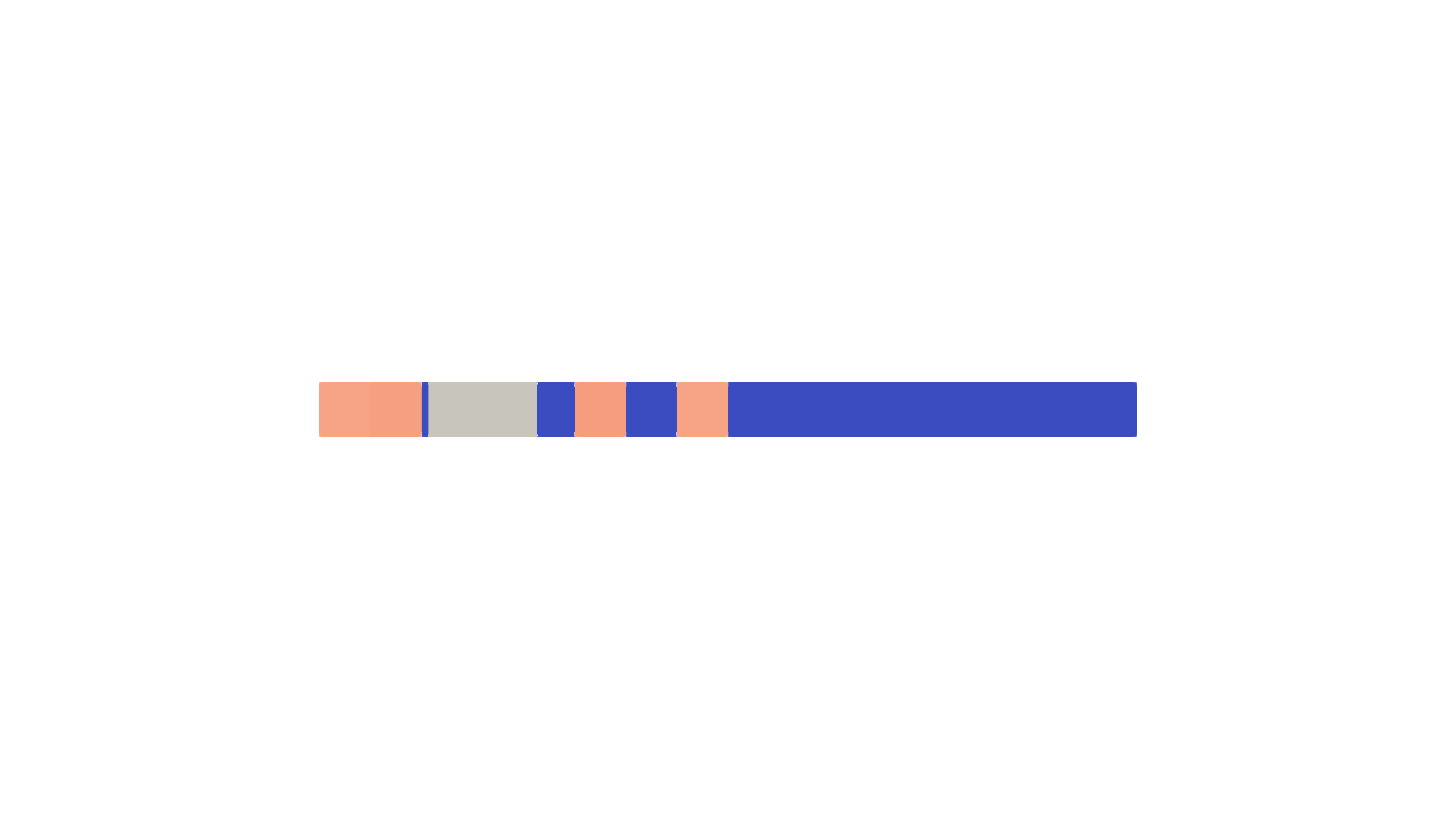}}%
	\hfill%
	\caption{Evolution of a \dq{1}{2} $4$-time step quantum circuit  system for $3$ repetitions. \label{fig:sp-3-1-results-d1q2-multistep-16}}    
\end{figure}

\paragraph{\dq{1}{2} -- multiple time steps and multiple obstacles.}
\Cref{fig:sp-3-1-results-d1q2-multistep-16} shows the simulations of a \dq{1}{2} system
with $16$ gridpoints,
where the gridpoints with indices $0$ and $4$ are initialized in the $\ket{11}$ state.
The solid domain is depicted in gray and
occupies the $2^{\text{nd}}$ and $3^{\text{rd}}$ gridpoints.
The edges of the fluid of the domain are prescribed periodic boundary conditions.
The quantum circuit uses $4$ qubits to encode the $16$ gridpoints,
 and an additional $18$ qubits to encode
the locality of each gridpoint for $4$ time steps.
Reinitialization is carried out every $4$ time steps by means of the pointwise method,
and since there are no non-trivial equivalence classes in \dq{1}{2},
reinitialization is straightforward and exact.
Darker shades of red indicate a higher particle occupancy ($2$), while lighter shades
indicate that a single particle is present.
Information is extracted from the quantum state by means of shots taken
at the end of each the circuit.
\Cref{fig:sp-3-1-results-d1q2-multistep-16-00-steps} shows the initial state
of the system, while Figures \ref{fig:sp-3-1-results-d1q2-multistep-16-01-steps},
\ref{fig:sp-3-1-results-d1q2-multistep-16-02-steps}, and \ref{fig:sp-3-1-results-d1q2-multistep-16-03-steps}
show the state of the system following $1$, $2$, and $3$ circuit repetitions,
with each repetition evolving the system for $4$ discrete time steps.

\begin{figure}
	\centering
	\hfill
	\subcaptionbox{The system in the $1^{\text{st}}$ time step. \label{fig:sp-3-1-results-d1q2-multiobject-16-00-steps}}{\includegraphics[width=0.45\linewidth,trim=1cm 12cm 1cm 5cm,clip]{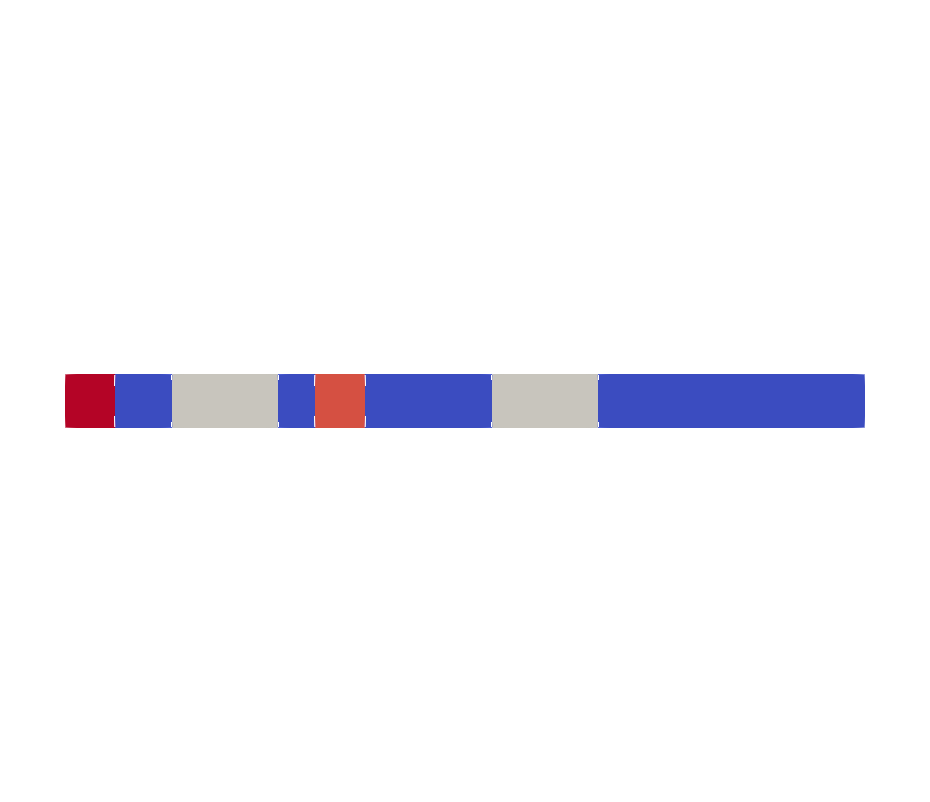}}%
	\hfill
	\subcaptionbox{The system in the $2^{\text{nd}}$ time step. \label{fig:sp-3-1-results-d1q2-multiobject-16-01-steps}}{\includegraphics[width=0.45\linewidth,trim=1cm 12cm 1cm 5cm,clip]{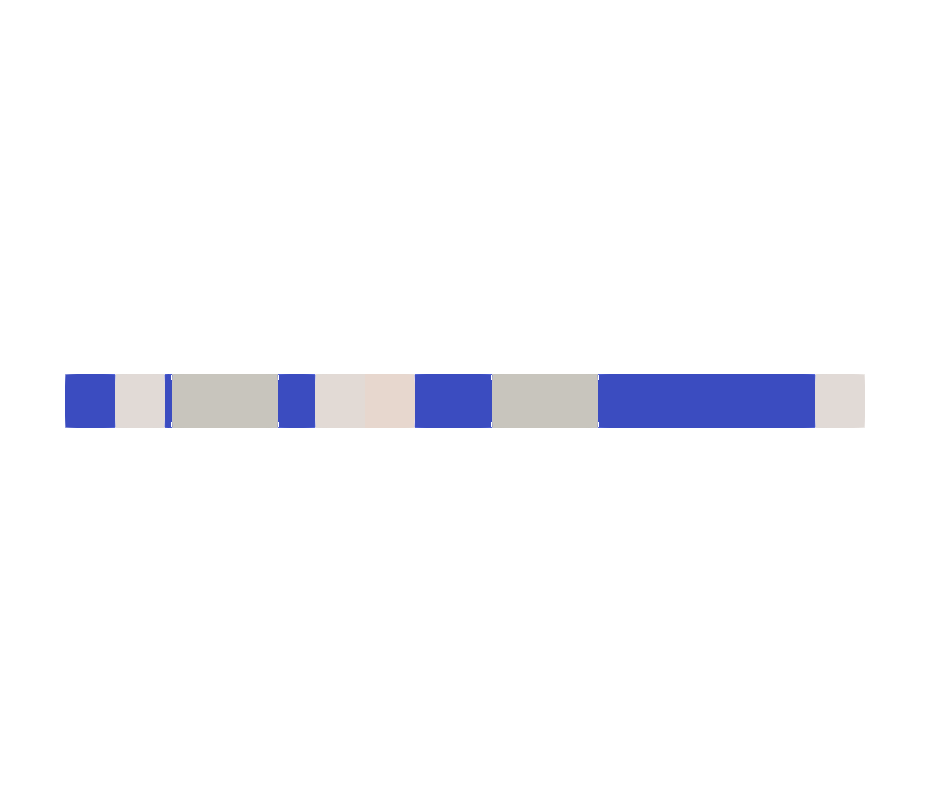}}%
	\hfill%
	\\
	\hfill
	\subcaptionbox{The system in the $3^{\text{rd}}$ time step. \label{fig:sp-3-1-results-d1q2-multiobject-16-02-steps}}{\includegraphics[width=0.45\linewidth,trim=1cm 12cm 1cm 10cm,clip]{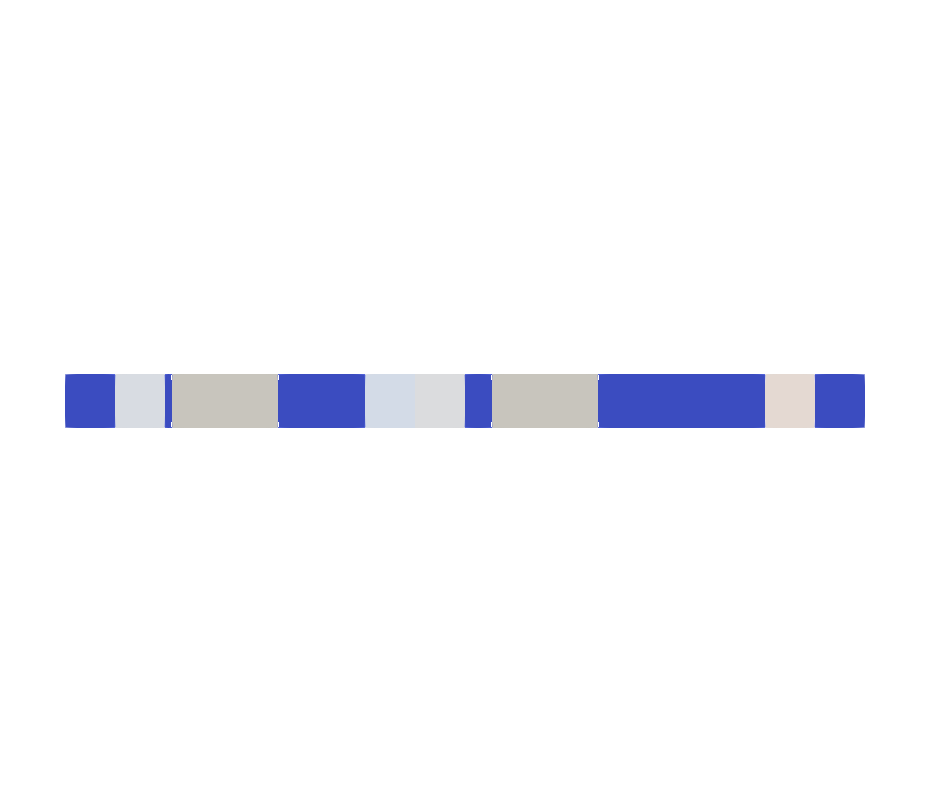}}%
	\hfill
	\subcaptionbox{The system in the $4^{\text{th}}$ time step. \label{fig:sp-3-1-results-d1q2-multiobject-16-03-steps}}{\includegraphics[width=0.45\linewidth,trim=1cm 12cm 1cm 10cm,clip]{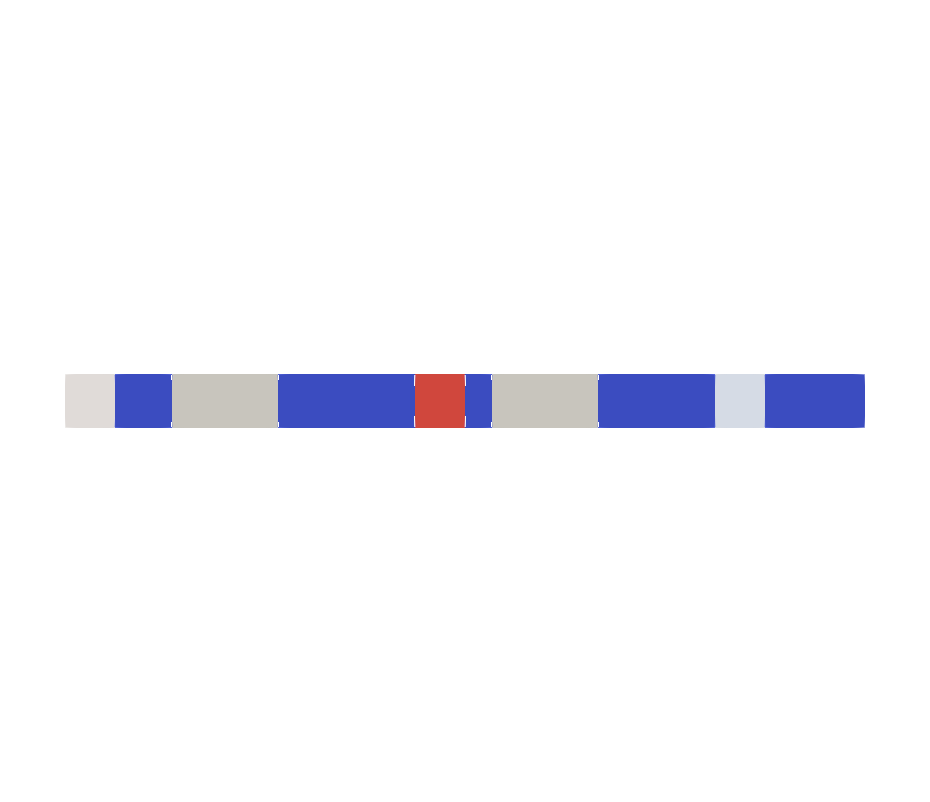}}%
	\hfill%
	\\
	\hfill
	\subcaptionbox{The system in the $5^{\text{th}}$ time step. \label{fig:sp-3-1-results-d1q2-multiobject-16-04-steps}}{\includegraphics[width=0.45\linewidth,trim=1cm 12cm 1cm 10cm,clip]{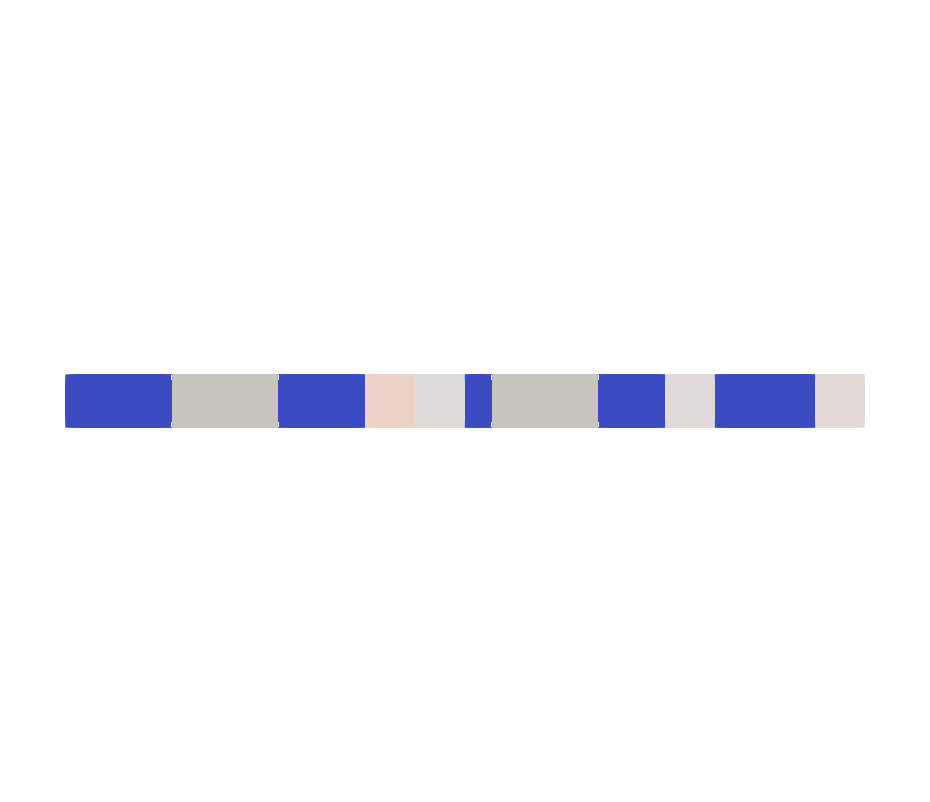}}%
	\hfill
	\subcaptionbox{The system in the $6^{\text{th}}$ time step. \label{fig:sp-3-1-results-d1q2-multiobject-16-05-steps}}{\includegraphics[width=0.45\linewidth,trim=1cm 12cm 1cm 10cm,clip]{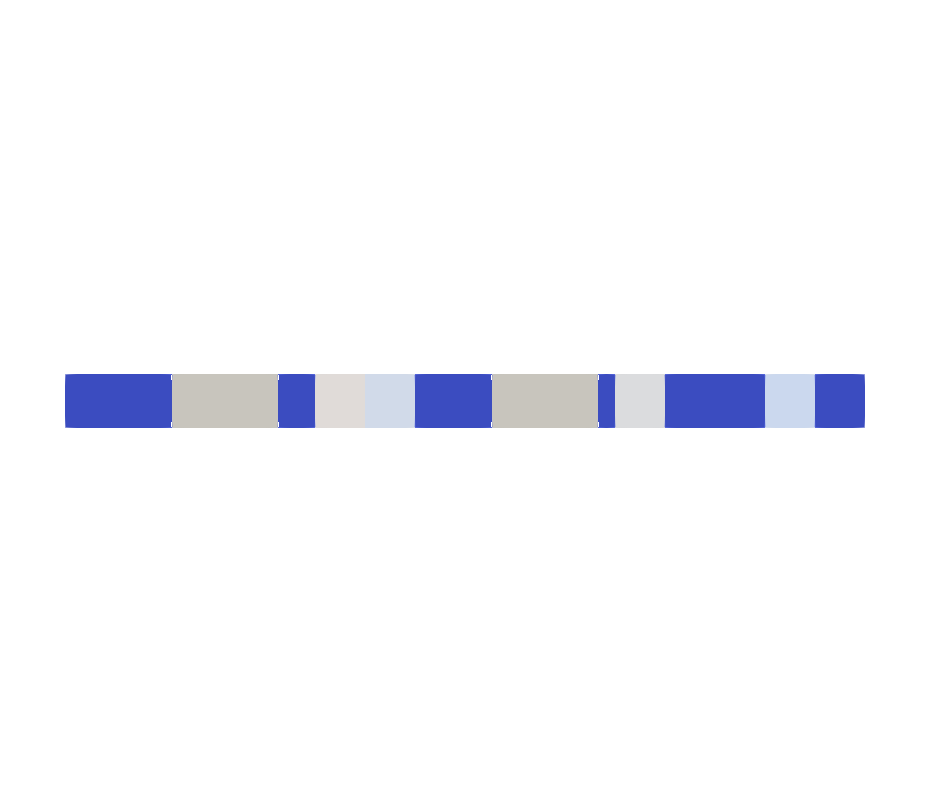}}%
	\hfill%
	\caption{Evolution of a \dq{1}{2} $1$-time step quantum circuit  system for $6$ time steps. \label{fig:sp-3-1-results-d1q2-multiobject-16}}    
\end{figure}

\Cref{fig:sp-3-1-results-d1q2-multiobject-16} shows the more granular evolution of a
similar $16$ gridpoint \dq{1}{2} system with two solid obstacles
placed spanning the gridpoints indexed $2, 3, 7$, and $8$.
The $0$ and $4$ gridpoints are initialized in the $\ket{11}$ state.
Unlike the previous example, the quantum circuit of this system
preforms restarts after every time step and as such only requires
$6$ qubits to encode the Space-Time stencil of one time step, and $10$ qubits in total.
Shades of red indicate a the presence of two particles at a gridpoint,
whereas lighter shades of blue and white indicate the presence of a single particle.
Numerical inaccuracies are again due to the postprocessing having been carried out
on shot, as opposed to expectation values.
Figures \ref{fig:sp-3-1-results-d1q2-multiobject-16-00-steps} through \ref{fig:sp-3-1-results-d1q2-multiobject-16-05-steps} show the evolution of
the system on a step-by-step basis, with the consistent application
of periodic boundary conditions at the edges of the fluid domain,
and bounce-back boundary conditions at the boundary of the solid objects.

\begin{figure}
	\centering
	\hfill
	\subcaptionbox{The system in the $1^{\text{st}}$ time step. \label{fig:sp-3-1-results-d2q4-circle-32x16-00-steps}}{\includegraphics[width=0.45\linewidth,trim=2cm 5cm 2cm 4cm,clip]{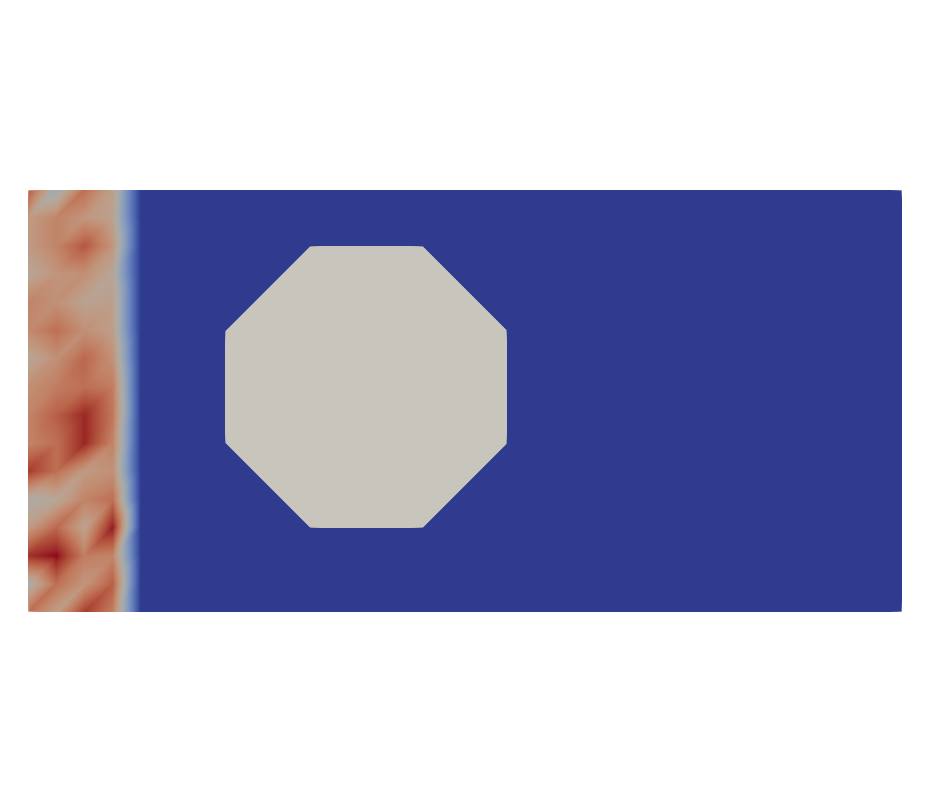}}%
	\hfill
	\subcaptionbox{The system in the $5^{\text{th}}$ time step. \label{fig:sp-3-1-results-d2q4-circle-32x16-04-steps}}{\includegraphics[width=0.45\linewidth,trim=2cm 5cm 2cm 4cm,clip]{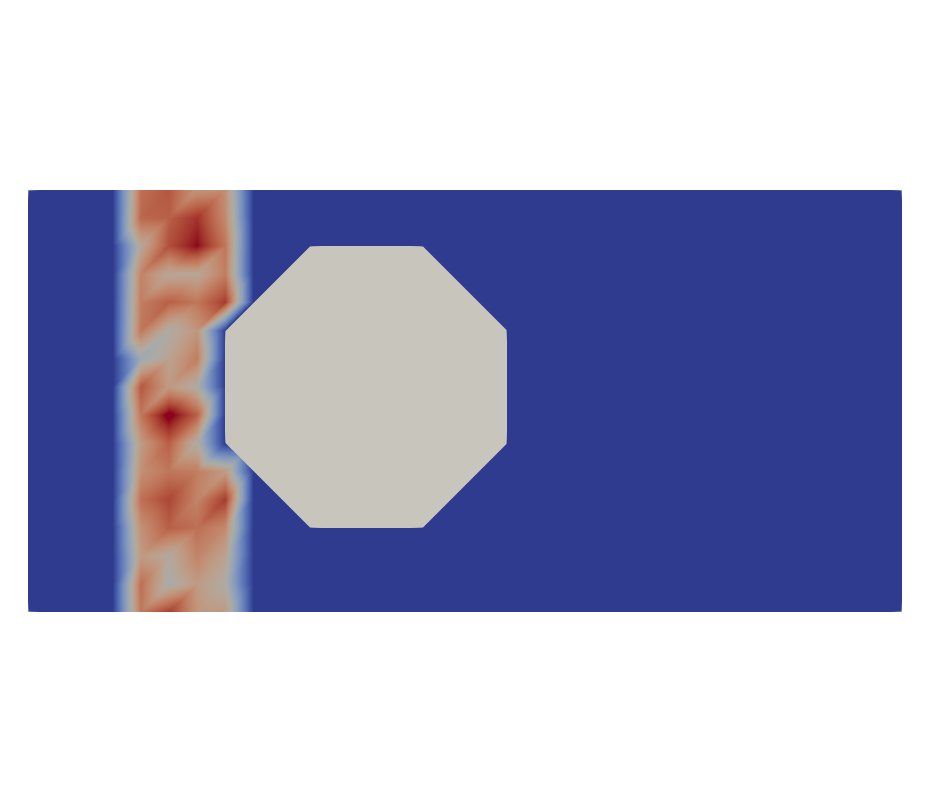}}%
	\hfill%
	\\
	\hfill
	\subcaptionbox{The system in the $10^{\text{th}}$ time step. \label{fig:sp-3-1-results-d2q4-circle-32x16-09-steps}}{\includegraphics[width=0.45\linewidth,trim=2cm 5cm 2cm 4cm,clip]{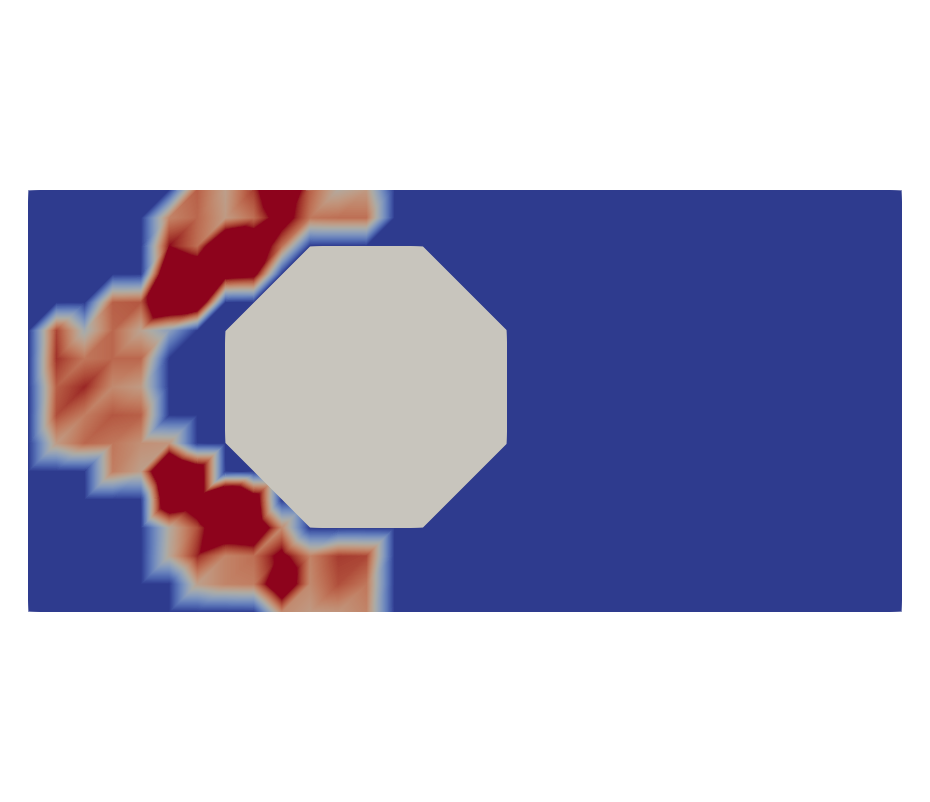}}%
	\hfill
	\subcaptionbox{The system in the $15^{\text{th}}$ time step. \label{fig:sp-3-1-results-d2q4-circle-32x16-14-steps}}{\includegraphics[width=0.45\linewidth,trim=2cm 5cm 2cm 4cm,clip]{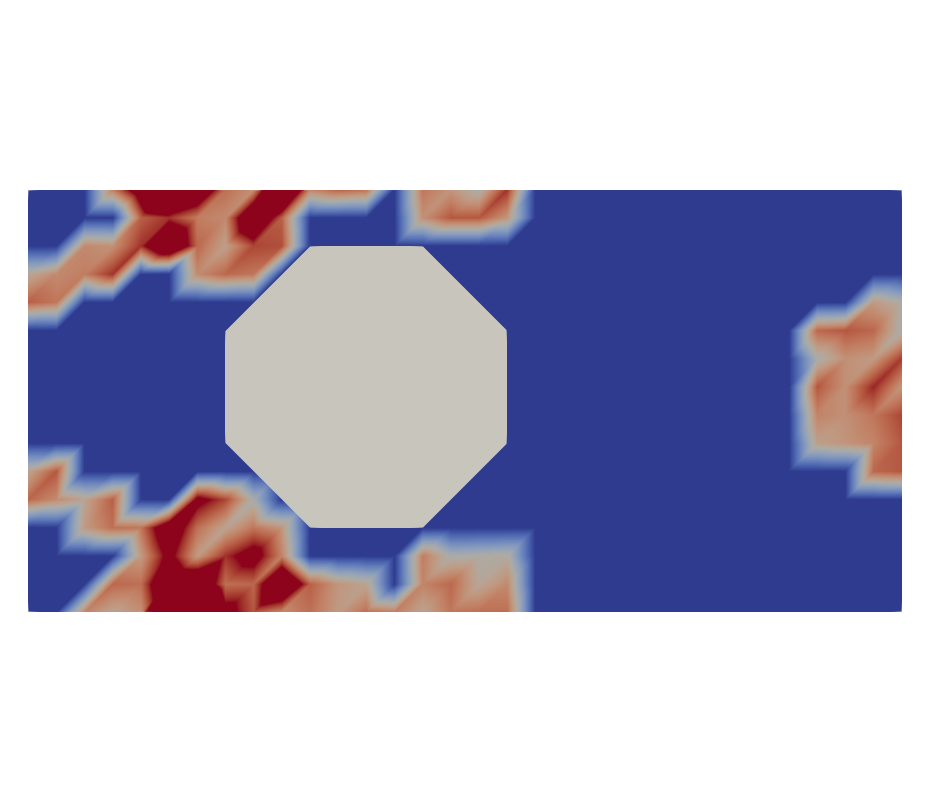}}%
	\hfill%
	\\
	\hfill
	\subcaptionbox{The system in the $20^{\text{th}}$ time step. \label{fig:sp-3-1-results-d2q4-circle-32x16-19-steps}}{\includegraphics[width=0.45\linewidth,trim=2cm 5cm 2cm 4cm,clip]{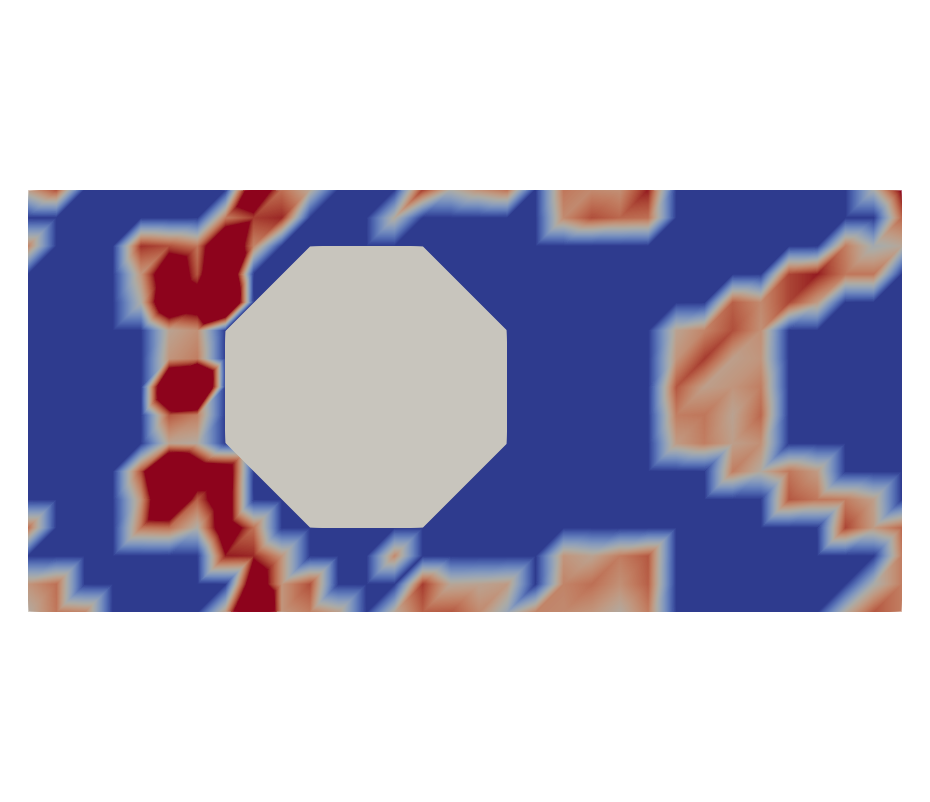}}%
	\hfill
	\subcaptionbox{The system in the $25^{\text{th}}$ time step. \label{fig:sp-3-1-results-d2q4-circle-32x16-24-steps}}{\includegraphics[width=0.45\linewidth,trim=2cm 5cm 2cm 4cm,clip]{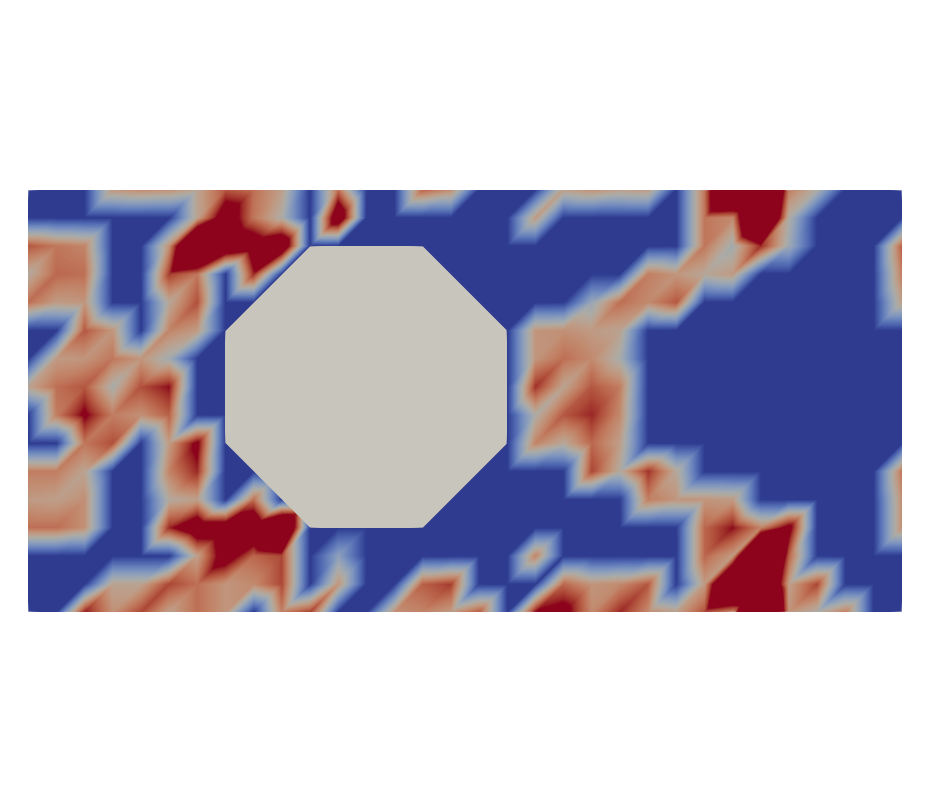}}%
	\hfill%
	\caption{Evolution of a \dq{2}{4} system for 25 time steps. \label{fig:sp-3-1-results-d2q4-circle-32x16}}
\end{figure}

\paragraph{\dq{2}{4} -- staircase approximation.} \Cref{fig:sp-3-1-results-d2q4-circle-32x16} shows the evolution of \dq{2}{4} system
that simulates a $32\times 16$ lattice, with a solid object in the shape of a circle
centered at $(12, 8)$ with a radius of $5$ gridpoints.
The initial conditions are set up such that the gridpoints within the volume spanning
$(0, 0) \times (2, 15)$ are occupied by a single particle moving in the positive $x$ direction.
To make the simulation feasible for commonplace classical hardware available today,
we perform reinitializations after each time step by reducing 
the superposition created by the collision operator to a deterministic state.
This way, no decomposition of large unitary matrices is required
and such simulations can be carried out efficiently.
This reinitialization mechanism is also implemented in the \qlbm~library.
The simulation required $29$ qubits: $20$ qubits to encode the extended computational
basis state that performs streaming and collision for $1$ time step,
and $\log_2 32 + \log_2 16 = 9$ qubits to entangle the velocity register to the physical grid.
The choice of simulating one time step is made for pragmatical reasons,
as the simulation of two time steps would require $52$ qubits,
which exceeds practical computational capabilities.

Figures \ref{fig:sp-3-1-results-d2q4-circle-32x16-00-steps} and
\ref{fig:sp-3-1-results-d2q4-circle-32x16-04-steps} show the collisionless
advection of particles through space, as the $\ket{1000}$ is preserved by streaming.
These steps are exact, even with the added reinitialization steps.
Following the interaction with the solid domain,
particles are bounced back, such that the collision-sensitive basis state $\ket{1010}$
occurs at the left boundary of the object.
Following several more time steps, the reinitialization approximation leads
to higher particle density areas (highlighted by the darker shade of red)
occurring near the boundary of the object, as shown in \Cref{fig:sp-3-1-results-d2q4-circle-32x16-09-steps}.
The process continues, in Figures \ref{fig:sp-3-1-results-d2q4-circle-32x16-14-steps} through
\ref{fig:sp-3-1-results-d2q4-circle-32x16-24-steps},
where the additional complexity of periodic boundary condition is accounted for.

\section{Conclusion\label{sec:sp-4-conclusion}}

This work introduced novel building blocks
for every step of the quantum lattice gas automata
loop, from initialization to measurement.
We described the implementation of expressive pointwise
methods and efficient volumetric alternatives for 
the application of initial and boundary conditions.
We extended the efficient volumetric boundary conditions to novel
geometric patterns that can be applied to efficiently
model more complex shapes.
We developed and generalized previous one-to-ne collision
operators to arbitrary velocity discretization
under a new, less restrictive interpretation of equivalence classes.
Finally, we outlined the properties of observables
that recover quantities of interest from the quantum state.
Each building block is accompanied by a complexity
analysis in terms of one- and two-qubit gates,
and all implementations of described building blocks are 
made available in an open-source library.

Although we emphasized the utility of the introduced building blocks
using the general form of the Space-Time data encoding \cite{schalkers2024importance},
only minimal modifications are required to implement them in the 
QLGA algorithms of, for instance, \citet{love2019quantum}, \citet{kocherla2024fully}, \citet{zamora2025float}, and Fonio et al. \cite{fonio2023quantum, fonio2025adaptive},
which we hope will benefit several  independent directions of QLGA research.
In the future, we aim to increase the computational efficiency
of QLGA and investigate its applicability to realistic use cases.
On the computational side,
reducing the complexity of each building
block and investigating more efficient decompositions could
improve the situations in which QLGA may become a feasible alternative
to classical solvers.
On the use case side, we aim to empirically investigate the implications of our novel
collision operator with respect to different flow regimes and governing
equations, and further tailor the presented building blocks
to specific real-world applications.
\section{Acknowledgements\label{sec:ack}}

We gratefully acknowledge support from the joint research program
\emph{Quantum Computational Fluid Dynamics} by Fujitsu Limited and Delft
University of Technology, co-funded by the Netherlands Enterprise
Agency under project number PPS23-3-03596728.
We thank Monica L\u{a}c\u{a}tu\c{s}, Alex Sturges, and Jingya Li
for many fruitful discussions regarding the QLGA algorithm.
We furthermore thank Daniela Toader for her valuable feedback on the contents of this manuscript.

%\newpage

\bibliographystyle{plainnat}
\bibliography{references}

@article{harrow2009quantum,
  title={Quantum algorithm for linear systems of equations},
  author={Harrow, Aram W and Hassidim, Avinatan and Lloyd, Seth},
  journal={Physical review letters},
  volume={103},
  number={15},
  pages={150502},
  year={2009},
  publisher={APS}
}

@article{montanaro2016quantum,
  title={Quantum algorithms and the finite element method},
  author={Montanaro, Ashley and Pallister, Sam},
  journal={Physical Review A},
  volume={93},
  number={3},
  pages={032324},
  year={2016},
  publisher={APS}
}

@article{cao2013quantum,
	title={Quantum algorithm and circuit design solving the Poisson equation},
	author={Cao, Yudong and Papageorgiou, Anargyros and Petras, Iasonas and Traub, Joseph and Kais, Sabre},
	journal={New Journal of Physics},
	volume={15},
	number={1},
	pages={013021},
	year={2013},
	publisher={IOP Publishing}
}

@article{tiwari2025algorithmic,
	title={Algorithmic Advances Towards a Realizable Quantum Lattice Boltzmann Method},
	author={Tiwari, Apurva and Iaconis, Jason and Jojo, Jezer and Ray, Sayonee and Roetteler, Martin and Hill, Chris and Pathak, Jay},
	journal={arXiv preprint arXiv:2504.10870},
	year={2025}
}

@article{xu2025improved,
	title={Improved quantum lattice Boltzmann method for advection-diffusion equations with a linear collision model},
	author={Xu, Li and Li, Ming and Zhang, Lei and Sun, Hai and Yao, Jun},
	journal={Physical Review E},
	volume={111},
	number={4},
	pages={045305},
	year={2025},
	publisher={APS}
}

@article{wawrzyniak2025dynamic,
	title={Dynamic Circuits for the Quantum Lattice-Boltzmann Method},
	author={Wawrzyniak, David and Winter, Josef and Schmidt, Steffen and Indiniger, Thomas and Jan{\ss}en, Christian F and Schramm, Uwe and Adams, Nikolaus A},
	journal={arXiv preprint arXiv:2502.02131},
	year={2025}
}

@article{fonio2023quantum,
	title = {Quantum collision circuit, quantum invariants and quantum phase estimation procedure for fluid dynamic lattice gas automata},
	journal = {Computers \& Fluids},
	volume = {299},
	pages = {106688},
	year = {2025},
	issn = {0045-7930},
	author = {Niccolò Fonio and Pierre Sagaut and Giuseppe {Di Molfetta}}
}

@article{chen2022quantum,
	title={Quantum approach to accelerate finite volume method on steady computational fluid dynamics problems},
	author={Chen, Zhao-Yun and Xue, Cheng and Chen, Si-Ming and Lu, Bing-Han and Wu, Yu-Chun and Ding, Ju-Chun and Huang, Sheng-Hong and Guo, Guo-Ping},
	journal={Quantum Information Processing},
	volume={21},
	number={4},
	pages={137},
	year={2022},
	publisher={Springer}
}

@article{aaronson2015read,
  title={Read the fine print},
  author={Aaronson, Scott},
  journal={Nature Physics},
  volume={11},
  number={4},
  pages={291--293},
  year={2015},
  publisher={Nature Publishing Group UK London}
}

@article{kyriienko2021solving,
  title={Solving nonlinear differential equations with differentiable quantum circuits},
  author={Kyriienko, Oleksandr and Paine, Annie E and Elfving, Vincent E},
  journal={Physical Review A},
  volume={103},
  number={5},
  pages={052416},
  year={2021},
  publisher={APS}
}

@article{demirdjian2022variational,
	title={Variational quantum solutions to the advection--diffusion equation for applications in fluid dynamics},
	author={Demirdjian, Reuben and Gunlycke, Daniel and Reynolds, Carolyn A and Doyle, James D and Tafur, Sergio},
	journal={Quantum Information Processing},
	volume={21},
	number={9},
	pages={322},
	year={2022},
	publisher={Springer}
}

@article{sato2021variational,
	title={Variational quantum algorithm based on the minimum potential energy for solving the Poisson equation},
	author={Sato, Yuki and Kondo, Ruho and Koide, Satoshi and Takamatsu, Hideki and Imoto, Nobuyuki},
	journal={Physical Review A},
	volume={104},
	number={5},
	pages={052409},
	year={2021},
	publisher={APS}
}

@book{nielsen2010quantum,
  title={Quantum computation and quantum information},
  author={Nielsen, Michael A and Chuang, Isaac L},
  year={2010},
  publisher={Cambridge university press}
}

@misc{qiskit2024,
      title={Quantum computing with {Q}iskit},
      author={Javadi-Abhari, Ali and Treinish, Matthew and Krsulich, Kevin and Wood, Christopher J. and Lishman, Jake and Gacon, Julien and Martiel, Simon and Nation, Paul D. and Bishop, Lev S. and Cross, Andrew W. and Johnson, Blake R. and Gambetta, Jay M.},
      year={2024},
      doi={10.48550/arXiv.2405.08810},
      eprint={2405.08810},
      archivePrefix={arXiv},
      primaryClass={quant-ph}
}

@article{kruger2017lattice,
  title={The lattice Boltzmann method},
  author={Kr{\"u}ger, Timm and Kusumaatmaja, Halim and Kuzmin, Alexandr and Shardt, Orest and Silva, Goncalo and Viggen, Erlend Magnus},
  journal={Springer International Publishing},
  volume={10},
  number={978-3},
  pages={4--15},
  year={2017},
  publisher={Springer}
}

@article{yepez2001quantum,
  title={Quantum lattice-gas model for computational fluid dynamics},
  author={Yepez, Jeffrey},
  journal={Physical Review E},
  volume={63},
  number={4},
  pages={046702},
  year={2001},
  publisher={APS}
}

@article{yepez2002efficient,
  title={An efficient and accurate quantum lattice-gas model for the many-body Schr{\"o}dinger wave equation},
  author={Yepez, Jeffrey and Boghosian, Bruce},
  journal={Computer Physics Communications},
  volume={146},
  number={3},
  pages={280--294},
  year={2002},
  publisher={Elsevier}
}

@article{yepez2002quantum,
  title={Quantum lattice-gas model for the Burgers equation},
  author={Yepez, Jeffrey},
  journal={Journal of Statistical Physics},
  volume={107},
  pages={203--224},
  year={2002},
  publisher={Springer}
}

@article{yepez2001quantumdiffusion,
	title={Quantum lattice-gas model for the diffusion equation},
	author={Yepez, Jeffrey},
	journal={International Journal of Modern Physics C},
	volume={12},
	number={09},
	pages={1285--1303},
	year={2001},
	publisher={World Scientific}
}

@article{todorova2020quantum,
  title={Quantum algorithm for the collisionless Boltzmann equation},
  author={Todorova, Blaga N and Steijl, Ren{\'e}},
  journal={Journal of Computational Physics},
  volume={409},
  pages={109347},
  year={2020},
  publisher={Elsevier}
}

@article{schalkers2024efficient,
  title={Efficient and fail-safe quantum algorithm for the transport equation},
  author={Schalkers, Merel A and M{\"o}ller, Matthias},
  journal={Journal of Computational Physics},
  volume={502},
  pages={112816},
  year={2024},
  publisher={Elsevier}
}

@article{steijl2020quantum,
  title={Quantum algorithms for nonlinear equations in fluid mechanics},
  author={Steijl, Rene},
  journal={Quantum computing and communications},
  year={2020},
  publisher={IntechOpen London}
}

@article{moawad2022investigating,
  title={Investigating hardware acceleration for simulation of CFD quantum circuits},
  author={Moawad, Youssef and Vanderbauwhede, Wim and Steijl, Ren{\'e}},
  journal={Frontiers in Mechanical Engineering},
  volume={8},
  pages={925637},
  year={2022},
  publisher={Frontiers Media SA}
}

@misc{itani2023qalb,
	title={Quantum Algorithm for Lattice Boltzmann (QALB) Simulation of Incompressible Fluids with a Nonlinear Collision Term}, 
	author={Wael Itani and Katepalli R. Sreenivasan and Sauro Succi},
	year={2023},
	eprint={2304.05915},
	archivePrefix={arXiv},
	primaryClass={quant-ph},
	url={https://arxiv.org/abs/2304.05915}, 
}

@misc{sanavio2025carleman,
	title={Carleman-lattice-Boltzmann quantum circuit with matrix access oracles}, 
	author={Claudio Sanavio and William A. Simon and Alexis Ralli and Peter Love and Sauro Succi},
	year={2025},
	eprint={2501.02582},
	archivePrefix={arXiv},
	primaryClass={comp-ph},
	url={https://arxiv.org/abs/2501.02582}, 
}

@article{kumar2024decomposition,
	title={Decomposition of nonlinear collision operator in quantum Lattice Boltzmann algorithm},
	author={Kumar, E Dinesh and Frankel, Steven H},
	journal={Europhysics Letters},
	volume={148},
	number={3},
	pages={38003},
	year={2024},
	publisher={IOP Publishing}
}

@article{sanavio2024lattice,
  title={Lattice Boltzmann--Carleman quantum algorithm and circuit for fluid flows at moderate Reynolds number},
  author={Sanavio, Claudio and Succi, Sauro},
  journal={AVS Quantum Science},
  volume={6},
  number={2},
  year={2024},
  publisher={AIP Publishing}
}

@article{itani2022analysis,
  title={Analysis of Carleman linearization of lattice Boltzmann},
  author={Itani, Wael and Succi, Sauro},
  journal={Fluids},
  volume={7},
  number={1},
  pages={24},
  year={2022},
  publisher={MDPI}
}

@article{budinski2021quantum,
  title={Quantum algorithm for the advection--diffusion equation simulated with the lattice Boltzmann method},
  author={Budinski, Ljubomir},
  journal={Quantum Information Processing},
  volume={20},
  number={2},
  pages={57},
  year={2021},
  publisher={Springer}
}

@article{ljubomir2022quantum,
  title={Quantum algorithm for the Navier--Stokes equations by using the streamfunction-vorticity formulation and the lattice Boltzmann method},
  author={Budinski, Ljubomir},
  journal={International Journal of Quantum Information},
  volume={20},
  number={02},
  pages={2150039},
  year={2022},
  publisher={World Scientific}
}

@article{schalkers2024importance,
  title={On the importance of data encoding in quantum Boltzmann methods},
  author={Schalkers, Merel A and M{\"o}ller, Matthias},
  journal={Quantum Information Processing},
  volume={23},
  number={1},
  pages={20},
  year={2024},
  publisher={Springer}
}

@article{schalkers2024momentum,
	title={Momentum exchange method for quantum Boltzmann methods},
	author={Schalkers, Merel A and M{\"o}ller, Matthias},
	journal={Computers \& Fluids},
	volume={285},
	pages={106453},
	year={2024},
	publisher={Elsevier}
}

@article{paraview,
  title={36-paraview: An end-user tool for large-data visualization},
  author={Ahrens, James and Geveci, Berk and Law, Charles and Hansen, C and Johnson, C},
  journal={The visualization handbook},
  volume={717},
  pages={50038--1},
  year={2005},
  publisher={Citeseer}
}

@article{wawrzyniak2024quantum,
	title = {A quantum algorithm for the lattice-Boltzmann method advection-diffusion equation},
	journal = {Computer Physics Communications},
	volume = {306},
	pages = {109373},
	year = {2025},
	issn = {0010-4655},
	doi = {https://doi.org/10.1016/j.cpc.2024.109373},
	url = {https://www.sciencedirect.com/science/article/pii/S0010465524002960},
	author = {David Wawrzyniak and Josef Winter and Steffen Schmidt and Thomas Indinger and Christian F. Janßen and Uwe Schramm and Nikolaus A. Adams},
}

@article{mcclean2018barren,
  title={Barren plateaus in quantum neural network training landscapes},
  author={McClean, Jarrod R and Boixo, Sergio and Smelyanskiy, Vadim N and Babbush, Ryan and Neven, Hartmut},
  journal={Nature communications},
  volume={9},
  number={1},
  pages={4812},
  year={2018},
  publisher={Nature Publishing Group UK London}
}

@article{childs2012hamiltonian,
  title={Hamiltonian simulation using linear combinations of unitary operations},
  author={Childs, Andrew M and Wiebe, Nathan},
  journal={arXiv preprint arXiv:1202.5822},
  year={2012}
}

@article{georgescu2025qlbm,
	title={qlbm--A Quantum Lattice Boltzmann Software Framework},
	author={Georgescu, C{\u{a}}lin A and Schalkers, Merel A and M{\"o}ller, Matthias},
	journal={Computer Physics Communications},
	pages={109699},
	year={2025},
	publisher={Elsevier}
}

@book{wolf2004lattice,
  title={Lattice-gas cellular automata and lattice Boltzmann models: an introduction},
  author={Wolf-Gladrow, Dieter A},
  year={2004},
  publisher={Springer}
}

@article{barenco1995elementary,
  title={Elementary gates for quantum computation},
  author={Barenco, Adriano and Bennett, Charles H and Cleve, Richard and DiVincenzo, David P and Margolus, Norman and Shor, Peter and Sleator, Tycho and Smolin, John A and Weinfurter, Harald},
  journal={Physical review A},
  volume={52},
  number={5},
  pages={3457},
  year={1995},
  publisher={APS}
}

@article{musk2020comparison,
  title={A comparison of quantum and traditional Fourier transform computations},
  author={Musk, Damian R},
  journal={Computing in Science \& Engineering},
  volume={22},
  number={6},
  pages={103--110},
  year={2020},
  publisher={IEEE}
}

@article{herbert2024almost,
  title={Almost-optimal computational-basis-state transpositions},
  author={Herbert, Steven and Sorci, Julien and Tang, Yao},
  journal={Physical Review A},
  volume={110},
  number={1},
  pages={012437},
  year={2024},
  publisher={APS}
}

@book{deza2012figurate,
  title={Figurate numbers},
  author={Deza, Elena and Deza, Michel},
  year={2012},
  publisher={World Scientific}
}

@article{heese2022representation,
  title={Representation of binary classification trees with binary features by quantum circuits},
  author={Heese, Raoul and Bickert, Patricia and Niederle, Astrid Elisa},
  journal={Quantum},
  volume={6},
  pages={676},
  year={2022},
  publisher={Verein zur F{\"o}rderung des Open Access Publizierens in den Quantenwissenschaften}
}

@article{nor2007three,
  title={Three-dimensional thermal lattice Boltzmann simulation of natural convection in a cubic cavity},
  author={Nor Azwadi, CS and Tanahashi, T},
  journal={International Journal of Modern Physics B},
  volume={21},
  number={01},
  pages={87--96},
  year={2007},
  publisher={World Scientific}
}

@article{frisch1986lattice,
	title={Lattice-gas automata for the Navier-Stokes equation},
	author={Frisch, Uriel and Hasslacher, Brosl and Pomeau, Yves},
	journal={Physical review letters},
	volume={56},
	number={14},
	pages={1505},
	year={1986},
	publisher={APS}
}

@article{frisch1987lattice,
author = {Frisch, Uriel and Dhumieres, Dominique and Hasslacher, B. and Lallemand, P. and Pomeau, Yves and Rivet, J.},
year = {1987},
month = {01},
pages = {},
title = {Lattice Gas Hydrodynamics in Two and Three Dimensions},
volume = {1},
journal = {Complex Systems}
}

@article{benzi1992lattice,
	title={The lattice Boltzmann equation: theory and applications},
	author={Benzi, Roberto and Succi, Sauro and Vergassola, Massimo},
	journal={Physics Reports},
	volume={222},
	number={3},
	pages={145--197},
	year={1992},
	publisher={Elsevier}
}

@article{boghosian1999lattice,
	title = {Lattice gases and cellular automata},
	journal = {Future Generation Computer Systems},
	volume = {16},
	number = {2},
	pages = {171-185},
	year = {1999},
	issn = {0167-739X},
	doi = {https://doi.org/10.1016/S0167-739X(99)00045-X},
	url = {https://www.sciencedirect.com/science/article/pii/S0167739X9900045X},
	author = {Bruce M. Boghosian},
}

@article{ladd1994numerical,
  title={Numerical simulations of particulate suspensions via a discretized Boltzmann equation. Part 1. Theoretical foundation},
  author={Ladd, Anthony JC},
  journal={Journal of fluid mechanics},
  volume={271},
  pages={285--309},
  year={1994},
  publisher={Cambridge University Press}
}

@article{ladd1994numericalp2,
  title={Numerical simulations of particulate suspensions via a discretized Boltzmann equation. Part 2. Numerical results},
  author={Ladd, Anthony JC},
  journal={Journal of fluid mechanics},
  volume={271},
  pages={311--339},
  year={1994},
  publisher={Cambridge University Press}
}

@article{ruiz2017quantum,
  title={Quantum arithmetic with the quantum Fourier transform},
  author={Ruiz-Perez, Lidia and Garcia-Escartin, Juan Carlos},
  journal={Quantum Information Processing},
  volume={16},
  pages={1--14},
  year={2017},
  publisher={Springer}
}

@article{draper2000addition,
  title={Addition on a quantum computer},
  author={Draper, Thomas G},
  journal={arXiv preprint quant-ph/0008033},
  year={2000}
}

@book{toffoli1987cellular,
	title={Cellular automata machines: a new environment for modeling},
	author={Toffoli, Tommaso and Margolus, Norman},
	year={1987},
	publisher={MIT press}
}

@article{wang2025quantum,
	title={Quantum lattice Boltzmann method for simulating nonlinear fluid dynamics},
	author={Wang, Boyuan and Meng, Zhaoyuan and Zhao, Yaomin and Yang, Yue},
	journal={arXiv preprint arXiv:2502.16568},
	year={2025}
}

@article{mani2023perspective,
	title={A perspective on the state of aerospace computational fluid dynamics technology},
	author={Mani, Mori and Dorgan, Andrew J},
	journal={Annual Review of Fluid Mechanics},
	volume={55},
	number={1},
	pages={431--457},
	year={2023},
	publisher={Annual Reviews}
}

@article{zingaro2024electromechanics,
	title={An electromechanics-driven fluid dynamics model for the simulation of the whole human heart},
	author={Zingaro, Alberto and Bucelli, Michele and Piersanti, Roberto and Regazzoni, Francesco and Quarteroni, Alfio and others},
	journal={Journal of Computational Physics},
	volume={504},
	pages={112885},
	year={2024},
	publisher={Elsevier}
}

@article{wijesooriya2023technical,
	title={A technical review of computational fluid dynamics (CFD) applications on wind design of tall buildings and structures: Past, present and future},
	author={Wijesooriya, Kasun and Mohotti, Damith and Lee, Chi-King and Mendis, Priyan},
	journal={Journal of Building Engineering},
	volume={74},
	pages={106828},
	year={2023},
	publisher={Elsevier}
}

@article{theis2017end,
	title={The end of moore's law: A new beginning for information technology},
	author={Theis, Thomas N and Wong, H-S Philip},
	journal={Computing in science \& engineering},
	volume={19},
	number={2},
	pages={41--50},
	year={2017},
	publisher={IEEE}
}

@book{v1966theory,
	title={Theory of self-reproducing automata},
	author={v Neumann, John and Burks, Arthur W},
	year={1966},
	publisher={University of Illinois Press Urbana}
}

@misc{chopard2012cellular,
	title={Cellular Automata and Lattice Boltzmann Modeling of Physical Systems.},
	author={Chopard, Bastien},
	year={2012}
}

@article{feynman1982simulating,
	title={Simulating physics with computers},
	author={Feynman, Richard P},
	journal={International Journal of Theoretical Physics},
	volume={21},
	pages={467–488},
	year={1982},
	doi={10.1007/BF02650179}
}

@article{meyer1996unitarity,
	title={Unitarity in one dimensional nonlinear quantum cellular automata},
	author={Meyer, David A},
	journal={arXiv preprint quant-ph/9605023},
	year={1996}
}

@article{meyer1996quantum,
	title={From quantum cellular automata to quantum lattice gases},
	author={Meyer, David A},
	journal={Journal of Statistical Physics},
	volume={85},
	pages={551--574},
	year={1996},
	publisher={Springer}
}

@article{meyer1997quantuma,
	title={Quantum mechanics of lattice gas automata: One-particle plane waves and potentials},
	author={Meyer, David A},
	journal={Physical Review E},
	volume={55},
	number={5},
	pages={5261},
	year={1997},
	publisher={APS}
}

@article{meyer1997quantumb,
	title={Quantum lattice gases and their invariants},
	author={Meyer, David A},
	journal={International Journal of Modern Physics C},
	volume={8},
	number={04},
	pages={717--735},
	year={1997},
	publisher={World Scientific}
}

@article{meyer1998quantum,
	title={Quantum mechanics of lattice gas automata: boundary conditions and other inhomogeneities},
	author={Meyer, David A},
	journal={Journal of Physics A: Mathematical and General},
	volume={31},
	number={10},
	pages={2321},
	year={1998},
	publisher={IOP Publishing}
}

@article{succi1993lattice,
	title={Lattice Boltzmann equation for quantum mechanics},
	author={Succi, Sauro and Benzi, Roberto},
	journal={Physica D: Nonlinear Phenomena},
	volume={69},
	number={3-4},
	pages={327--332},
	year={1993},
	publisher={Elsevier}
}

@article{succi1996numerical,
	title = {Numerical solution of the Schr\"odinger equation using discrete kinetic theory},
	author = {Succi, Sauro},
	journal = {Phys. Rev. E},
	volume = {53},
	issue = {2},
	pages = {1969--1975},
	numpages = {0},
	year = {1996},
	month = {Feb},
	publisher = {American Physical Society},
	doi = {10.1103/PhysRevE.53.1969},
	url = {https://link.aps.org/doi/10.1103/PhysRevE.53.1969}
}

@inproceedings{yepez1998quantum,
	title={Quantum computation of fluid dynamics},
	author={Yepez, Jeffrey},
	booktitle={NASA International Conference on Quantum Computing and Quantum Communications},
	pages={34--60},
	year={1998},
	organization={Springer}
}

@article{love2019quantum,
	title={On quantum extensions of hydrodynamic lattice gas automata},
	author={Love, Peter},
	journal={Condensed Matter},
	volume={4},
	number={2},
	pages={48},
	year={2019},
	publisher={MDPI}
}

@article{zamora2025efficient,
	title={Efficient quantum lattice gas automata},
	author={Zamora, Antonio David Bastida and Budinski, Ljubomir and Niemim{\"a}ki, Ossi and Lahtinen, Valtteri},
	journal={Computers \& Fluids},
	volume={286},
	pages={106476},
	year={2025},
	publisher={Elsevier}
}

@article{zamora2025float,
  title={Lattice gas automata with floating-point numbers: A connection between molecular dynamics and lattice Boltzmann method for quantum computers},
  author={Zamora, Antonio David Bastida and Budinski, Ljubomir and Sagaut, Pierre and Lahtinen, Valtteri},
  journal={Physical Review E},
  volume={112},
  number={1},
  pages={015305},
  year={2025},
  publisher={APS}
}

@article{blommel2018integer,
	title={Integer lattice gas with Monte Carlo collision operator recovers the lattice Boltzmann method with Poisson-distributed fluctuations},
	author={Blommel, Thomas and Wagner, Alexander J},
	journal={Physical Review E},
	volume={97},
	number={2},
	pages={023310},
	year={2018},
	publisher={APS}
}

@article{fonio2025adaptive,
  title={Adaptive lattice-gas algorithm: Classical and quantum implementations},
  author={Fonio, Niccol{\`o} and Sagaut, Pierre and Budinski, Ljubomir and Lahtinen, Valtteri},
  journal={Physical Review E},
  volume={112},
  number={3},
  pages={035302},
  year={2025},
  publisher={APS}
}

@article{kocherla2024fully,
	title={Fully quantum algorithm for mesoscale fluid simulations with application to partial differential equations},
	author={Kocherla, Sriharsha and Song, Zhixin and Chrit, Fatima Ezahra and Gard, Bryan and Dumitrescu, Eugene F and Alexeev, Alexander and Bryngelson, Spencer H},
	journal={AVS Quantum Science},
	volume={6},
	number={3},
	year={2024},
	publisher={AIP Publishing}
}

@article{hardy1973time,
	title={Time evolution of a two-dimensional model system. I. Invariant states and time correlation functions},
	author={Hardy, J and Pomeau, Yves and De Pazzis, O},
	journal={Journal of Mathematical Physics},
	volume={14},
	number={12},
	pages={1746--1759},
	year={1973},
	publisher={American Institute of Physics}
}

@article{hardy1976molecular,
	title={Molecular dynamics of a classical lattice gas: Transport properties and time correlation functions},
	author={Hardy, J and De Pazzis, O and Pomeau, Yves},
	journal={Physical review A},
	volume={13},
	number={5},
	pages={1949},
	year={1976},
	publisher={APS}
}

@article{shakeel2013quantum,
	title={When is a quantum cellular automaton (QCA) a quantum lattice gas automaton (QLGA)?},
	author={Shakeel, Asif and Love, Peter J},
	journal={Journal of Mathematical Physics},
	volume={54},
	number={9},
	year={2013},
	publisher={AIP Publishing}
}

@article{farrelly2020review,
	title={A review of quantum cellular automata},
	author={Farrelly, Terry},
	journal={Quantum},
	volume={4},
	pages={368},
	year={2020},
	publisher={Verein zur F{\"o}rderung des Open Access Publizierens in den Quantenwissenschaften}
}

@article{budinski2023efficient,
	title={Efficient parallelization of quantum basis state shift},
	author={Budinski, Ljubomir and Niemim{\"a}ki, Ossi and Zamora-Zamora, Roberto and Lahtinen, Valtteri},
	journal={Quantum Science and Technology},
	volume={8},
	number={4},
	pages={045031},
	year={2023},
	publisher={IOP Publishing}
}

@article{raissi2019physics,
	title={Physics-informed neural networks: A deep learning framework for solving forward and inverse problems involving nonlinear partial differential equations},
	author={Raissi, Maziar and Perdikaris, Paris and Karniadakis, George E},
	journal={Journal of Computational physics},
	volume={378},
	pages={686--707},
	year={2019},
	publisher={Elsevier}
}

@article{berger2025trainable,
	title={Trainable embedding quantum physics informed neural networks for solving nonlinear PDEs},
	author={Berger, Stefan and Hosters, Norbert and M{\"o}ller, Matthias},
	journal={Scientific Reports},
	volume={15},
	number={1},
	pages={1--14},
	year={2025},
	publisher={Nature Publishing Group}
}

@article{panichi2025quantum,
	title={Quantum physics informed neural networks for multi-variable partial differential equations},
	author={Panichi, Giorgio and Corli, Sebastiano and Prati, Enrico},
	journal={arXiv preprint arXiv:2503.12244},
	year={2025}
}

@article{boghosian1997quantum,
	title={Quantum lattice-gas models for the many-body Schr{\"o}dinger equation},
	author={Boghosian, Bruce M and Taylor, Washington},
	journal={International Journal of Modern Physics C},
	volume={8},
	number={04},
	pages={705--716},
	year={1997},
	publisher={World Scientific}
}

@article{boghosian1998simulating,
	title={Simulating quantum mechanics on a quantum computer},
	author={Boghosian, Bruce M and Taylor IV, Washington},
	journal={Physica D: Nonlinear Phenomena},
	volume={120},
	number={1-2},
	pages={30--42},
	year={1998},
	publisher={Elsevier}
}

@software{georgescu2025qlgabuildingblocksreplication,
	author       = {Georgescu, Calin A. and
	Schalkers, Merel A. and
	Möller, Matthias},
	title        = {Replication Package for "Fully Quantum Lattice Gas
	Automata Building Blocks for Computational Basis
	State Encodings"
	},
	month        = jun,
	year         = 2025,
	publisher    = {Zenodo},
	doi          = {10.5281/zenodo.15636196},
	url          = {https://doi.org/10.5281/zenodo.15636196},
}

@article{kitaev1995quantum,
  title={Quantum measurements and the Abelian stabilizer problem},
  author={Kitaev, A Yu},
  journal={arXiv preprint quant-ph/9511026},
  year={1995}
}

@article{mosca2004exact,
  title={Exact quantum Fourier transforms and discrete logarithm algorithms},
  author={Mosca, Michele and Zalka, Christof},
  journal={International Journal of Quantum Information},
  volume={2},
  number={01},
  pages={91--100},
  year={2004},
  publisher={World Scientific}
}

@article{kempe2003quantum,
  title={Quantum random walks: an introductory overview},
  author={Kempe, Julia},
  journal={Contemporary Physics},
  volume={44},
  number={4},
  pages={307--327},
  year={2003},
  publisher={Taylor \& Francis}
}
\end{document}